\documentclass[iop]{emulateapj}
\usepackage{natbib,apjfonts,graphicx}
\newcommand{\HI}{\ion{H}{1}}
\newcommand{\HII}{\ion{H}{2}}

\newcommand{\kms}{\mbox{km~s$^{-1}$}}
\newcommand{\Msol}{\mbox{M$_\odot$}}

\newcommand{\kkms}{\mbox{K~km~s$^{-1}$}}
\newcommand{\kkmsas}{\mbox{K~km~s$^{-1}$~arcsec$^2$}}
\newcommand{\lcou}{\mbox{K~km~s$^{-1}$~pc$^2$}}
\newcommand{\xcou}{\mbox{cm$^{-2}$~(K~km~s$^{-1}$)$^{-1}$}}

\newcommand{\xco}{\mbox{$X_{\rm CO}$}}

\shorttitle{Magellanic Mopra Assessment. I.}
\slugcomment{Updated \today}
\journalinfo{To appear in the Astrophysical Journal Supplement}

\begin{document}

\title{The Magellanic Mopra Assessment (MAGMA). I. The Molecular Cloud Population of the Large Magellanic Cloud}

\author{
Tony Wong\altaffilmark{1,2}, 
Annie Hughes\altaffilmark{3,4,5}, 
J\"urgen Ott\altaffilmark{6}, 
Erik Muller\altaffilmark{7,8}, 
Jorge L. Pineda\altaffilmark{9}, 
Jean-Philippe Bernard\altaffilmark{10,11}, 
You-Hua Chu\altaffilmark{1}, 
Yasuo Fukui\altaffilmark{7}, 
Robert A. Gruendl\altaffilmark{1}, 
Christian Henkel\altaffilmark{12,13}, 
Akiko Kawamura\altaffilmark{7,8}, 
Ulrich Klein\altaffilmark{14}, 
Leslie W. Looney\altaffilmark{1}, 
Sarah Maddison\altaffilmark{3}, 
Yoji Mizuno\altaffilmark{7}, 
Deborah Paradis\altaffilmark{10,11},
Jonathan Seale\altaffilmark{1,15}, 
\&
Daniel E. Welty\altaffilmark{1}
}
\altaffiltext{1}{Astronomy Department, University of Illinois, Urbana, IL 61801, USA}
\altaffiltext{2}{E-mail: wongt@astro.illinois.edu}
\altaffiltext{3}{Centre for Astrophysics and Supercomputing, Swinburne University of Technology, Hawthorn, VIC 3122, Australia}
\altaffiltext{4}{CSIRO Astronomy and Space Science, PO Box 76, Epping, NSW 1710, Australia}
\altaffiltext{5}{Max-Planck-Institut f\"ur Astronomie, K\"onigstuhl 17, D-69117, Heidelberg, Germany}
\altaffiltext{6}{National Radio Astronomy Observatory, PO Box O, 1003 Lopezville Road, Socorro, NM 87801, USA}
\altaffiltext{7}{Department of Astrophysics, Nagoya University, Furo-cho, Chikusa-ku, Nagoya 464-8602, Japan}
\altaffiltext{8}{ALMA-J Project Office, National Astronomical Observatory of Japan, 2-21-1 Osawa, Mitaka, Tokyo 181-8588, Japan}
\altaffiltext{9}{Jet Propulsion Laboratory, California Institute of Technology, 4800 Oak Grove Drive, Pasadena, CA 91109-8099}
\altaffiltext{10}{CNRS, IRAP, 9 Av.\ Colonel Roche, BP 44346, F-31028 Toulouse cedex 4, France}
\altaffiltext{11}{Universit\'{e} de Toulouse, UPS-OMP, IRAP, F-31028 Toulouse cedex 4, France}
\altaffiltext{12}{Max-Planck-Institut f\"ur Radioastronomie, Auf dem H\"ugel 69, D-53121 Bonn, Germany}
\altaffiltext{13}{Astronomy Department, Faculty of Science, King Abdulaziz University, P.O. Box 80203, Jeddah, Saudi Arabia}
\altaffiltext{14}{Argelander Institut f\"ur Astronomie, Universit\"at Bonn, Auf dem H\"ugel 71, D-53121 Bonn, Germany}
\altaffiltext{15}{Space Telescope Science Institute, 3700 San Martin Drive, Baltimore, MD 21218, USA}

\begin{abstract}

We present the properties of an extensive sample of molecular clouds in the Large Magellanic Cloud (LMC) mapped at 11 pc resolution in the CO(1--0) line.  Targets were chosen based on a limiting CO flux and peak brightness as measured by the NANTEN survey.  The observations were conducted with the ATNF Mopra Telescope as part of the Magellanic Mopra Assessment (MAGMA).  We identify clouds as regions of connected CO emission, and find that the distributions of cloud sizes, fluxes and masses are sensitive to the choice of decomposition parameters.  In all cases, however, the luminosity function of CO clouds is steeper than $dN/dL \propto L^{-2}$, suggesting that a substantial fraction of mass is in low-mass clouds.  A correlation between size and linewidth, while apparent for the largest emission structures, breaks down when those structures are decomposed into smaller structures.  We argue that the correlation between virial mass and CO luminosity is the result of comparing two covariant quantities, with the correlation appearing tighter on larger scales where a size-linewidth relation holds.  The virial parameter (the ratio of a cloud's kinetic to self-gravitational energy) shows a wide range of values and exhibits no clear trends with the CO luminosity or the likelihood of hosting young stellar object (YSO) candidates, casting further doubt on the assumption of virialization for molecular clouds in the LMC.  Higher CO luminosity increases the likelihood of a cloud harboring a YSO candidate, and more luminous YSOs are more likely to be coincident with detectable CO emission, confirming the close link between giant molecular clouds and massive star formation.

\end{abstract}


\section{Introduction}

Recent surveys of CO in Local Group galaxies have achieved sufficient resolution to resolve individual molecular clouds \citep[see reviews by][]{Blitz:07,Fukui:10}.  These studies represent an important step towards connecting our understanding of star formation in the Milky Way to more distant and less evolved systems.  Among the questions that these studies seek to address are whether molecular cloud characteristics, and the conditions for their formation, follow universal patterns, and what the resulting implications for star formation may be.

The Large Magellanic Cloud (LMC) is the nearest star-forming galaxy to our own ($d \approx 50$ kpc), yet the distribution of its CO emission has been revealed in detail only recently, due to the LMC's large angular size and the paucity of millimeter-wave facilities in the southern hemisphere.  The first complete CO map, at a (smoothed) resolution of 12\arcmin, was published by \citet{Cohen:88}, based on observations with the 1.2-m Columbia Millimeter-Wave Telescope.  Similar spatial coverage was later obtained by the 4-m NANTEN telescope, at a resolution of 2\farcm6; results of the initial NANTEN survey were reported by \citet{Fukui:99} and \citet{Mizuno:01}, and results of a second, more sensitive survey have been recently published by \citet[][hereafter Fu08]{Fukui:08}.  In addition to the large-scale surveys, detailed mapping of individual cloud complexes (at $\sim$1\arcmin\ resolution, corresponding to 15 pc) has been pursued with the Swedish-ESO Submillimetre Telescope (SEST) to explore the sizes, dynamical state, and molecular and isotopic abundances of individual clouds \citep[e.g.,][]{Israel:93,Kutner:97,Johansson:98,Israel:03}.

The new data set presented in this paper arises from a natural extension of the NANTEN surveys: the targeted mapping of known CO complexes in the LMC at improved angular resolution (45\arcsec, or 11 pc at a distance of 50 kpc), adequate to resolve the largest giant molecular clouds (GMCs).  These maps are a principal outcome of the Magellanic Mopra Assessment (MAGMA), a long-term project of CO mapping of both the LMC and the Small Magellanic Cloud (SMC) with the Mopra Telescope.  Initial results from MAGMA have been presented by \citet{Pineda:09} and \citet[][hereafter Hu10]{Hughes:10} for the LMC and \citet{Muller:10} for the SMC.  In the current work we present the completed LMC data set and three nested catalogs of GMCs, obtained using different assumptions for identifying discrete clouds.  We compare the cloud properties and scaling relations derived from these catalogs with previous work and discuss the evidence for and against clouds being in a state of virial equilibrium.  Although our final catalogs differ from the one used in Hu10, the distributions of cloud properties do not change dramatically, and in the present paper we do not revisit the relationships between GMC properties and environmental conditions (such as stellar surface densities, interstellar pressure, and far-ultraviolet radiation field) examined in Hu10.  We expect substantially improved ancillary (non-CO) data to become available in the near future, at which point we intend to re-examine those relationships.

As one of the initial applications of the MAGMA data set, we examine the correspondence between CO clouds and known or suspected young stellar objects (YSOs).  While stars are believed to be formed in molecular clouds, the relationship between molecular gas and star formation has been difficult to analyze in external galaxies because of observational biases.  For example, while H$\alpha$ is commonly used to trace recent star formation, it is susceptible to extinction and therefore the correlation between CO and H$\alpha$ may break down on scales approaching that of molecular cloud complexes.  At infrared wavelengths, even deeply embedded young stars are detectable if they are sufficiently massive, leading to two recent attempts to obtain a census of massive YSOs in the LMC based on {\it Spitzer Space Telescope} data \citep{Whitney:08,Gruendl:09}.  As the reliability and completeness of such catalogs improve, it should be possible to measure the star formation efficiency and lifetimes of molecular clouds based on number counts of YSOs rather than integrated fluxes of dust or gas \citep[e.g.,][]{Indebetouw:08,Chen:10}.

We emphasize that, while we will generally refer to the sources we detect as ``molecular clouds,'' the results presented in this paper refer only to the CO-emitting molecular clouds.  The reliability of CO as a tracer of H$_2$ has been a topic of active discussion \citep[e.g.,][]{Leroy:07,Bot:07}.  Unfortunately, the bulk of the H$_2$ is at temperatures too low to produce detectable emission, so it can be probed only through absorption or through indirect emission tracers such as CO, dust and gamma rays.  {\it Far Ultraviolet Spectroscopic Explorer} ({\it FUSE}) observations of H$_2$ in absorption towards LMC targets suggest a total {\it diffuse} H$_2$ mass of $8\times10^{6}$\,\Msol\ \citep{Tumlinson:02}, $\lesssim2$\% of the LMC's atomic gas mass.  However, the presence of a significant mass of H$_2$ residing in CO-dark envelopes surrounding LMC molecular clouds is not strongly excluded by this result, as the \citet{Tumlinson:02} sample contains predominantly sightlines with low extinction.  Indeed, the ratio of CO emission to H$_2$ mass is likely to decrease on larger spatial scales \citep{Israel:00}, and studies of LMC dust emission at far-infrared wavelengths \citep[e.g.,][]{Bernard:08,Roman:10} suggest the presence of more dust than can be accounted for using standard conversion factors for \HI\ and CO emission.  A problem for these studies is whether to attribute variations in the {\it inferred} dust-to-gas ratio to variations in dust abundance (or emissivity) or to incorrect estimation of gas masses.  In any case, there is general agreement that CO emission traces regions where hydrogen is predominantly molecular, although it may not identify all such regions.

We begin with an overview of the survey strategy, the observational parameters, and data reduction procedures in Section \ref{sec:obs}.  
The identification of clouds, their derived properties and a comparison with YSO catalogs are presented in Section \ref{sec:results}.  Finally, Section \ref{sec:disc} presents a discussion and Section \ref{sec:conc} a summary of our results.

\section{The MAGMA Survey of the LMC}\label{sec:obs}

\subsection{Survey Design}

The MAGMA survey area was limited in total to roughly 3.6 deg$^2$ by the need to staff the observations and complete them in a reasonable amount of time.
Thus, we chose to map regions coincident with, or close to, known CO clouds detected by NANTEN.  
The second NANTEN survey (Fu08) detected 272 clouds, of which 230 were detected in more than 2 grid positions (spaced 2\arcmin\ apart) and termed ``giant molecular clouds.''
Hereafter we refer to these 230 objects as NANTEN {\it GMCs}, and the larger sample of 272 clouds simply as NANTEN {\it clouds}.
When the scope of MAGMA was expanded from limited regions in the eastern and central part of the LMC to the entire galaxy, we adopted both flux and brightness criteria for NANTEN GMCs to be included in our survey.
A minimum extrapolated CO flux of $1.2 \times 10^5$ \kkmsas\ (equivalent to a CO luminosity of 7000 \kkms\ pc$^2$ at the distance of the LMC) was chosen in order to ensure that the largest GMCs were included.
(Details of the flux extrapolation used by Fu08 and this paper are given in Section~\ref{sec:sigdet}.)
In addition, a minimum peak CO brightness of 1 \kkms\ was chosen to exclude a handful of clouds (Nos.\ 34, 38, 78, \& 138 in Fu08) which met the flux requirement due to a large extrapolated size but were deemed unlikely to be detected by Mopra.
Our fiducial sample was thus limited to 114 NANTEN GMCs, although in fact many of the remaining GMCs ended up in the MAGMA field because of their proximity to survey targets.
A detailed comparison of the MAGMA and NANTEN samples is given in Section~\ref{sec:nanten}.

\begin{figure*}
\begin{center}
\includegraphics[angle=-90,width=\textwidth]{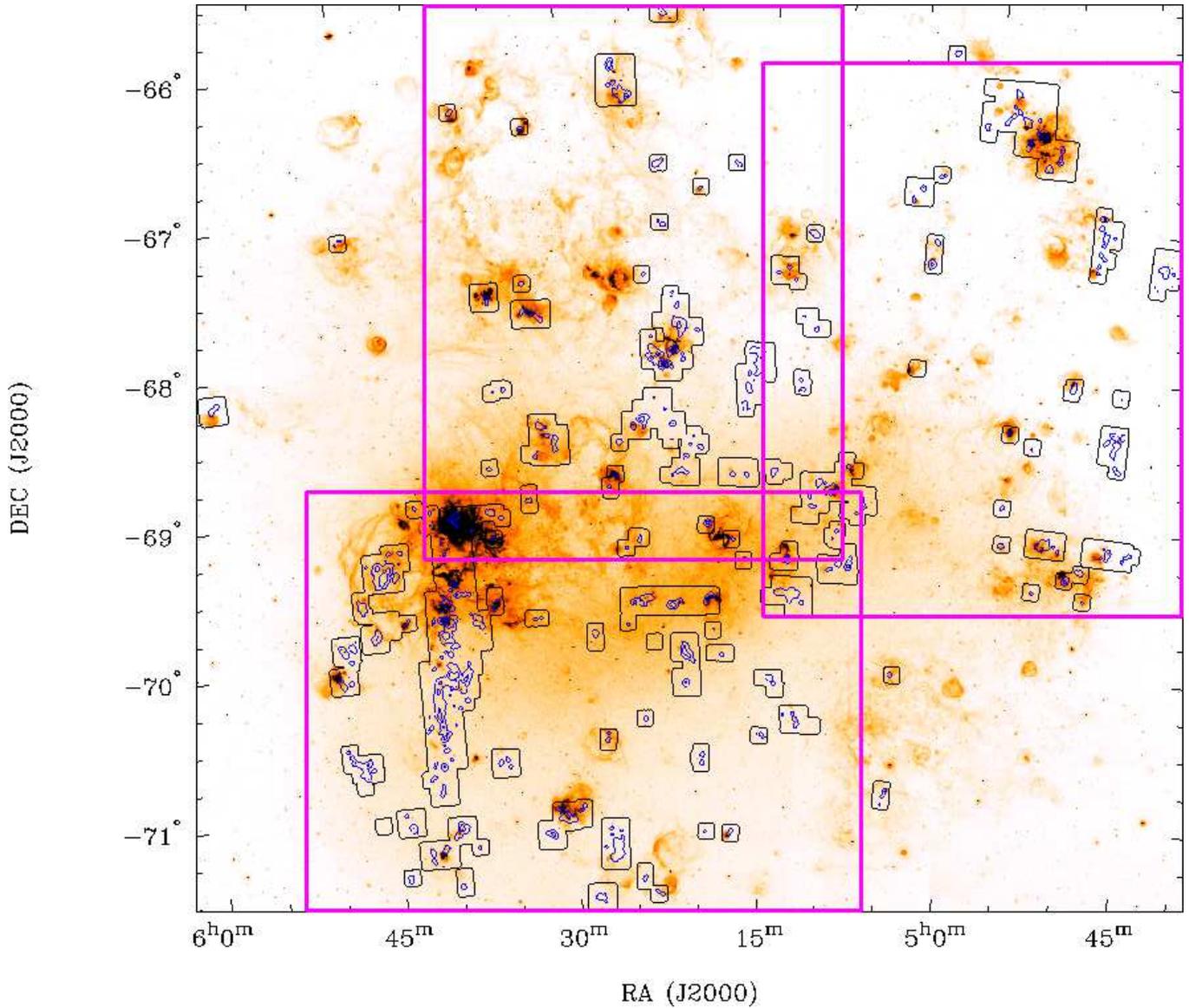}
\end{center}
\caption{
CO intensity contours from MAGMA, at levels of 2 and 10 K \kms, overlaid on an H$\alpha$ image from MCELS \citep{Smith:99}, shown on a logarithmic stretch to bring out fainter emission.  The CO image was derived by smoothing the MAGMA cube to 1\arcmin\ resolution and applying a blanking mask corresponding to the 3$\sigma$ contour of a 1.5\arcmin-resolution smoothed cube.  Multiple small rectangular contours show the regions mapped in CO with Mopra. Large rectangular boxes show the regions magnified in Figures~\ref{fig:zoom1} and \ref{fig:zoom2}.  The H$\alpha$ image has not been continuum subtracted, although bright point sources have been removed by a moving boxcar median filter.
\label{fig:magmaha}}
\end{figure*}

\subsection{Observations and Data Processing}

Observations were performed with the ATNF\footnote{The Australia Telescope is funded by the Commonwealth of Australia for operation as a National Facility managed by CSIRO.} Mopra 22-m telescope from 2005 May to 2010 October.
We divided our target regions into square 5\farcm2 $\times$ 5\farcm2 patches whose centers were separated by 4\farcm75, ensuring significant overlap between adjacent patches.
Each patch was mapped at least twice in the On-The-Fly (OTF) observing mode, in which spectra are taken continuously as the telescope is scanned across the sky.
Scanning is conducted in either right ascension or declination; to minimize scanning artifacts we alternated the scan direction when repeating a patch.
Along each row of an OTF field, individual spectra are recorded every 7\arcsec, so that the 33\arcsec\ (FWHM) telescope beam is oversampled in the scanning direction. 
The spacing between rows is 10\arcsec, also oversampling the beam.
In our standard configuration, an OTF map consists of 31 rows with 44 2-second samples taken per row, so each map contains about 2700 spectra (counting both polarizations separately).  
A 20-second off-source spectrum, taken towards a region within 1\arcdeg\ that shows little \HI\ emission ($\int T_{\rm B}\,dv \lesssim 280$ K \kms, corresponding to $N_{\rm H} \lesssim 5 \times 10^{20}$ cm$^{-2}$), is observed before each row, and the above-atmosphere system temperature is estimated every 25 minutes using an ambient (hot) load.
Typical system temperatures for the survey observations are between 500 and 600 K; observing is abandoned when system temperatures exceed $\sim$1000 K.
Fluctuations in system temperature between hot load measurements are monitored using a rapidly switched noise source injected at the front end of the receiver.
A single OTF map takes about 75 minutes to complete.
After each map, the pointing of the antenna is verified and updated by observing the nearby SiO maser R Dor. Prior to correction, the pointing errors are typically below 10\arcsec.
The final MAGMA CO data cube includes data from 1163 OTF maps with approximately 3\,100\,000 individual spectra.  
Figure~\ref{fig:magmaha} shows the spatial coverage of the MAGMA data set in comparison to an H$\alpha$ image from the Magellanic Clouds Emission-Line Survey (MCELS; \citealt{Smith:99}). 

Data reduction was performed using the ATNF's MIRIAD, {\it Livedata}, and {\it Gridzilla} software packages.
{\it Livedata} performs bandpass calibration using off-source spectra, fits and subtracts a linear baseline, and applies additional calibration factors based on assumed telescope efficiencies, which we derived from an analysis of standard spectra taken towards Orion KL \citep[cf.][]{Ladd:05}.
The efficiencies we have assumed for CO are listed in Table~\ref{tbl:eff}, and represent the ``extended beam'' efficiency $\eta_{\rm xb}$ discussed by \citet{Ladd:05}, which includes the effect of coupling to the inner error beam for sources larger than $\sim$2\arcmin.
We determined $\eta_{\rm xb}$ for each standard spectrum by dividing the observed peak antenna temperature by 100~K, which is the corrected peak CO brightness temperature for Orion KL given by \citet{Ladd:05}.
An average value for $\eta_{\rm xb}$ was then calculated for each observing season with a typical rms deviation of $\sim$10\%.
Note that some of the apparent changes in $\eta_{\rm xb}$ likely reflect changes in the system temperature scale due to instrument modifications rather than changes in the aperture efficiency.
{\it Gridzilla} takes the calibrated spectra and grids them onto a data cube using a Gaussian smoothing function of FWHM 33\arcsec, comparable to that of the Mopra primary beam.
We weight the spectra by the inverse of the system temperature when gridding.
Based on simulations of the gridding process we estimate that the final resolution of our maps is close to 45\arcsec.

Changes in the spectrometers available at Mopra between 2005 and 2006 have resulted in varying spectral resolution and bandwidth over the period of our observations.  
The spectral resolution varied from 33.6 to 62.4 kHz in 2005, so data from that year were resampled by spline interpolation to the resolution of 33.68 kHz (0.09 \kms) available in later years using the MIRIAD task {\sc mopfix} before processing with {\it Livedata} and {\it Gridzilla}.
Since all of our analysis is performed at a much coarser resolution (0.5 \kms), this resampling has little effect on the final data products.
{\it Gridzilla} generated a data cube with 1599 channels spanning radial velocities of 180 to 320 \kms\ in the LSR frame; we subsequently binned the data in velocity, reducing the number of channels by a factor of 6.
To summarize, our final output cube, produced using an orthographic projection, has an effective spatial resolution of 45\arcsec, sampled onto a grid spacing of 15\arcsec, and a channel spacing of 0.526 \kms.

\begin{table}
\begin{center}
\caption{Adopted Efficiencies for the Mopra Telescope\label{tbl:eff}}
\begin{tabular}{lccc}
\hline
Start Date & End Date & $\eta_{\rm xb}$\\\hline
2005-05-29 & 2005-10-10 & 0.54\\
2005-10-20 & 2005-10-29 & 0.38\\
2006-06-25 & 2006-10-27 & 0.38\\
2007-07-01 & 2007-09-30 & 0.45\\
2008-06-19 & 2008-06-29 & 0.39\\
2008-08-17 & 2008-09-30 & 0.48\\
2009-06-15 & 2009-09-30 & 0.48\\
2010-07-26 & 2010-10-10 & 0.43\\
\hline
\end{tabular}
\end{center}
\end{table}

%

\begin{figure*}
\begin{center}
\includegraphics[width=\textwidth]{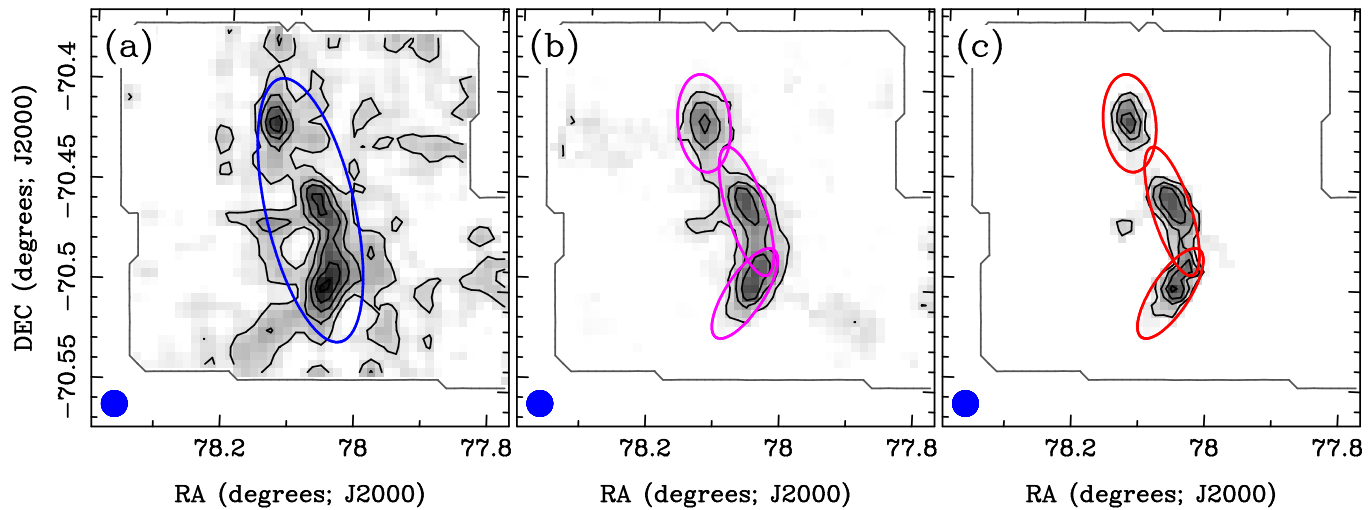}
\end{center}
\caption{
Integrated CO intensity maps for a relatively isolated GMC (No.\ 69 in Fu08) produced using three different methods: (a) Direct integration of a 10 \kms\ window centered on the emission; (b) masking using a 3$\sigma$ threshold applied to a smoothed cube; (c) signal detection with CPROPS as described in the text.  Contour levels range from 2 to 10 K \kms\ in increments of 2 K \kms.  The ellipses in each panel (left to right) indicate the extrapolated sizes and orientations of structures identified by CPROPS using the ``islands'', ``physical'', and ``data-based'' decomposition parameters respectively; note that these structures have been identified in the data cube and not in the CO intensity maps shown.  The Mopra beam size (Gaussian FWHM) is represented by the blue circle at the lower left of each panel.
\label{fig:nt69}}
\end{figure*}

\subsection{Integrated Intensity Images}\label{sec:mom0}

We explored three different methods to construct CO intensity images.  The first, and most straightforward, is {\it direct integration} over the expected velocity range.
This results in excessive noise when applied to the entire galaxy, although limited application to small regions of the LMC is generally more successful [see Figure~\ref{fig:nt69}(a)], because smaller velocity ranges can be chosen.
The second and third methods involve integration over a mask that isolates significant emission from noise.
For the second method, which we call the {\it smooth-and-mask} method, the original cube is convolved with a Gaussian in the spatial domain to achieve a coarser resolution (90\arcsec), and then a 3$\sigma$ threshold level is chosen in the smoothed cube to generate a mask which is applied to the original cube.
For the third method, which we call the {\it dilated mask} method, we applied the dilated CPROPS mask described in Section~\ref{sec:sigdet} below.  This begins by identifying regions of high significance ($>$3$\sigma$) and expanding to connected regions of lower significance ($>$2$\sigma$).
Examples of applying the second and third methods to the region around a particular GMC (No.\ 69 in Fu08, also depicted in Fig.\ 2 of Hu10) are shown in Figures~\ref{fig:nt69}(b) and (c) respectively.
As discussed below (Section~\ref{sec:nanten}), the smooth-and-mask method provides the best agreement with the CO flux measured in the NANTEN cube.
We show the CO intensity derived from this method as contours in Figure~\ref{fig:magmaha}.
On the other hand, the dilated mask method, being more conservative, may be preferred in situations where the reliability of signal identification is the primary concern.

At the 45\arcsec\ resolution of our final channel maps, the RMS brightness temperature in a 0.5 \kms\ channel ranges from 0.2 to 0.5 K with a mean value of 0.3 K.  
For a typical equivalent line width of $\sqrt{2\pi}\sigma_v \sim 3$ \kms, or 6 spectral channels, this corresponds to an RMS integrated intensity of 0.4 \kkms.
Thus, for a Galactic \xco-factor of $2 \times 10^{20}\,\xcou$ \citep{Strong:96}, the 3$\sigma$ sensitivity limit is $\sim 1.2\;\kkms \approx$ 5 \Msol\ pc$^{-2}$.
This sensitivity limit is optimistic as it assumes one can correctly identify the mean CO velocity; without such prior information one would need to integrate over a wider velocity range, thus increasing the map noise.
Moreover, if the appropriate value of \xco\ is larger than the Galactic value, as is commonly assumed, the sensitivity limit would scale upward accordingly.

\begin{table}
\begin{center}
\caption{Comparison of Global CO Fluxes From Various Methods\label{tbl:coflux}}
\begin{tabular}{lccc}
\hline
Method & NANTEN data & NANTEN data & MAGMA\\
 & (full) & (MAGMA region) & data\\
\hline
Direct integration & 8.2 & 6.3 & 14\\
Smoothed 3$\sigma$ cut & 7.2 & 5.8 & 4.7\\
CPROPS islands & 5.0 & 4.7 & 3.0\\
\hline
\end{tabular}
\tablecomments{Units are $10^7$ \kkmsas.}
\end{center}
\end{table}

\begin{figure*}
\includegraphics[width=0.7\textwidth]{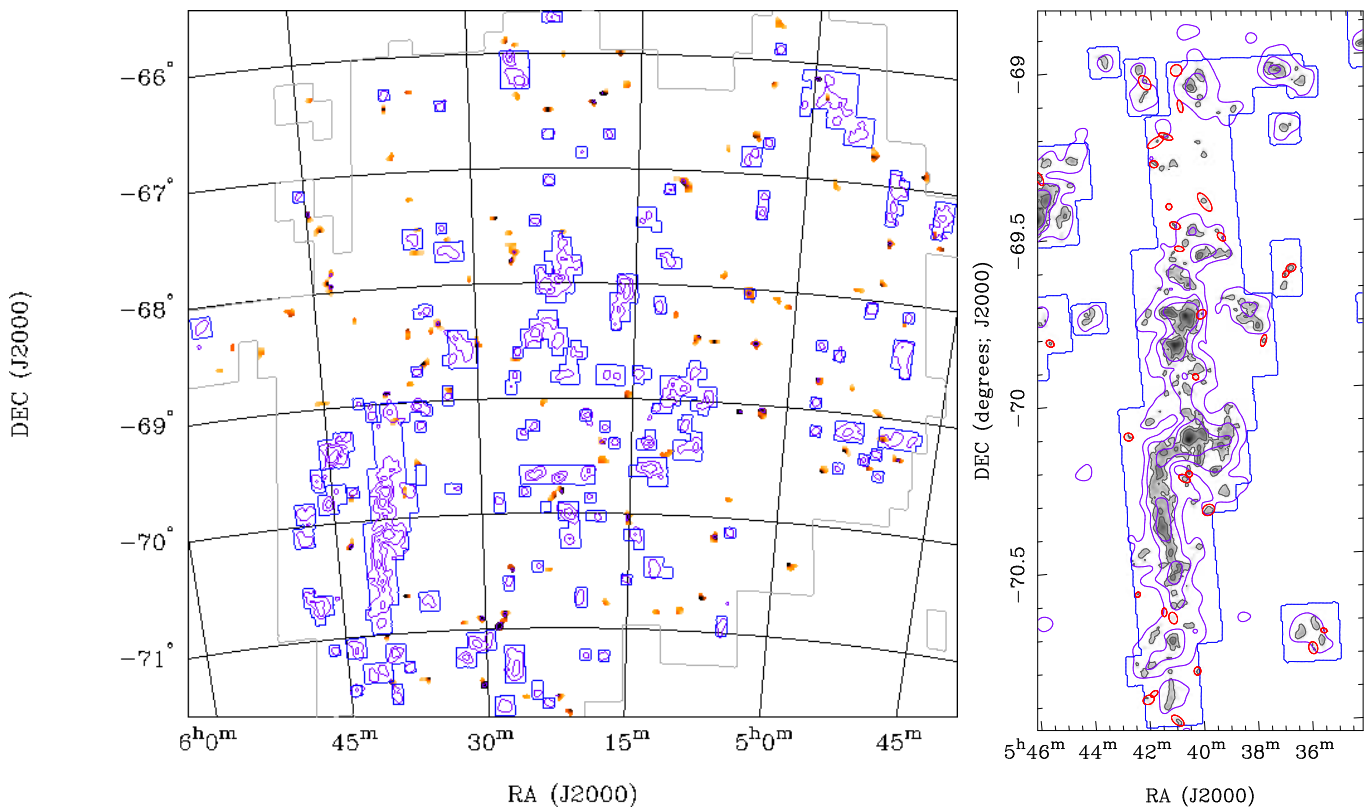}\hfill
\includegraphics[width=0.29\textwidth]{f3b.eps}
\caption{
{\it Left:} Map showing as shaded regions the NANTEN CO clouds that do not overlap with MAGMA clouds in position and velocity.  These constitute 41\% (112/272) of the NANTEN cloud sample, and primarily lie outside the regions observed with Mopra (shown as blue rectangular contours).  Violet contours show the NANTEN CO integrated intensity (at levels of 1, 4, 9, and 16 K \kms), with a light gray contour at the periphery outlining the region observed with NANTEN.\@
{\it Right:} Map showing as red ellipses the locations of MAGMA clouds near the 30 Doradus region in the southeast of the LMC that do not overlap with NANTEN clouds in position and velocity.  The MAGMA CO intensity map is shown in grayscale at contour levels of 2 and 10 K \kms, overlaid with the NANTEN CO intensity (violet contours) at levels of 0.75, 3, and 6.75 K \kms.
\label{fig:ntmpcomp}}
\end{figure*}

\subsection{Comparison with NANTEN}\label{sec:nanten}

The NANTEN CO cube presented by Fu08 contains a total flux of $8.2 \times 10^7$ \kkmsas\ when integrated directly, or $7.2 \times 10^7$ \kkmsas\ after applying a smoothed (to 6\arcmin) 3$\sigma$-contour mask.\footnote{For CO(1--0), 1 \kkmsas\ = $9.6 \times 10^{-3}$ Jy \kms\ = $3.7 \times 10^{-23}$ W m$^{-2}$.}
The corresponding fluxes, when restricted to the regions surveyed by Mopra, are $6.3 \times 10^7$ \kkmsas\ and $5.8 \times 10^7$ \kkmsas\ respectively.
Thus, both approaches show that the MAGMA field contains $\sim$80\% of the total CO flux observed by NANTEN in the LMC.\@
The Mopra CO cube contains a total flux of $1.4 \times 10^8$ \kkmsas\ when integrated directly, or $4.7 \times 10^7$ \kkmsas\ when applying a smoothed (to 1\farcm5) 3$\sigma$-contour mask.  
The surprisingly large flux obtained by direct integration of the Mopra cube appears to result from systematic errors in the spectral baseline which accumulate when summing large numbers of noise channels.
We consider that measurement unreliable, and conclude that integrating the spectra over their full velocity range will not yield reliable line fluxes.
On the other hand, the fluxes measured by NANTEN and Mopra using the smooth-and-mask technique are in reasonable agreement (to within $\sim$20\%; see Table~\ref{tbl:coflux}).

In Section~\ref{sec:sigdet} we discuss how we used a dilated mask to identify regions (``islands'') of significant emission.  
The total CO flux summed over these 450 islands is $3.0 \times 10^7$ \kkmsas.  
This represents only 64\% of the total CO flux as determined by the smooth-and-mask technique, suggesting that a significant fraction of the CO flux lies near or below the limit imposed by our CO detection criteria (cf.\ Figure~\ref{fig:nt69}).
We explored this possibility by relaxing the threshold for including emission that is contiguous with strong emission (i.e., further dilating the mask).
Lowering the threshold from 2$\sigma$ to 1.5$\sigma$ increases the flux contained with the mask by 26\%, implying that CO emission is indeed present just below our detection limit.
For comparison, Fu08 identified significant CO emission in the NANTEN cube using a dilated mask as well, with a total masked CO flux, restricted to the MAGMA observed region, of $4.7 \times 10^7$ \kkmsas.
This is 81\% of the total NANTEN-measured flux over this region, when determined using the smooth-and-mask technique, and 75\% of the NANTEN flux when determined using direct integration (Table~\ref{tbl:coflux}).

While the NANTEN survey catalogued 272 clouds, we stress again that the MAGMA survey covers a much more restricted region of sky.
For 85 of the ``islands'' identified in the MAGMA cube, there is a one-to-one correspondence with a NANTEN cloud.
An additional 275 ``islands'' correspond to 72 NANTEN clouds in such a way that more than one ``island'' overlaps with a single NANTEN cloud.
Finally there were a handful of matches (3 islands, 6 NANTEN clouds) where two or more NANTEN clouds overlap a single island.
Removing clouds common to the latter two groups, we find a total of 160 NANTEN clouds associated in both position and velocity with MAGMA islands, underscoring our earlier point that while only 114 NANTEN clouds were originally targeted, a larger number were ultimately observed.
The additional clouds are mostly lower-luminosity GMCs, although 4 are considered ``small'' clouds in Fu08.

The left panel of Figure~\ref{fig:ntmpcomp} shows in color scale the population of NANTEN clouds not observed or not detected by MAGMA, while the right panel shows as red ellipses the population of MAGMA clouds in the ``molecular ridge'' south of 30 Doradus which were not detected by NANTEN.
The effective sensitivity of the two surveys appears to be quite comparable, as evidenced by the fact that most of the NANTEN clouds not detected by MAGMA lie outside the MAGMA observed regions, and most of the MAGMA clouds not detected by NANTEN are small clouds at the periphery of NANTEN-detected clouds.
Of the original 114 targeted NANTEN GMCs, only one (No.\ 46 in Fu08) was not formally detected by MAGMA; this can be seen near $\alpha_{2000}$=5$^{\rm h}$4$^{\rm m}$,
$\delta_{2000}$=$-68\arcdeg$.
Thus, even though the MAGMA survey is a factor of $\sim$2 worse in brightness sensitivity, this is compensated by the fact that most (if not all) GMCs have unresolved structure at the NANTEN resolution and therefore appear brighter when observed with Mopra.

To summarize, based on the NANTEN data, the observed MAGMA field covers a large fraction ($\sim$80\%) of the total CO emission from the LMC, and an even larger fraction ($\sim$94\%) of the CO flux found in catalogued clouds.  
This is despite the much smaller area covered by MAGMA (3.6 deg$^2$ rather than 30 deg$^2$).  
Total CO fluxes over the MAGMA field, as determined using the smooth-and-mask method, are consistent between the NANTEN and MAGMA cubes.  
However, the fraction of the total flux that is found in CO-detected ``clouds'' is lower for MAGMA (64\%) than for NANTEN (81\%) over this same area.  
This may reflect the poorer brightness sensitivity of the Mopra observations, resulting in a significant fraction of emission lying just outside the cloud boundaries, as well as the fact that the MAGMA fields enclose the NANTEN clouds quite tightly, so that the clouds contribute a disproportionate fraction of the NANTEN flux in these regions.  
Indeed, the fraction of flux in clouds for the NANTEN survey as a whole is only $\sim$70\%.  
We therefore stress that the MAGMA survey's strength lies in the characterization of the bright CO clouds, and that due to limited spatial coverage and sensitivity it provides relatively weak constraints on the CO emission flux (let alone H$_2$ mass) of the LMC as a whole.

\section{Results}\label{sec:results}

\begin{figure}
\begin{center}
\includegraphics[width=0.45\textwidth]{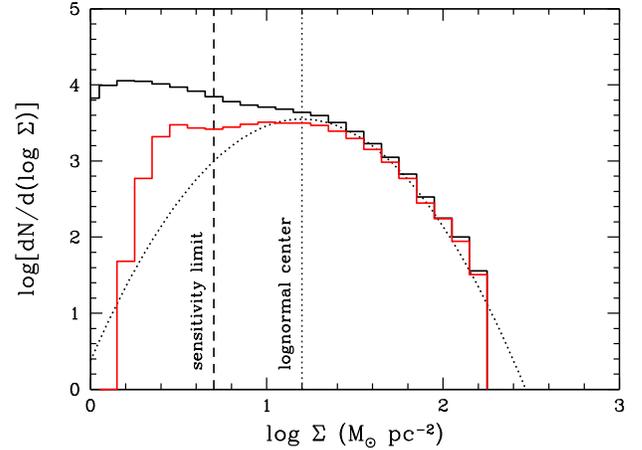}
\end{center}
\caption{
Distribution of the logarithm of the molecular gas surface density, $\Sigma$, (derived from the integrated CO intensity using a constant \xco) in the observed MAGMA field.  The black histogram shows values obtained when using the smooth-and-mask method to derive the CO intensity, while the red histogram shows values obtained using the dilated mask method.  A vertical dashed line represents our nominal 3$\sigma$ sensitivity limit of $1.2\;\kkms \approx$ 5 \Msol\ pc$^{-2}$.  The dotted parabola represents a lognormal distribution of surface density with a mean value of 16 \Msol\ pc$^{-2}$ and a dispersion above this value that is consistent with the data in the black histogram.
\label{fig:nhist}}
\end{figure}

\subsection{Column Density Distribution}\label{sec:nhist}

A histogram of CO integrated intensity across the MAGMA field can be converted into an H$_2$ column density distribution under the assumption of a constant CO-to-H$_2$ conversion factor (\xco).  Figure~\ref{fig:nhist} shows the CO intensity distribution, converted to mass surface density $\Sigma$ using $\xco = 2.0 \times 10^{20}$\,\xcou, derived from two different methods of computing the integrated CO intensity, the smooth-and-mask approach and the dilated mask approach.  The dilated mask values, given by the red histogram, are clearly more conservative as discussed in Section~\ref{sec:mom0}, and as expected are strongly suppressed below the sensitivity limit.  A lognormal function with peak position chosen by eye and a width determined by the RMS value above the peak is shown for comparison.  In the high column density regime, the distribution can be reasonably approximated as lognormal, although there may be a cutoff at the high end around $\Sigma \sim 200$ \Msol\ pc$^{-2}$.  A lognormal distribution in column density is predicted by numerical simulations of the turbulent interstellar medium \citep[e.g.,][]{Ballesteros:11}, although an excess at low column densities, such as we find in Figure~\ref{fig:nhist}, is often seen as well, representing warmer or unperturbed gas \citep{Wada:07,Ballesteros:11}.  Given the limited sensitivity and biased spatial coverage of MAGMA, the amount of mass in the low column density regime is not well constrained.

\subsection{Cloud Identification}\label{sec:sigdet}

To identify significant CO emission in the MAGMA data cubes, we have used the algorithms presented by \citet[][hereafter RL06]{Rosolowsky:06}, implemented in \textsc{IDL} as part of the \textsc{CPROPS} package. 
\textsc{CPROPS} uses a dilated mask technique to isolate regions of significant emission within spectral line cubes, and a modified watershed algorithm to assign the emission into individual clouds.  
Regions of significant emission within the MAGMA data cubes are initially identified by finding volume pixels (``voxels'') with emission greater than a $3\sigma$ threshold across two contiguous velocity channels, where the RMS noise $\sigma$ is estimated at each sky position using the median absolute deviation of the spectrum. 
The mask is then expanded to include voxels that have $T_b > 2\sigma$ in at least two adjacent velocity channels, and in addition can be joined to the initial mask through a path that only passes through valid (unmasked) voxels.  
If a region defined this way (referred to as an ``island'') has a projected area less than a single telescope beam, or spans a velocity range of less than 4 channels, it is assumed to be a noise peak and is removed from the mask.
Note that Hu10 used slightly different masking criteria, requiring a 4$\sigma$ peak and expanding to the 1.5$\sigma$ edge, and adopted a single value of $\sigma$ for the entire cube.

Once ``islands'' of significant emission have been identified, CPROPS decomposes the emission into individual ``cloud'' structures.  
Moments of the emission along the spatial and spectral axes are used to determine the size, linewidth and flux of the clouds, and corrections for the finite map sensitivity and instrumental resolution are applied to the measured cloud properties. 
To explore the effects of tunable parameters on the results, the decomposition procedure is performed under three different sets of assumptions:

\begin{table}
\begin{center}
\caption{Properties of Clouds Derived from CPROPS Analyses\label{tbl:cprops_summ}}
\begin{tabular}{lccc}
\hline
& Islands & Physical & Data-based\\\hline
No.\ of clouds & 450 & 543 & 694\\
No.\ of isolated clouds  & 243 & 222 & 190\\
Tot.\ $M_{\rm lum}/M_{\odot}$ (unextrap.) & $7.4 \times 10^6$ & $5.4 \times 10^6$ & $4.1 \times 10^6$ \\
Total $M_{\rm lum}/M_{\odot}$ & $1.2 \times 10^7$ & $1.1 \times 10^7$ & $1.1 \times 10^7$ \\
$M_{\rm lum}(<M_{\rm min})/M_{\rm lum}$\tablenotemark{a} & 0.27 & 0.42 & 0.55 \\
Mean(rms) $\log M_{\rm lum,\,\rm M_{\odot}}$ & 4.00(0.54) & 4.02(0.49) & 3.97(0.44)\\
Mean(rms) $\log R_{\rm pc}$ & 1.13(0.27) & 1.09(0.22) & 1.05(0.20) \\
Mean(rms) $\log \sigma_{v,\,\rm km\,s^{-1}}$ & 0.18(0.20) & 0.17(0.19) & 0.14(0.19)\\
\hline
\end{tabular}
\tablenotetext{1}{Fraction of the total $M_{\rm lum}$ in identified clouds that is in clouds with masses below the completeness limit of $3 \times 10^4$ \Msol.}
\end{center}
\end{table}

\begin{figure*}
\begin{center}
\includegraphics[angle=-90,width=0.7\textwidth]{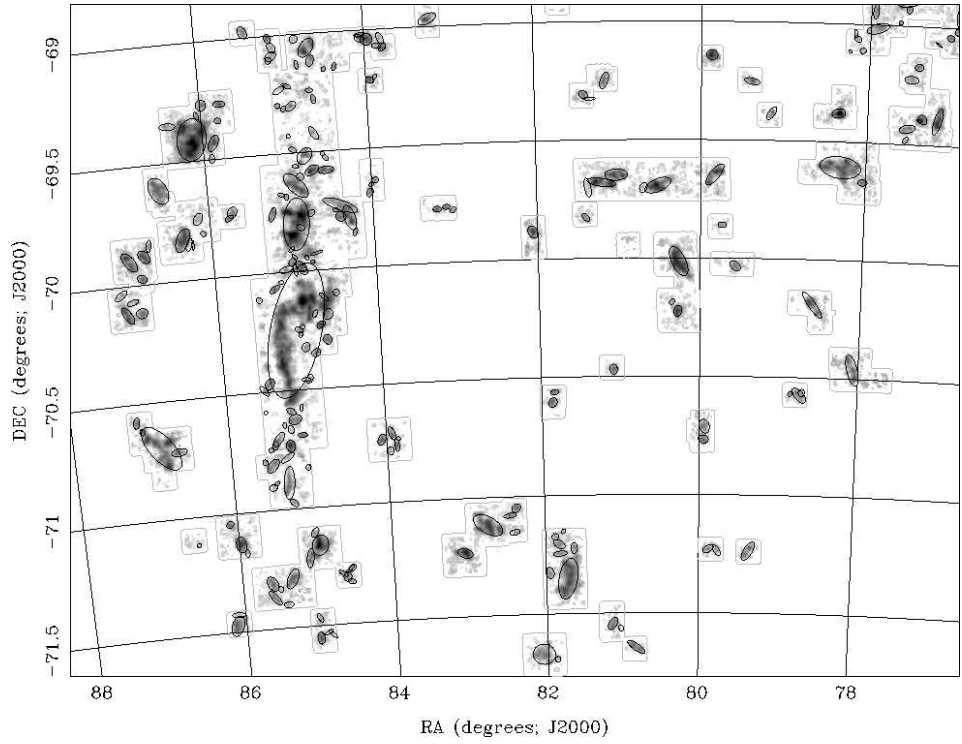}
\end{center}
\caption{
Zoom-in of the southeastern portion of the LMC, showing the integrated CO intensity from Figure~\ref{fig:magmaha} in grayscale and the sizes and orientations of the CO ``islands'' as black ellipses.  The major and minor axes of the ellipses as drawn have been extrapolated to the limit of perfect sensitivity but have not been corrected for the finite resolution of the telescope beam.
\label{fig:zoom1}}
\end{figure*}

\begin{figure*}
\begin{center}
\includegraphics[angle=-90,width=0.45\textwidth]{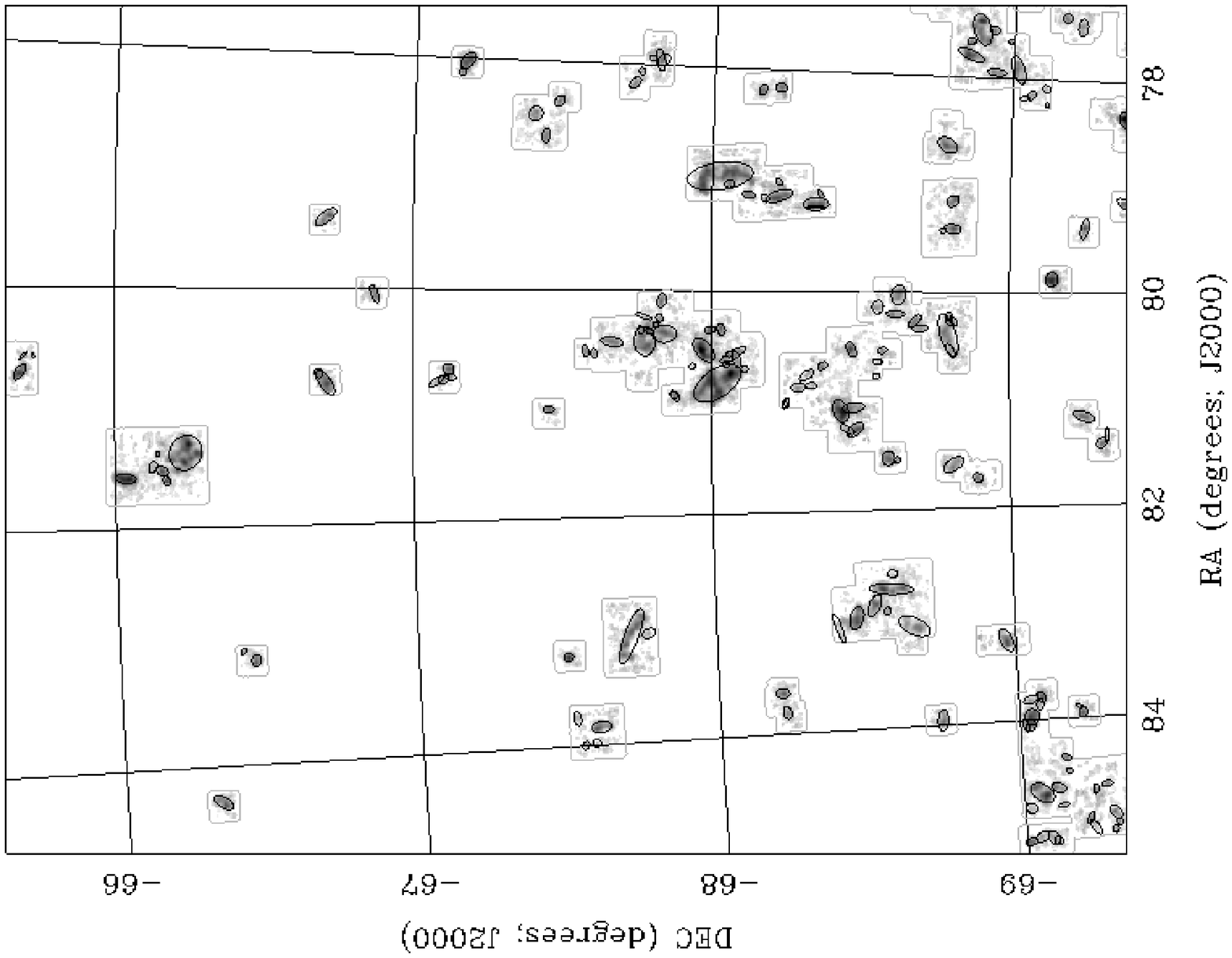}\quad
\includegraphics[angle=-90,width=0.45\textwidth]{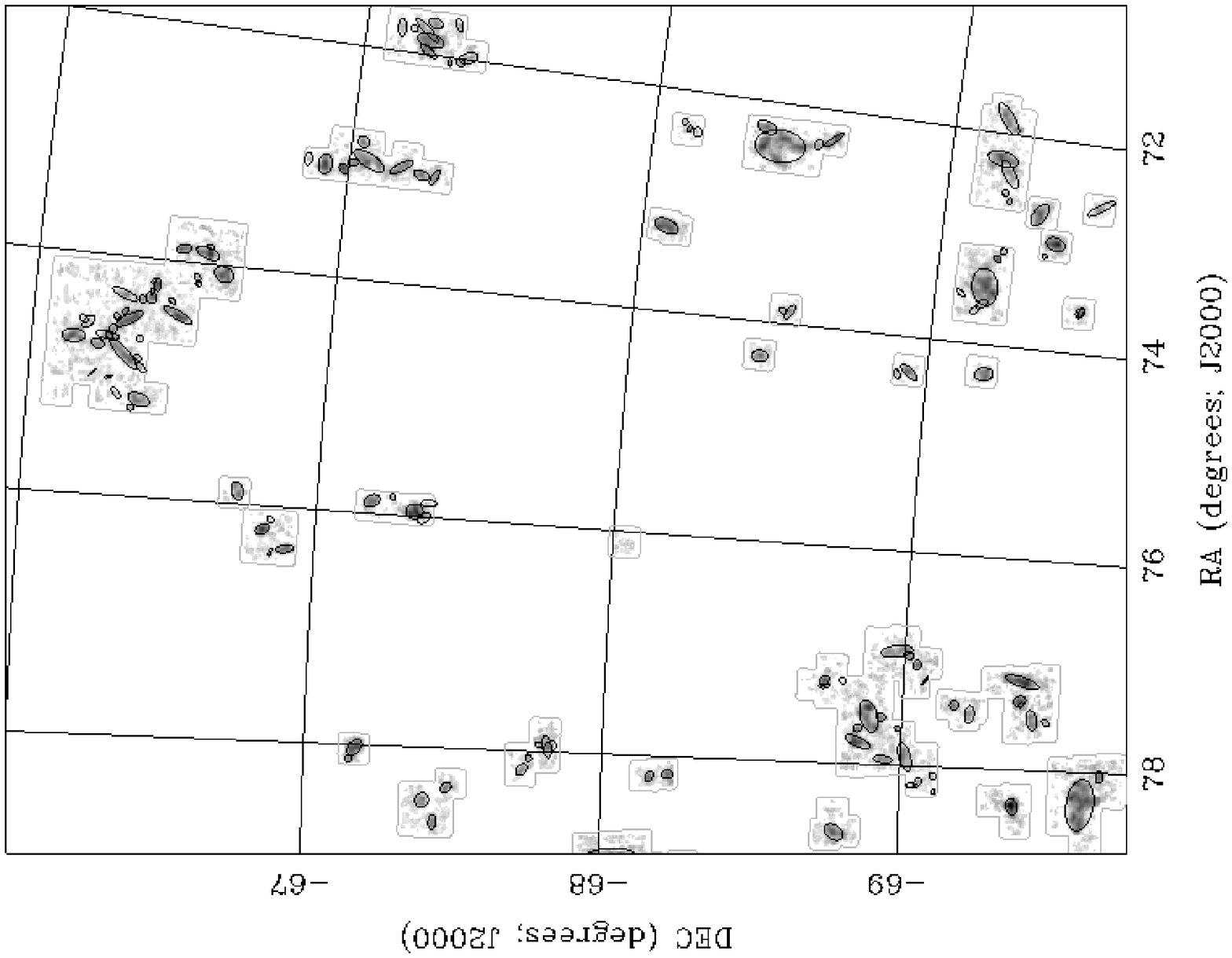}
\end{center}
\caption{
Same as Figure~\ref{fig:zoom1}, but for the ({\it left}) northern and ({\it right}) western portions of the LMC.
\label{fig:zoom2}}
\end{figure*}

\begin{enumerate}

\item {\it ``Islands'' parameters}: \textsc{CPROPS} simply catalogs all contiguous emission structures (``islands'') in the mask, regardless of size (but recall that islands must span an area of at least one telescope beam).  The size, linewidth, and flux of each island is determined by the standard RL06 algorithm, involving extrapolation to a zero-intensity boundary and correction for finite resolution of the telescope and spectrometer (see below).

\item {\it ``Physical'' parameters}: Islands are decomposed into clouds, defined as local maxima within a moving box of size 5 $\times$ 5 pixels ($\approx$18 pc) and velocity width 9 channels ($\approx$4.7 \kms).  These scales are chosen to be sensitive to structures that observers commonly define as GMCs in the Galaxy.  The algorithm requires that local maxima meet certain distinctness criteria (they must lie at least 1~K above the merge level with any other maximum, and yield significantly different moments when merged with adjacent maxima), and assigns to each maximum the surrounding emission that lies above the merge level with all other maxima.  Sensitivity and resolution corrections are then applied.

\item {\it ``Data-based'' parameters}: Clouds are identified as local maxima within a moving box of size 5 $\times$ 5 pixels and 9 channels, as in the ``physical'' case.  Each local maximum is required to lie at least 2$\sigma$ ($\approx$0.6 K) above the merge level with any other maximum.  Sensitivity and resolution corrections are then applied.

\end{enumerate}

The results of the three decompositions are summarized in Table~\ref{tbl:cprops_summ}.  It is clear that each of these approaches has advantages and disadvantages.  The ``islands'' approach ignores substructure, and defines clouds based on boundaries which may depend on the sensitivity of the observations.  The ``data-based'' approach identifies all structures which would be easily distinguished by eye, but the resulting sizes tend to lie near the resolution limit, hindering comparison with studies performed at different spatial resolution.  The ``physical'' approach is similar to the data-based approach, with the main difference being the stricter constrast requirement for distinctness (1 K $\approx$ 3$\sigma$ rather than 2$\sigma$).  We note that the decomposition of islands into clouds (whether ``physical'' or ``data-based'') is not flux-preserving: emission which cannot be uniquely assigned to a cloud is discarded.

\begin{table*}
\caption{Cloud Properties for Islands Decomposition\label{tbl:cprops_isl}}
\resizebox{\textwidth}{!} {
\begin{tabular}{lcccrrrccccrcrcrcccc}
\hline
ID & $\alpha$ (2000) & $\delta$ (2000) & $V_{\rm lsr}$ & Maj. & Min. & P.A. & $R$ & $\delta R$\tablenotemark{a} & $\Delta v$ & $\delta(\Delta v)$ & $L_{\rm CO}$\tablenotemark{b} & $\delta L$ & $M_{\rm vir}$\tablenotemark{c} & $\delta M$ & $N_{\rm vox}$ & $T_{\rm pk}$ & Isol? & NT\tablenotemark{d} & Hen\tablenotemark{e} \\
& (h:m:s) & (\arcdeg:\arcmin:\arcsec) & (km/s) & (pc) & (pc) & ($^\circ$) & (pc) &  & (km/s) & & & & & & & (K) & & ID & ID\\\hline
 A1 & 04:46:50.2 & $-$67:18:22 & 252.6 &  10.6 &   4.8 & 109 &   6.3 &  0.53 &  1.20 &  0.41 &   8.3 &  0.44 &   9.4 &  1.13 &    76 & 1.03 & Y &\nodata &\nodata \\
 A2 & 04:46:53.0 & $-$67:13:34 & 249.4 &  11.2 &   7.8 & 163 &  15.2 &  0.42 &  1.33 &  0.22 &  27.6 &  0.16 &  28.1 &  0.65 &   273 & 1.40 & N & 2 &\nodata \\
 A3 & 04:47:09.6 & $-$67:12:00 & 249.3 &  24.4 &   5.6 & 140 &  16.4 &  0.23 &  1.94 &  0.21 &  30.2 &  0.21 &  64.1 &  0.53 &   295 & 1.30 & N & 2 &\nodata \\
 A4 & 04:47:09.8 & $-$67:09:25 & 265.4 &   5.9 &   5.3 & 126 &   5.9 &  1.33 &  0.47 &  1.18 &   3.6 &  1.14 &   1.4 &  2.68 &    32 & 0.74 & Y &\nodata &\nodata \\
 A5 & 04:47:11.5 & $-$67:06:58 & 261.9 &  14.5 &   5.8 &  97 &  13.2 &  0.43 &  0.59 &  0.34 &  10.5 &  0.26 &   4.8 &  0.92 &    79 & 1.44 & Y & 2 &\nodata \\
 A6 & 04:47:26.4 & $-$69:10:19 & 239.8 &  28.0 &   9.3 & 127 &  28.6 &  0.09 &  1.06 &  0.10 &  80.5 &  0.03 &  33.4 &  0.23 &   842 & 2.78 & Y & 1 &\nodata \\
 A7 & 04:47:30.2 & $-$67:13:01 & 257.3 &  21.0 &   9.8 & 171 &  25.5 &  0.20 &  2.61 &  0.08 & 125.5 &  0.06 & 180.9 &  0.28 &  1247 & 1.91 & N & 2 &\nodata \\
 A8 & 04:47:53.0 & $-$67:12:50 & 265.0 &  15.7 &   5.3 & 146 &  12.0 &  0.26 &  1.08 &  0.15 &  34.5 &  0.10 &  14.5 &  0.39 &   325 & 2.09 & N & 2 &\nodata \\
 A9 & 04:47:55.7 & $-$67:21:04 & 256.9 &  14.3 &   8.1 &  20 &  18.1 &  0.52 &  1.72 &  0.68 &  26.8 &  0.45 &  55.6 &  1.73 &   121 & 1.22 & N & 2 &\nodata \\
A10 & 04:48:07.2 & $-$67:19:30 & 258.3 &   7.3 &   5.8 & 116 &   8.5 &  0.87 &  1.14 &  0.52 &  11.3 &  0.45 &  11.5 &  1.45 &    68 & 1.50 & N & 2 &\nodata \\
A11 & 04:48:10.8 & $-$67:17:28 & 257.8 &   5.4 &   2.4 &  99 & \nodata & \nodata &  1.00 &  0.75 &   6.3 &  0.49 & \nodata & \nodata &    41 & 1.24 & Y & 2 &\nodata \\
A12 & 04:48:56.9 & $-$68:22:23 & 250.9 &  15.9 &   8.9 & 162 &  20.6 &  0.26 &  2.11 &  0.13 &  58.7 &  0.16 &  95.4 &  0.39 &   401 & 2.13 & N & 5 & N76 \\
A13 & 04:48:57.1 & $-$69:10:16 & 239.9 &  23.8 &  11.1 & 171 &  29.4 &  0.11 &  1.68 &  0.08 & 130.9 &  0.03 &  86.1 &  0.18 &  1249 & 2.74 & N & 4 & N77D \\
\hline
\end{tabular}
}
\tablenotetext{1}{$\delta x$ is used to denote the fractional uncertainty in $x$.}
\tablenotetext{2}{Units are $10^2$ K km s$^{-1}$ pc$^2$.}
\tablenotetext{3}{Units are $10^3$ $M_\odot$.}
\tablenotetext{4}{Overlapping cloud(s) in \citet{Fukui:08}.}
\tablenotetext{5}{Associated \HII\ region from \citet{Henize:56}.}
\tablecomments{The full table is available as a machine-readable file.}
\end{table*}

\begin{table*}
\caption{Cloud Properties for Physical Decomposition\label{tbl:cprops_phy}}
\resizebox{\textwidth}{!} {
\begin{tabular}{lcccrrrccccrcrcrccc}
\hline
ID & $\alpha$ (2000) & $\delta$ (2000) & $V_{\rm lsr}$ & Maj. & Min. & P.A. & $R$ & $\delta R$\tablenotemark{a} & $\Delta v$ & $\delta(\Delta v)$ & $L_{\rm CO}$\tablenotemark{b} & $\delta L$ & $M_{\rm vir}$\tablenotemark{c} & $\delta M$ & $N_{\rm vox}$ & $T_{\rm pk}$ & Isol? & Isl.\\
& (h:m:s) & (\arcdeg:\arcmin:\arcsec) & (km/s) & (pc) & (pc) & ($^\circ$) & (pc) &  & (km/s) & & & & & & & (K) & & ID \\\hline
 B1 & 04:46:50.2 & $-$67:18:22 & 252.6 &  10.6 &   4.8 & 109 &   6.3 &  0.60 &  1.20 &  0.38 &   8.3 &  0.49 &   9.4 &  1.06 &    76 & 1.03 & Y & A1 \\
 B2 & 04:46:53.0 & $-$67:13:34 & 249.4 &  11.2 &   7.8 & 163 &  15.2 &  0.42 &  1.33 &  0.22 &  27.6 &  0.16 &  28.1 &  0.66 &   273 & 1.40 & N & A2 \\
 B3 & 04:47:09.6 & $-$67:12:00 & 249.3 &  24.4 &   5.6 & 140 &  16.4 &  0.24 &  1.94 &  0.19 &  30.2 &  0.21 &  64.1 &  0.53 &   295 & 1.30 & N & A3 \\
 B4 & 04:47:09.8 & $-$67:09:25 & 265.4 &   5.9 &   5.3 & 126 &   5.9 &  1.34 &  0.47 &  1.12 &   3.6 &  1.09 &   1.4 &  2.69 &    32 & 0.74 & Y & A4 \\
 B5 & 04:47:11.5 & $-$67:06:58 & 261.9 &  14.5 &   5.8 &  97 &  13.2 &  0.44 &  0.59 &  0.33 &  10.5 &  0.27 &   4.8 &  0.96 &    79 & 1.44 & Y & A5 \\
 B6 & 04:47:26.4 & $-$69:10:19 & 239.8 &  28.0 &   9.3 & 127 &  28.6 &  0.09 &  1.06 &  0.10 &  80.5 &  0.04 &  33.4 &  0.22 &   842 & 2.78 & Y & A6 \\
 B7 & 04:47:26.6 & $-$67:11:42 & 259.0 &   9.3 &   4.4 & 172 & \nodata & \nodata &  1.55 &  0.28 &  37.4 &  0.32 & \nodata & \nodata &   141 & 1.91 & N & A7 \\
 B8 & 04:47:30.7 & $-$67:13:34 & 256.3 &  14.7 &  11.0 &  18 &  22.6 &  0.41 &  1.45 &  0.26 &  75.6 &  0.31 &  49.1 &  0.68 &   293 & 1.85 & N & A7 \\
 B9 & 04:47:53.0 & $-$67:12:50 & 265.0 &  15.7 &   5.3 & 146 &  12.0 &  0.25 &  1.08 &  0.15 &  34.5 &  0.10 &  14.5 &  0.39 &   325 & 2.09 & N & A8 \\
B10 & 04:47:55.7 & $-$67:21:04 & 256.9 &  14.3 &   8.1 &  20 &  18.1 &  0.53 &  1.72 &  0.72 &  26.8 &  0.44 &  55.6 &  1.72 &   121 & 1.22 & N & A9 \\
B11 & 04:48:07.2 & $-$67:19:30 & 258.3 &   7.3 &   5.8 & 116 &   8.5 &  0.86 &  1.14 &  0.51 &  11.3 &  0.44 &  11.5 &  1.52 &    68 & 1.50 & N & A10 \\
B12 & 04:48:10.8 & $-$67:17:28 & 257.8 &   5.4 &   2.4 &  99 & \nodata & \nodata &  1.00 &  0.71 &   6.3 &  0.48 & \nodata & \nodata &    41 & 1.24 & Y & A11 \\
B13 & 04:48:53.8 & $-$69:10:05 & 239.7 &  16.5 &   7.8 & 157 &  19.1 &  0.25 &  1.69 &  0.13 &  86.4 &  0.09 &  57.0 &  0.40 &   431 & 2.74 & N & A13 \\
\hline
\end{tabular}
}
\end{table*}

\begin{table*}
\caption{Cloud Properties for Data-Based Decomposition\label{tbl:cprops_def}}
\resizebox{\textwidth}{!} {
\begin{tabular}{lcccrrrccccrcrcrcccc}
\hline
ID & $\alpha$ (2000) & $\delta$ (2000) & $V_{\rm lsr}$ & Maj. & Min. & P.A. & $R$ & $\delta R$\tablenotemark{a} & $\Delta v$ & $\delta(\Delta v)$ & $L_{\rm CO}$\tablenotemark{b} & $\delta L$ & $M_{\rm vir}$\tablenotemark{c} & $\delta M$ & $N_{\rm vox}$ & $T_{\rm pk}$ & Isol? & Isl. & Phy.\\
& (h:m:s) & (\arcdeg:\arcmin:\arcsec) & (km/s) & (pc) & (pc) & ($^\circ$) & (pc) &  & (km/s) & & & & & & & (K) & & ID & ID\\\hline
 C1 & 04:46:50.2 & $-$67:18:22 & 252.6 &  10.6 &   4.8 & 109 &   6.3 &  0.61 &  1.20 &  0.40 &   8.3 &  0.47 &   9.4 &  1.10 &    76 & 1.03 & Y &   A1 &   B1 \\
 C2 & 04:46:53.0 & $-$67:13:34 & 249.4 &  11.2 &   7.8 & 163 &  15.2 &  0.38 &  1.33 &  0.21 &  27.6 &  0.15 &  28.1 &  0.63 &   273 & 1.40 & N &   A2 &   B2 \\
 C3 & 04:47:01.7 & $-$67:10:52 & 247.9 &  10.8 &   6.7 & 136 &  13.1 &  0.57 &  1.20 &  0.32 &  16.0 &  0.45 &  19.5 &  0.86 &    95 & 1.30 & N &   A3 &   B3 \\
 C4 & 04:47:09.8 & $-$67:09:25 & 265.4 &   5.9 &   5.3 & 126 &   5.9 &  1.39 &  0.47 &  1.16 &   3.6 &  1.14 &   1.4 &  2.69 &    32 & 0.74 & Y &   A4 &   B4 \\
 C5 & 04:47:11.5 & $-$67:06:58 & 261.9 &  14.5 &   5.8 &  97 &  13.2 &  0.40 &  0.59 &  0.36 &  10.5 &  0.25 &   4.8 &  0.95 &    79 & 1.44 & Y &   A5 &   B5 \\
 C6 & 04:47:16.6 & $-$67:13:05 & 250.4 &  14.0 &   5.8 & 145 &  13.0 &  0.58 &  1.73 &  0.33 &  21.0 &  0.61 &  40.6 &  0.90 &   100 & 1.05 & N &   A3 &   B3 \\
 C7 & 04:47:22.8 & $-$69:09:54 & 240.0 &  17.8 &   7.5 & 112 &  19.2 &  0.14 &  0.80 &  0.13 &  52.9 &  0.07 &  12.9 &  0.29 &   410 & 2.78 & N &   A6 &   B6 \\
 C8 & 04:47:26.2 & $-$67:14:13 & 255.9 &  14.2 &   6.9 & 169 &  16.0 &  0.64 &  1.66 &  0.47 &  37.1 &  0.82 &  46.0 &  1.07 &   112 & 1.45 & N &   A7 &   B8 \\
 C9 & 04:47:26.6 & $-$67:11:42 & 259.0 &   9.3 &   4.4 & 172 & \nodata & \nodata &  1.55 &  0.27 &  37.4 &  0.31 & \nodata & \nodata &   141 & 1.91 & N &   A7 &   B7 \\
C10 & 04:47:37.4 & $-$67:12:32 & 256.9 &   7.6 &   5.7 &  42 &   8.6 &  1.17 &  1.19 &  0.46 &  28.3 &  0.51 &  12.6 &  1.58 &    80 & 1.85 & N &   A7 &   B8 \\
C11 & 04:47:46.8 & $-$69:12:25 & 238.6 &  12.7 &   6.0 &   9 &  12.9 &  0.40 &  1.07 &  0.39 &  17.6 &  0.44 &  15.4 &  0.94 &    83 & 1.77 & N &   A6 &   B6 \\
C12 & 04:47:53.0 & $-$67:12:50 & 265.0 &  15.7 &   5.3 & 146 &  12.0 &  0.24 &  1.08 &  0.15 &  34.5 &  0.10 &  14.5 &  0.41 &   325 & 2.09 & N &   A8 &   B9 \\
C13 & 04:47:55.7 & $-$67:21:04 & 256.9 &  14.3 &   8.1 &  20 &  18.1 &  0.52 &  1.72 &  0.71 &  26.8 &  0.44 &  55.6 &  1.73 &   121 & 1.22 & N &   A9 &  B10 \\
\hline
\end{tabular}
}
\end{table*}

As in Hu10, we adopt the standard \textsc{CPROPS} definitions to derive the basic physical properties of GMCs in the MAGMA cloud catalogue. The cloud radius is defined as $R = 1.91 \sigma_{R}$\,pc, where $\sigma_{R}$ is the geometric mean of the second moments of the emission along the cloud's major and minor axes. The velocity dispersion $\sigma_{\rm v}$ is the second moment of the emission distribution along the velocity axis, which for a Gaussian line profile is related to the FWHM linewidth, $\Delta v$, by $\Delta v = \sqrt{8 \ln 2}\sigma_{\rm v}$. The CO luminosity of the cloud $L_{\rm CO}$ is the integrated flux scaled by the square of the distance, i.e.
\begin{equation} 
L_{\rm CO} \; [\lcou] = D^{2}\; (\Sigma T_{i})\;\delta v\,\delta x\,\delta y
\label{eqn:lcodef}
\end{equation}
where $T_i$ is the brightness temperature of an individual voxel, $D$ is the distance to the LMC in parsecs (taken to be $5 \times 10^4$), $\delta x$ and $\delta y$ are the angular pixel dimensions in radians, and $\delta v$ is the width of one channel in \kms. The mass of molecular gas estimated from the GMC's CO luminosity, $M_{\rm lum}$ is
calculated as
\begin{equation}
M_{\rm lum} \; [\Msol] \equiv 4.4 \frac{\xco}{2 \times 10^{20}\;\xcou} L_{\rm CO}\; ,
\label{eqn:mcodef}
\end{equation}
where \xco\ is the assumed CO-to-H$_2$ conversion factor, and a factor of 1.36 is included to account for the mass contribution of helium. Note that the fiducial value of \xco\ used by \textsc{CPROPS} is $\xco = 2.0 \times 10^{20}$\,\xcou.  Although larger values have been derived for \xco\ based on the assumption of virial equilibrium (Fu08, Hu10), in this work we do not assume virialization, and so for simplicity we calculate $M_{\rm lum}$ based on this fiducial estimate of \xco.  \citet{Leroy:11} use a dust-based method to derive $\xco = 3.0 \times 10^{20}$\,\xcou\ for the LMC, based on comparison of {\it Spitzer} IR, NANTEN CO, and ATCA+Parkes \HI\ maps.  However, they caution that their estimate may be biased high by their assumption that the ratio of 160-$\mu$m dust opacity to gas column density ($\tau_{160}/N_{\rm H}$) is the same for \HI- and H$_2$-dominated regions.
There are a number of indications that the FIR emissivity of dust grains is enhanced in regions of higher gas density \citep[e.g.,][]{Bernard:99,Stepnik:03,Paradis:09,Planck:2037}, which would cause the dust-based approach to overestimate the amount of molecular gas.

The virial mass is estimated as 
\begin{equation}
M_{\rm vir} \; [\Msol] = 1040 \sigma_{\rm v}^{2}R\;,
\end{equation}
which assumes that molecular clouds are spherical with truncated $\rho \propto r^{-1}$ density profiles \citep{Maclaren:88}.  Naturally, no estimate of $M_{\rm vir}$ is possible for clouds which are not spatially resolved in both dimensions.  \textsc{CPROPS} estimates the error associated with a cloud property measurement using a bootstrapping method, which is described in Section 2.5 of RL06.

As emphasized by RL06, the resolution and sensitivity of a dataset influence the derived cloud properties. In order to reduce these observational biases, they recommend extrapolating the cloud property measurements to values that would be expected in the limiting case of perfect sensitivity (i.e. a brightness temperature threshold of 0\,K), and correcting for finite resolution in the spatial and spectral domains by deconvolving the telescope beam and width of a spectral channel from the measured cloud size and linewidth respectively. The procedures that \textsc{CPROPS} uses to apply these corrections are described in RL06 (see especially their Figure~2 and Sections 2.2 and 2.3); briefly stated, a linear extrapolation is used for $\sigma_R$ and $\sigma_v$ while a quadratic extrapolation is used for $L_{\rm CO}$.  Unless otherwise noted, the cloud property measurements used in the present work have been corrected for resolution and sensitivity bias.

Figures~\ref{fig:zoom1} and \ref{fig:zoom2} provide more detailed views of the CO intensity image with the positions and sizes of the CPROPS ``islands'' overlaid.  The major and minor axes of the ellipses as drawn have been extrapolated to the limit of perfect sensitivity but have not been corrected for finite spatial resolution.

\subsection{Catalog Reliability and Completeness}\label{sec:reliability}

The overall reliability of the cloud catalogs was checked by examining spectra in the CO data cube; in nearly all cases there was little doubt that CO emission had been detected.
Similarly, as discussed in Section~\ref{sec:nanten}, we found relatively few discrepancies with the NANTEN cloud catalog.
However, clouds identified as being distinct may still be connected at a lower brightness level.
We attempted to distinguish ``isolated'' islands of emission by excluding islands which connect to neighboring islands when contoured down to the 1.5$\sigma$ level.
To identify such isolated islands more strictly, we disqualified neighboring islands that can be connected through a vertex, in addition to islands that connect via a shared face, which is the more conventional definition for connected emission.
The number of isolated islands is tabulated in Table~\ref{tbl:cprops_summ}; isolated clouds are defined in the same way for the other two decompositions.
Of course, even the ``isolated'' regions of emission may join with other regions at lower contour levels that we are insensitive to.

Clouds may also be difficult to distinguish because they lie along the same line of sight.  Because of the low filling fraction of CO clouds and the face-on aspect of the LMC, we believe that chance superpositions of clouds are unlikely.  In particular, we find very few multi-peaked CO profiles, which would be indicative of overlapping clouds at distinct velocities.  Such an overlap occurs in only 1.6\% of positions for the ``islands'' catalog, 2.5\% for the ``physical'' decomposition and 4.2\% for the ``data-based'' decomposition.  Since only strong, well-separated components would be identified as distinct, we caution that these statistics depend on the signal-to-noise in the data cube.

While the MAGMA maps reveal many previously unknown low-luminosity clouds, the exclusion of NANTEN clouds with fluxes less than $1.2 \times 10^5$ \kkms\ arcsec$^2$ from the sample definition (corresponding to a luminous mass of $M_{\rm min} = 3 \times 10^4$ \Msol, assuming a Galactic conversion factor) provides a minimum completeness limit for the sample.
This limit corresponds to about twice the completeness limit quoted by Fu08 [corrected for a different assumed \xco-factor of $7 \times 10^{20}\,\xcou$], suggesting that all clouds with mass above $M_{\rm min}$ should have been detected.
However, it should be kept in mind that both the NANTEN and Mopra surveys are brightness temperature limited, so it is not possible to determine completeness limits for either survey with absolute certainty: clouds covering larger areas and/or having larger linewidths may still escape detection.
The minimum detectable cloud flux (assuming a 3$\sigma$ brightness temperature in at least 3 spectral channels) depends on the cloud size, but is roughly $10^{2.5}$ \kkms\ pc$^{-2}$ for a radius of $\sim$10 pc (comparable to our map resolution).
As this corresponds to 1400 \Msol, a factor of 20 below our assumed completeness limit, we consider it unlikely that the completeness limit is strongly affected by our map sensitivity.
We note that while the NANTEN clouds that form the parent sample of our survey lie above the completeness limit, their resolved sub-components are likely responsible for populating the regime between the detection and completeness limits.

\begin{figure*}
\begin{center}
\includegraphics[height=0.33\textwidth]{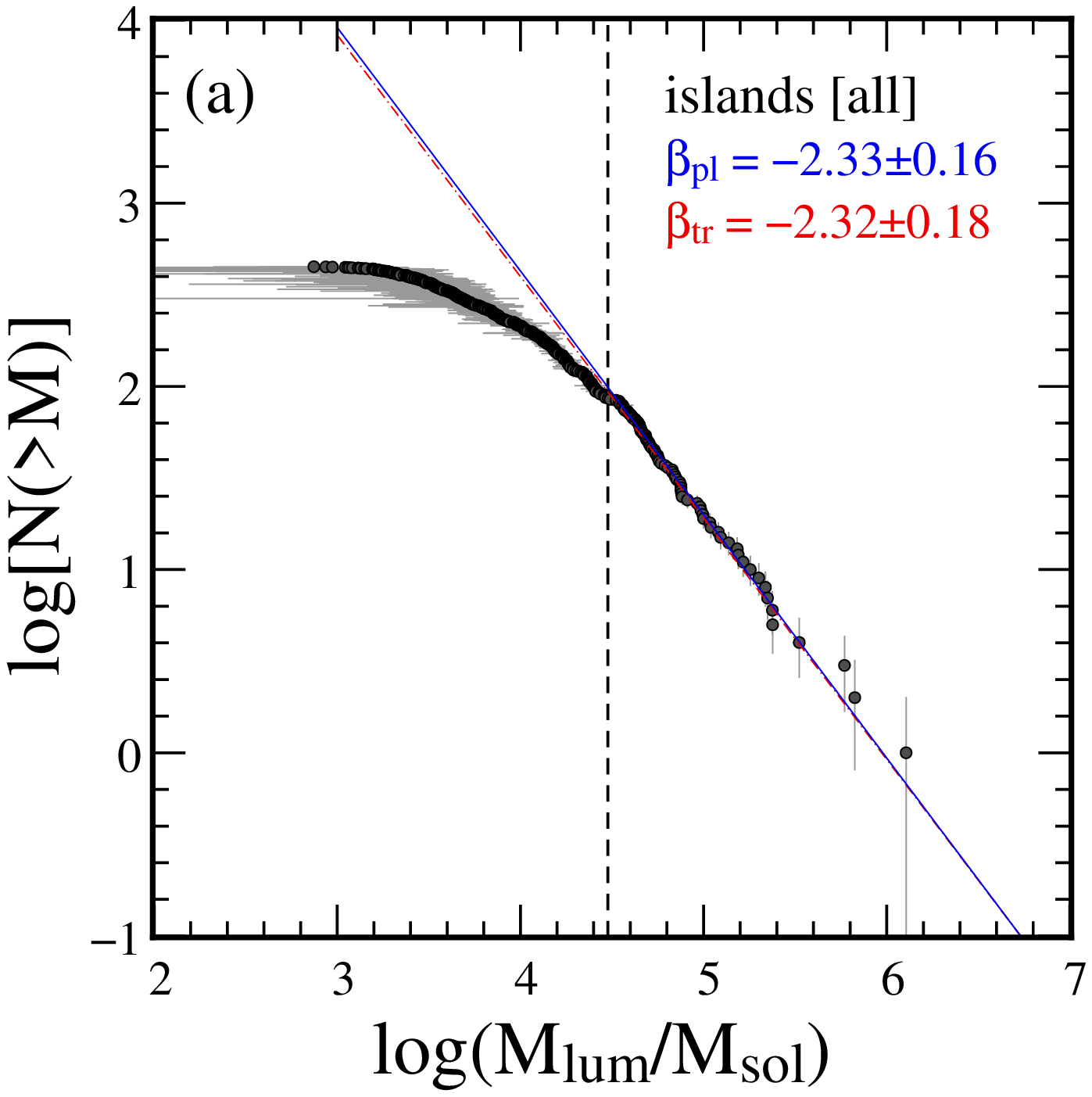}\quad
\includegraphics[height=0.33\textwidth]{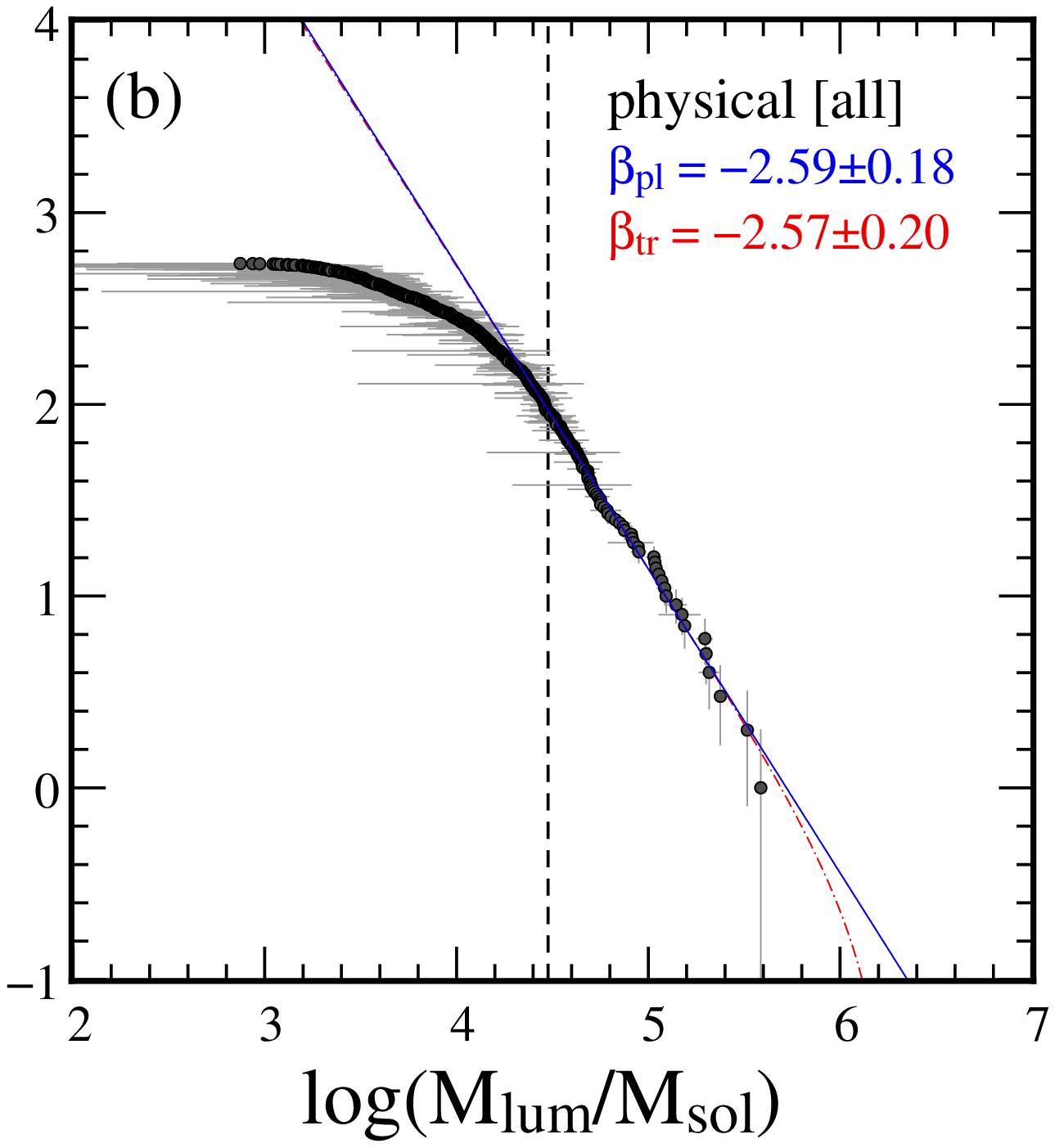}\quad
\includegraphics[height=0.33\textwidth]{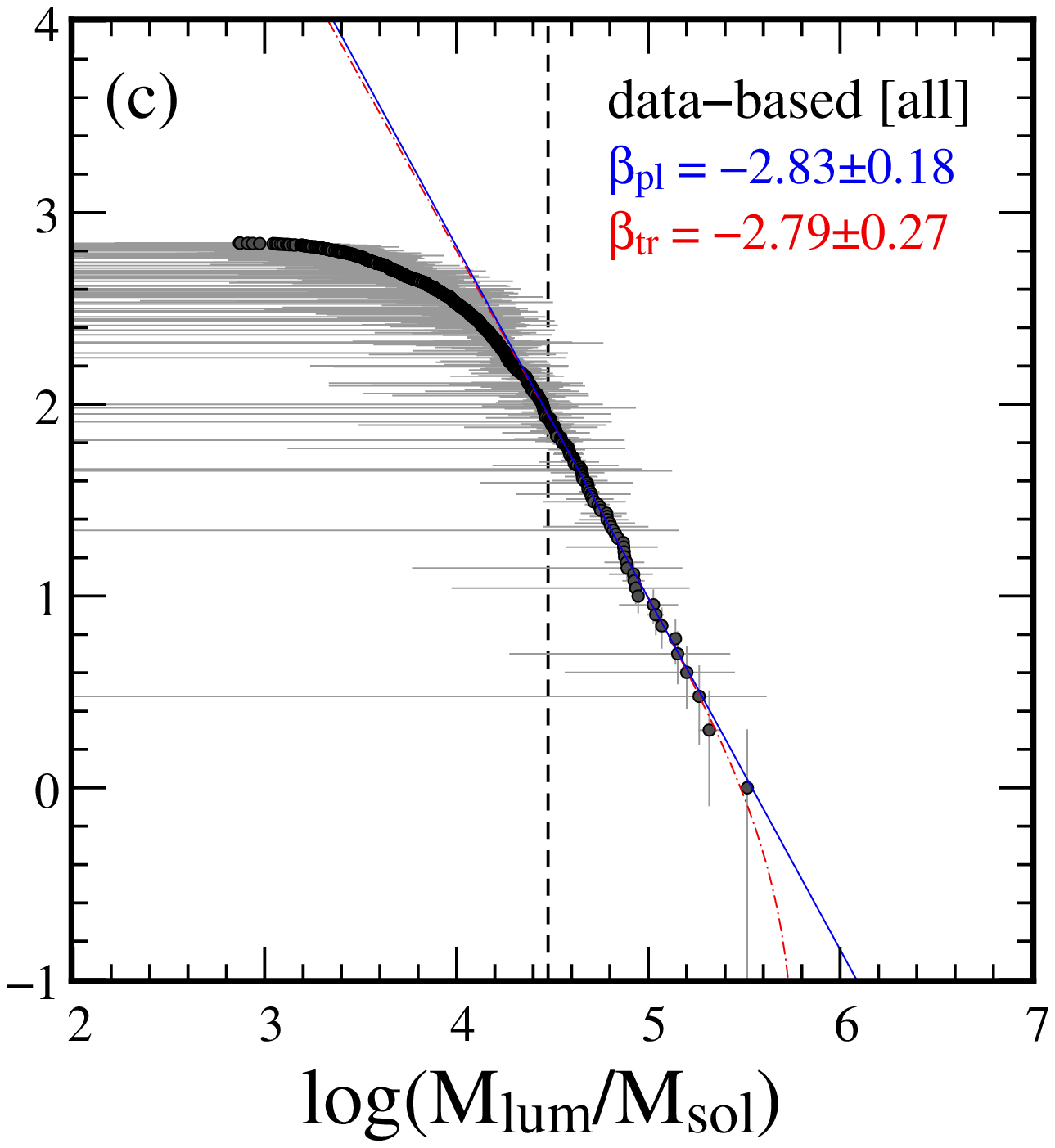}
\end{center}
\caption{
Cumulative luminous mass spectra for clouds defined with the islands ({\it left}), physical ({\it middle}), and data-based ({\it right}) parameter sets (see Sec.~\ref{sec:sigdet} for definitions).  Blue solid lines are simple power-law fits to points above a completeness limit of $3 \times 10^4$ \Msol\ (vertical dashed line).  Red dot-dashed lines are truncated power-law fits over the same range.
\label{fig:mlcum}}
\end{figure*}

\begin{figure*}
\begin{center}
\includegraphics[height=0.33\textwidth]{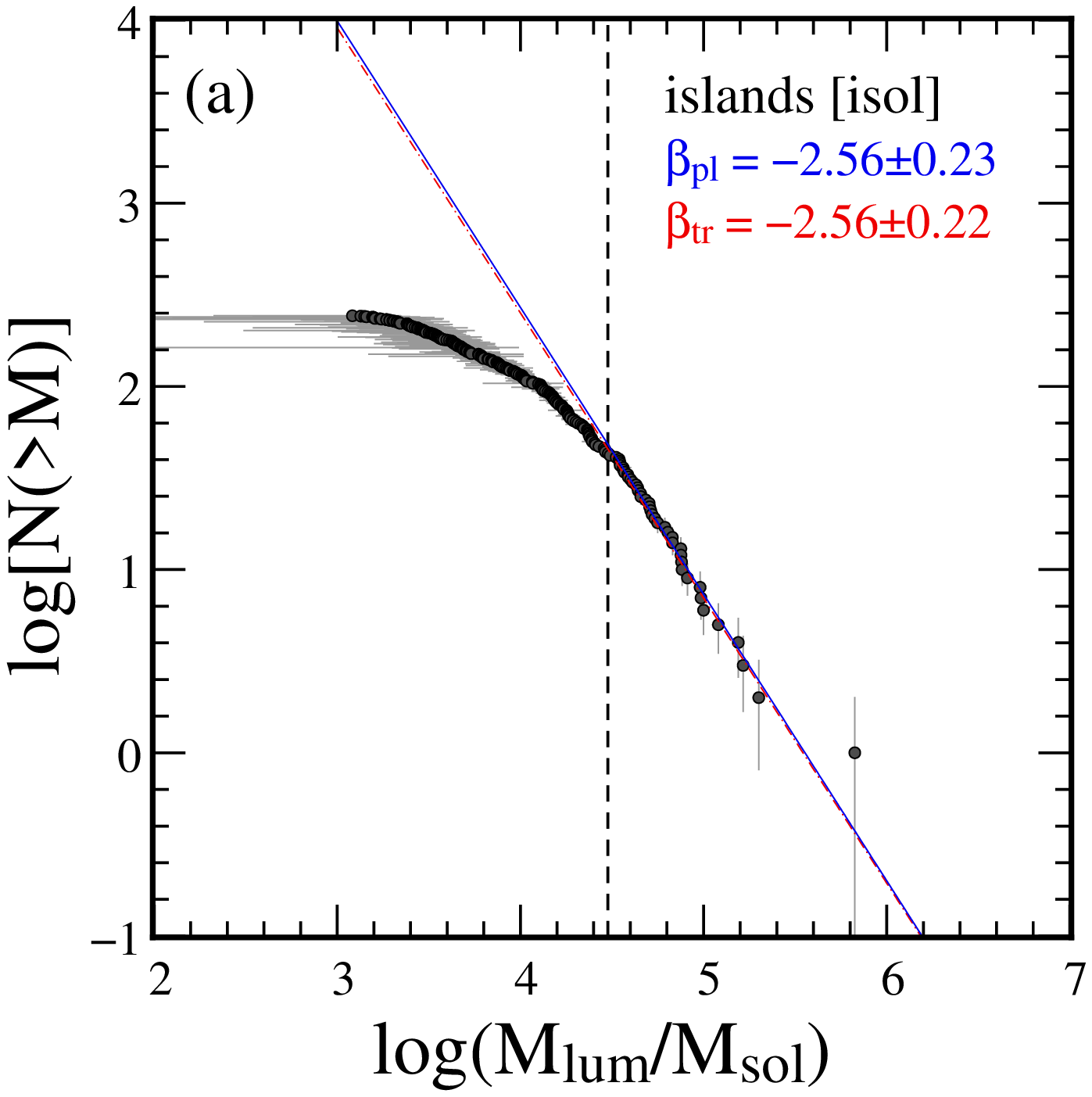}\quad
\includegraphics[height=0.33\textwidth]{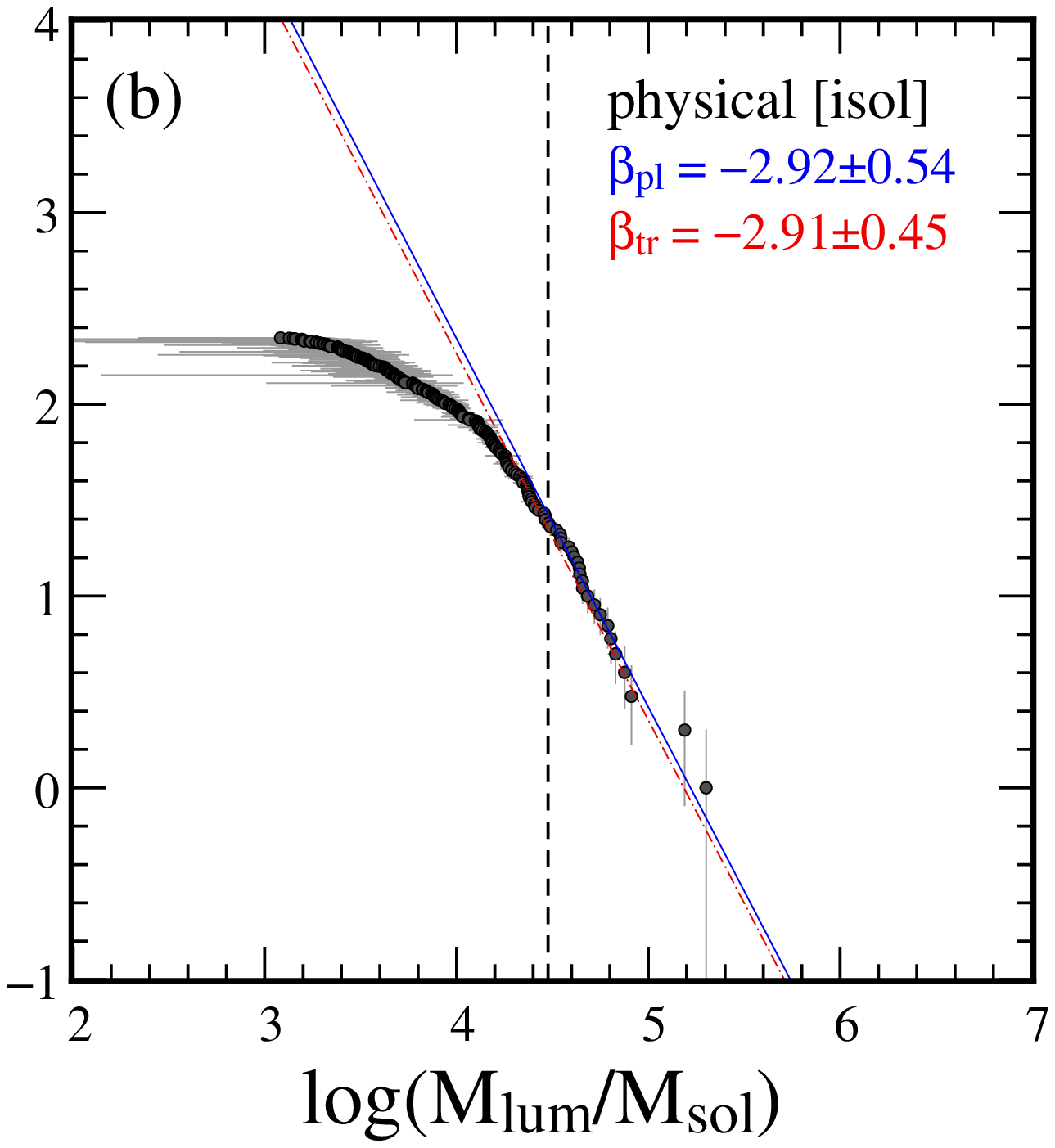}\quad
\includegraphics[height=0.33\textwidth]{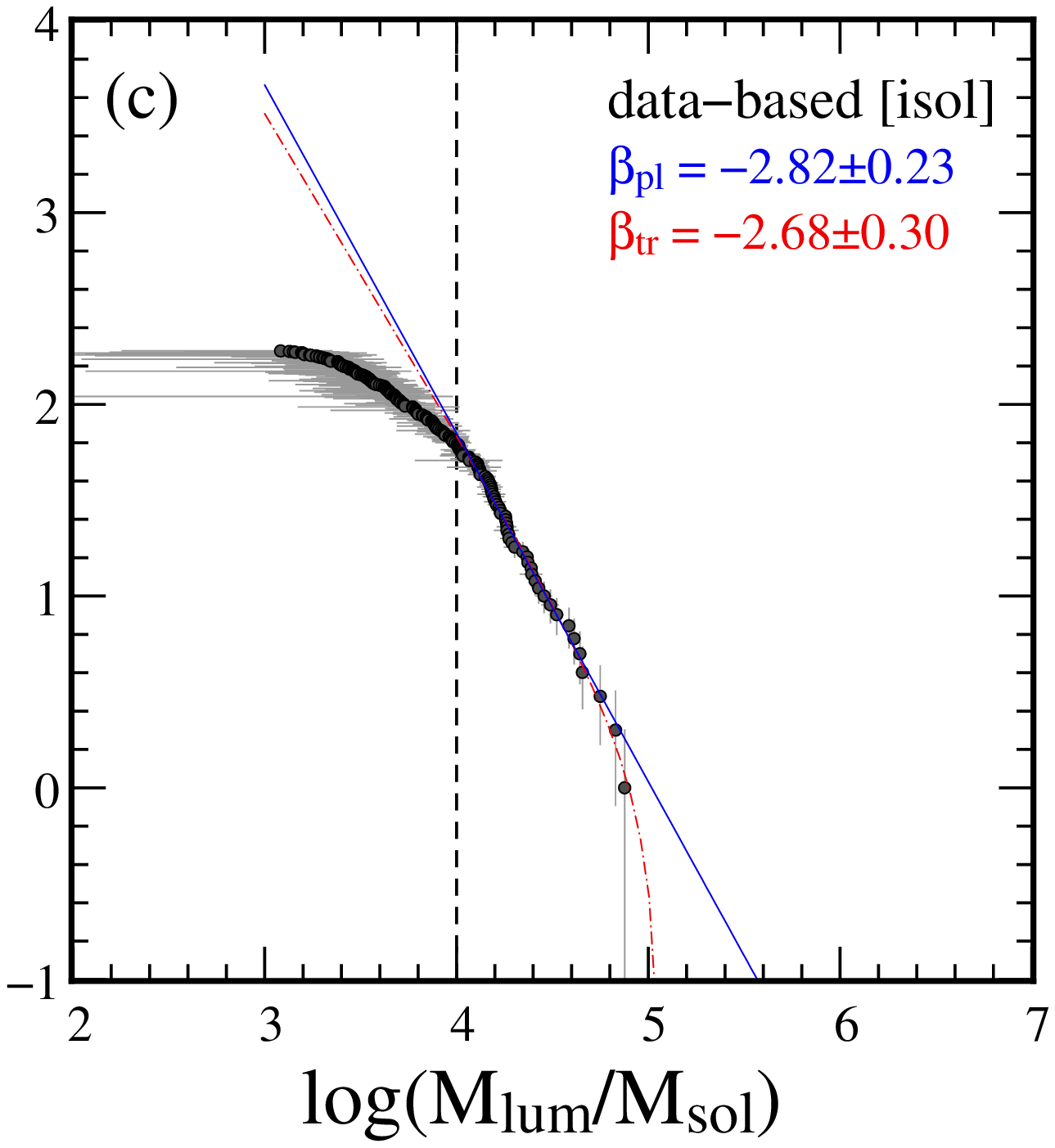}
\end{center}
\caption{
Cumulative luminous mass spectra for {\it isolated} clouds defined with the islands ({\it left}), physical ({\it middle}), and data-based ({\it right}) parameter sets.  Blue solid lines are simple power-law fits to points above a completeness limit of $3 \times 10^4$ \Msol\ (vertical dashed line), or $10^4$ \Msol\ for the data-based parameter set.  Red dot-dashed lines lines are truncated power-law fits.
\label{fig:mlcumiso}}
\end{figure*}

\begin{figure*}
\begin{center}
\includegraphics[height=0.33\textwidth]{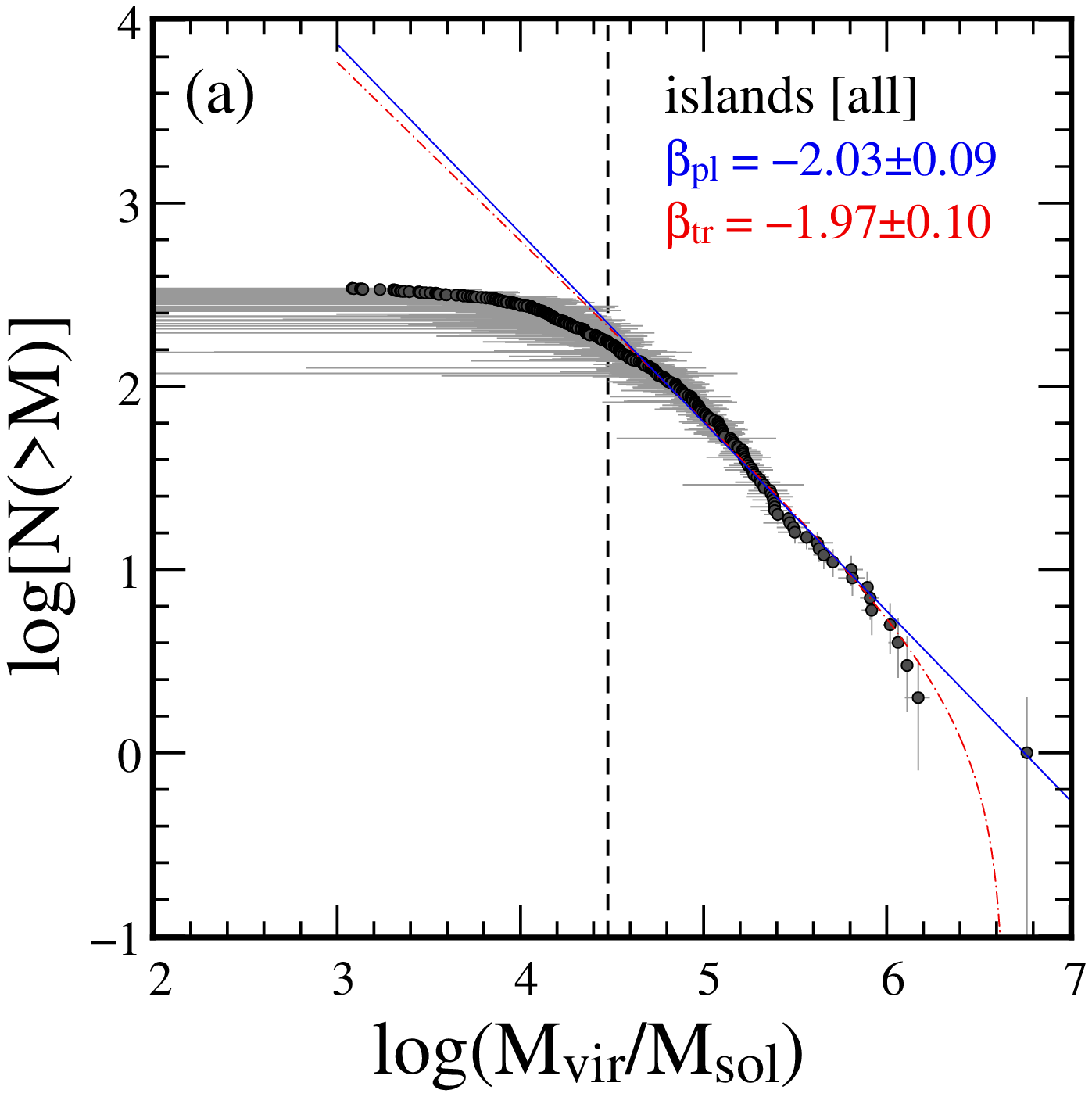}\quad
\includegraphics[height=0.33\textwidth]{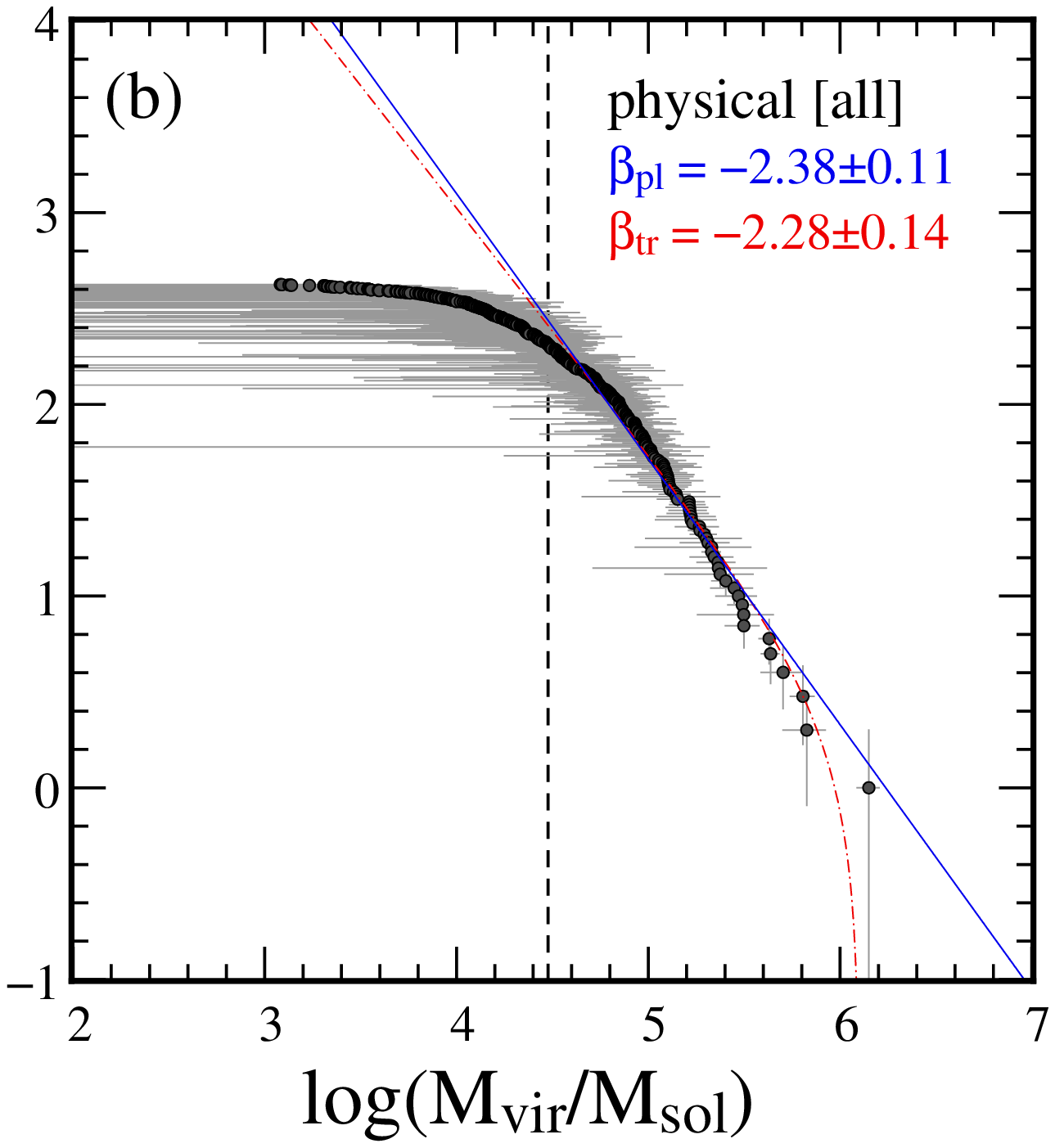}\quad
\includegraphics[height=0.33\textwidth]{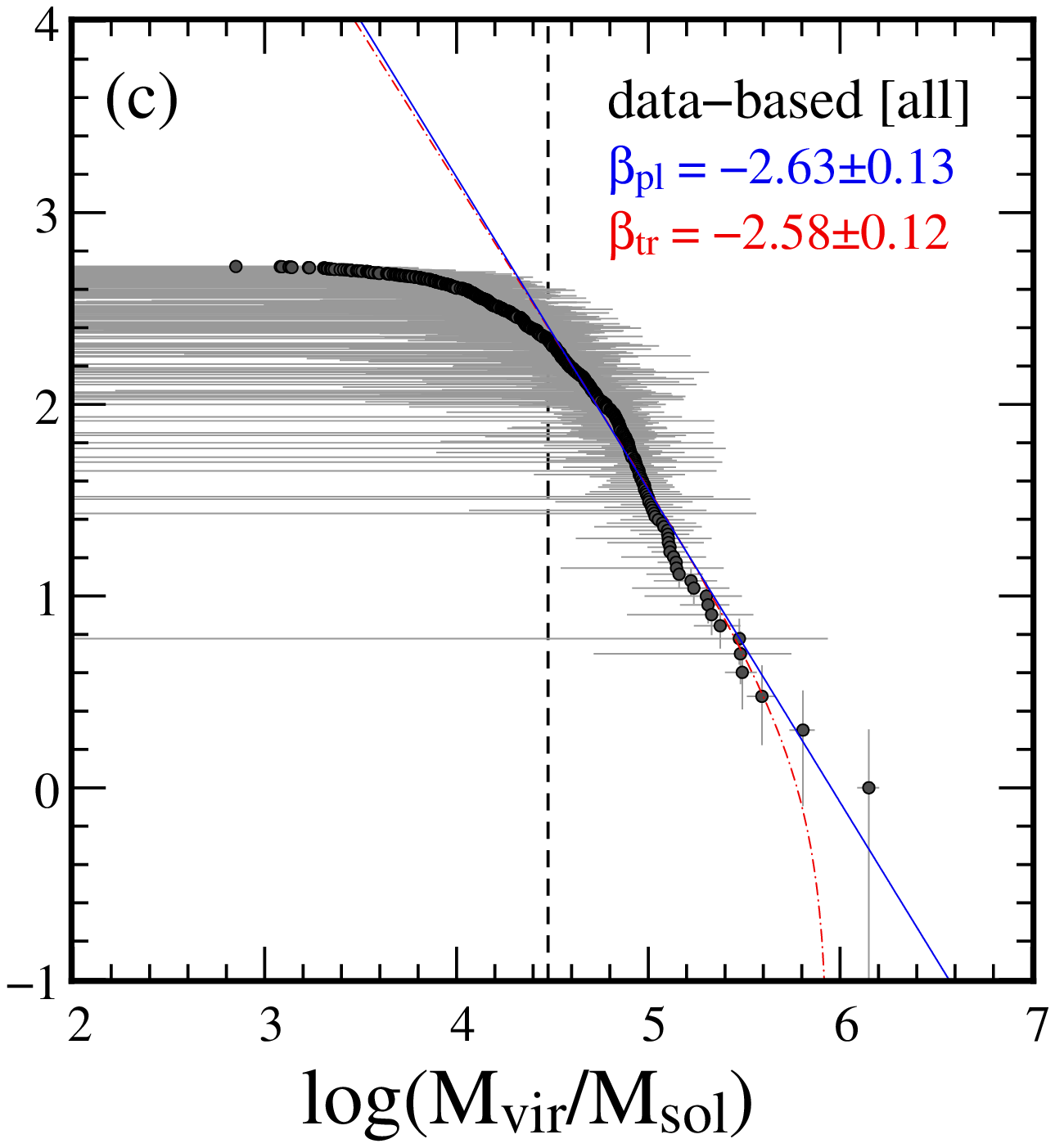}
\end{center}
\caption{
Cumulative virial mass spectra for clouds defined with the islands ({\it left}), physical ({\it middle}), and data-based ({\it right}) parameter sets.  Blue solid lines are simple power-law fits to points above a completeness limit of $3 \times 10^4$ \Msol\ (vertical dashed line).  Red dot-dashed lines lines are truncated power-law fits.
\label{fig:mvcum}}
\end{figure*}

\subsection{Tables of Cloud Properties}

The cloud properties derived by CPROPS are summarized in Tables~\ref{tbl:cprops_isl}, \ref{tbl:cprops_phy}, and \ref{tbl:cprops_def} for the ``islands'' catalog, ``physical'' decomposition, and ``data-based'' decomposition repectively.  
Only the first 13 lines of each table are shown; full tables are accessible from the electronic journal.
Each cloud is numbered in order of right ascension and given a prefix of A, B, or C to distinguish the three cloud sets.
We tabulate in columns (1)--(4) the ID number, center of each cloud in degrees of R.A. and Dec.\ and in LSR velocity (corrected for our subtraction of a galactic rotation model).  
Columns (5)--(7) give the major and minor axes of the cloud, corrected for sensitivity but not resolution effects, and the position angle of the major axis (measured in the usual sense from north towards east).  
We then tabulate the size $R$ [col.\ (8)--(9)], linewidth $\Delta v$ [col.\ (10)--(11)], luminosity $L_{\rm CO}$ [col.\ (12)--(13)], and virial mass $M_{\rm vir}$ [col.\ (14)--(15)] of each cloud, along with their respective fractional uncertainties ($\sigma_x/x$).  
Columns (16) and (17) give the number of significant voxels in the cloud and the peak brightness temperature among those voxels.  
Column (18) indicates whether a cloud is ``isolated'' or not.  
The remaining columns are different for different tables.
In Table~\ref{tbl:cprops_isl}, Column (19) lists cross-identifications with the NANTEN catalog (Fu08), and Column (20) lists cross-identifications with the \HII\ nebulae of \citet{Henize:56}.  The Henize cross-identifications are crude, requiring the center of an island to be within 5\arcmin\ of the nebula center (or within the nebula's radius if larger than 5\arcmin), and should be treated with caution.
In Table~\ref{tbl:cprops_phy}, Column (19) indicates the ID number of the island in Table~\ref{tbl:cprops_isl} which contains the cloud.
Similarly, in Table~\ref{tbl:cprops_def}, Columns (19) and (20) indicate the ID numbers of the islands and the ``physical'' clouds in Tables~\ref{tbl:cprops_isl} and \ref{tbl:cprops_phy} which contain the cloud.

\begin{table*}
\begin{center}
\caption{Mass Spectrum Fit Parameters\label{tbl:mspec}}
\resizebox{\textwidth}{!} {
\begin{tabular}{lccc|cc|ccc}
\hline
 & & & & \multicolumn{2}{c}{Simple power law} & \multicolumn{3}{c}{Truncated power law}\\
Sample & $M_{\rm min}/\Msol$ & Type & $N(>M_{\rm min})$ & $M_0/(10^6\,M_\odot)$ & $\beta$ & $N_0$ & $M_0/(10^6\,\Msol)$ & $\beta$ \\\hline
Islands (all)& $3 \times 10^4$ & $M_{\rm lum}$ & 86 & $ 0.95 \pm 0.37 $ & $ -2.33 \pm 0.16 $ & $ 0.00 \pm 0.83 $ & \nodata & $ -2.32 \pm 0.18 $ \\
    & $3 \times 10^4$ & $M_{\rm vir}$ & 174 & $ 5.63 \pm 2.00 $ & $ -2.03 \pm 0.09 $ & $ 1.66 \pm 1.81 $ & $ 4.41 \pm 4.10 $ & $ -1.97 \pm 0.10 $ \\
Islands (isol.)& $3 \times 10^4$ & $M_{\rm lum}$ & 43 & $ 0.36 \pm 0.13 $ & $ -2.56 \pm 0.23 $ & $ 0.00 \pm 2.29 $ & \nodata & $ -2.56 \pm 0.22 $ \\
    & $3 \times 10^4$ & $M_{\rm vir}$ & 88 & $ 1.74 \pm 0.59 $ & $ -2.16 \pm 0.13 $ & $ 1.58 \pm 1.65 $ & $ 1.61 \pm 0.71 $ & $ -2.06 \pm 0.17 $ \\
Physical (all)& $3 \times 10^4$ & $M_{\rm lum}$ & 92 & $ 0.53 \pm 0.15 $ & $ -2.59 \pm 0.18 $ & $ 0.15 \pm 1.48 $ & $ 1.81 \pm 0.25 $ & $ -2.57 \pm 0.20 $ \\
    & $3 \times 10^4$ & $M_{\rm vir}$ & 205 & $ 1.73 \pm 0.41 $ & $ -2.38 \pm 0.11 $ & $ 2.20 \pm 2.30 $ & $ 1.26 \pm 0.64 $ & $ -2.28 \pm 0.14 $ \\
Physical (isol.)& $3 \times 10^4$ & $M_{\rm lum}$ & 24 & $ 0.17 \pm 0.05 $ & $ -2.92 \pm 0.54 $ & $ 0.00 \pm 0.86 $ & \nodata & $ -2.91 \pm 0.45 $ \\
    & $3 \times 10^4$ & $M_{\rm vir}$ & 67 & $ 0.63 \pm 0.18 $ & $ -2.46 \pm 0.17 $ & $ 2.31 \pm 2.49 $ & $ 0.53 \pm 0.18 $ & $ -2.23 \pm 0.22 $ \\
Data-based (all)& $3 \times 10^4$ & $M_{\rm lum}$ & 84 & $ 0.35 \pm 0.08 $ & $ -2.83 \pm 0.18 $ & $ 0.40 \pm 1.78 $ & $ 0.61 \pm 0.21 $ & $ -2.79 \pm 0.27 $ \\
    & $3 \times 10^4$ & $M_{\rm vir}$ & 216 & $ 0.90 \pm 0.21 $ & $ -2.63 \pm 0.13 $ & $ 1.22 \pm 1.44 $ & $ 0.87 \pm 0.46 $ & $ -2.58 \pm 0.12 $ \\
Data-based (isol.)& $10^4$ & $M_{\rm lum}$ & 62 & $ 0.10 \pm 0.03 $ & $ -2.82 \pm 0.23 $ & $ 1.17 \pm 1.56 $ & $ 0.11 \pm 0.04 $ & $ -2.68 \pm 0.30 $ \\
    & $10^4$ & $M_{\rm vir}$ & 96 & $ 0.47 \pm 0.15 $ & $ -2.29 \pm 0.13 $ & $ 3.00 \pm 4.09 $ & $ 0.33 \pm 0.17 $ & $ -2.06 \pm 0.25 $ \\
\hline
\end{tabular}
}
\end{center}
\end{table*}

\begin{figure*}
\begin{center}
\includegraphics[height=0.33\textwidth]{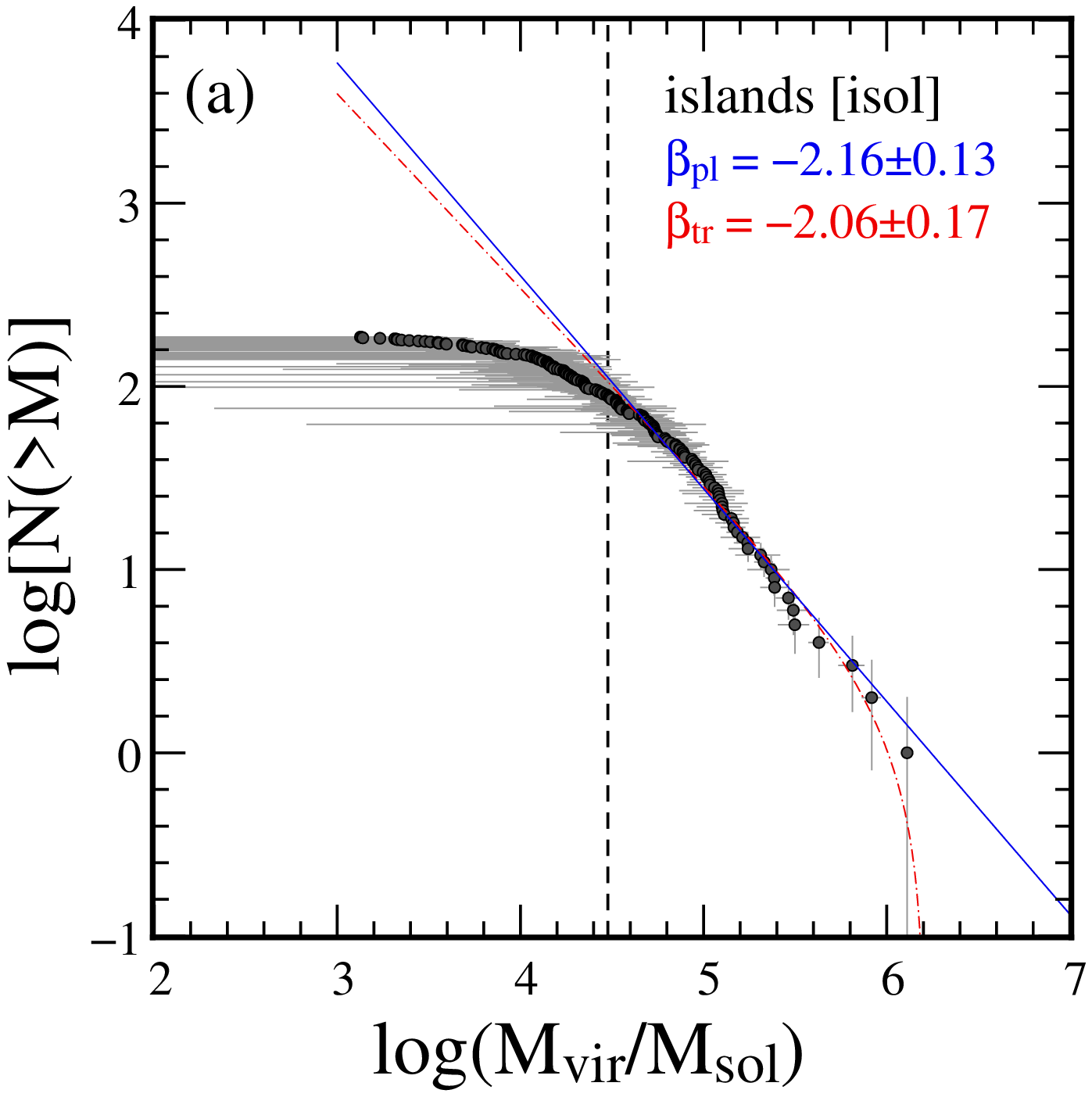}\quad
\includegraphics[height=0.33\textwidth]{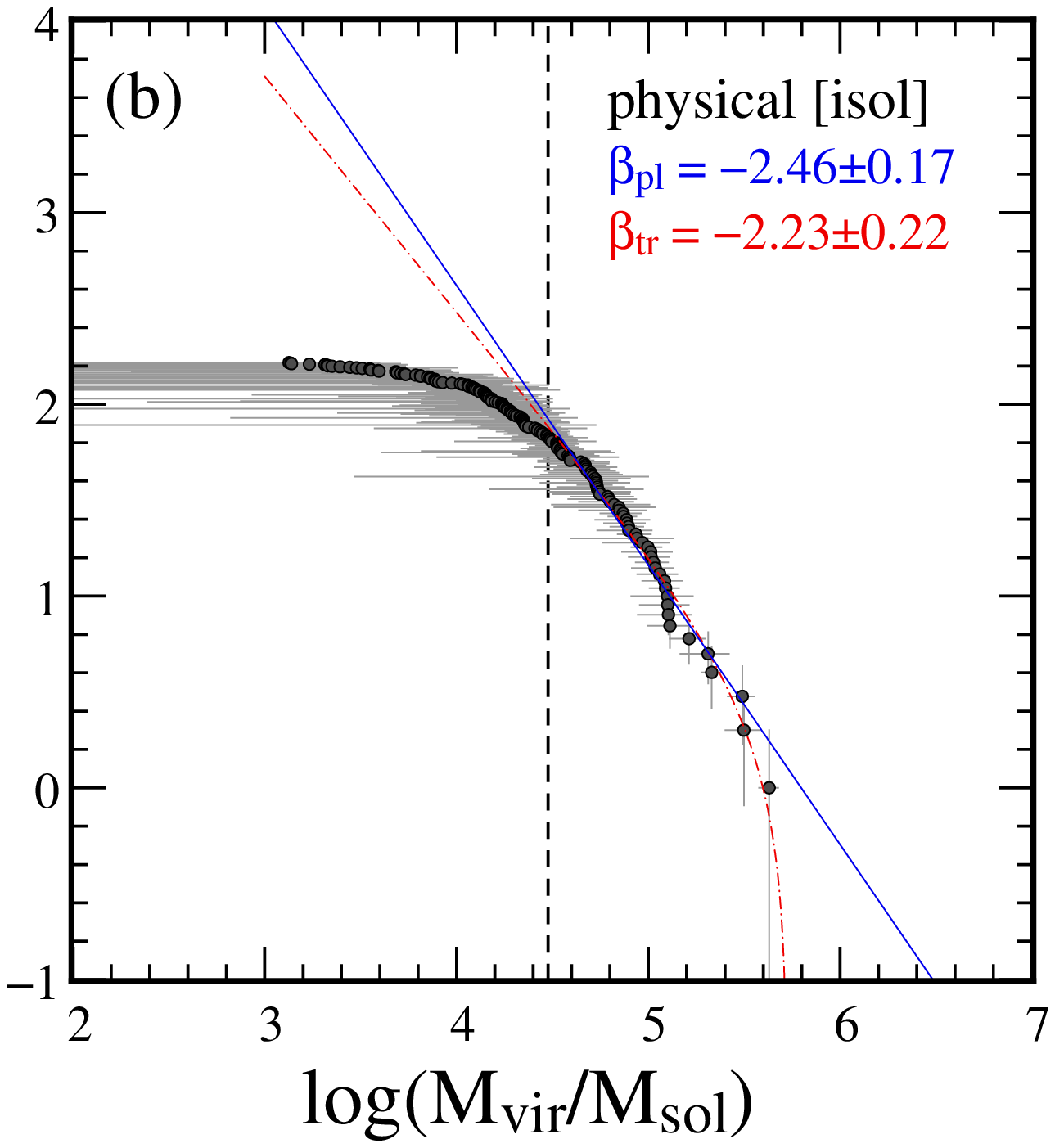}\quad
\includegraphics[height=0.33\textwidth]{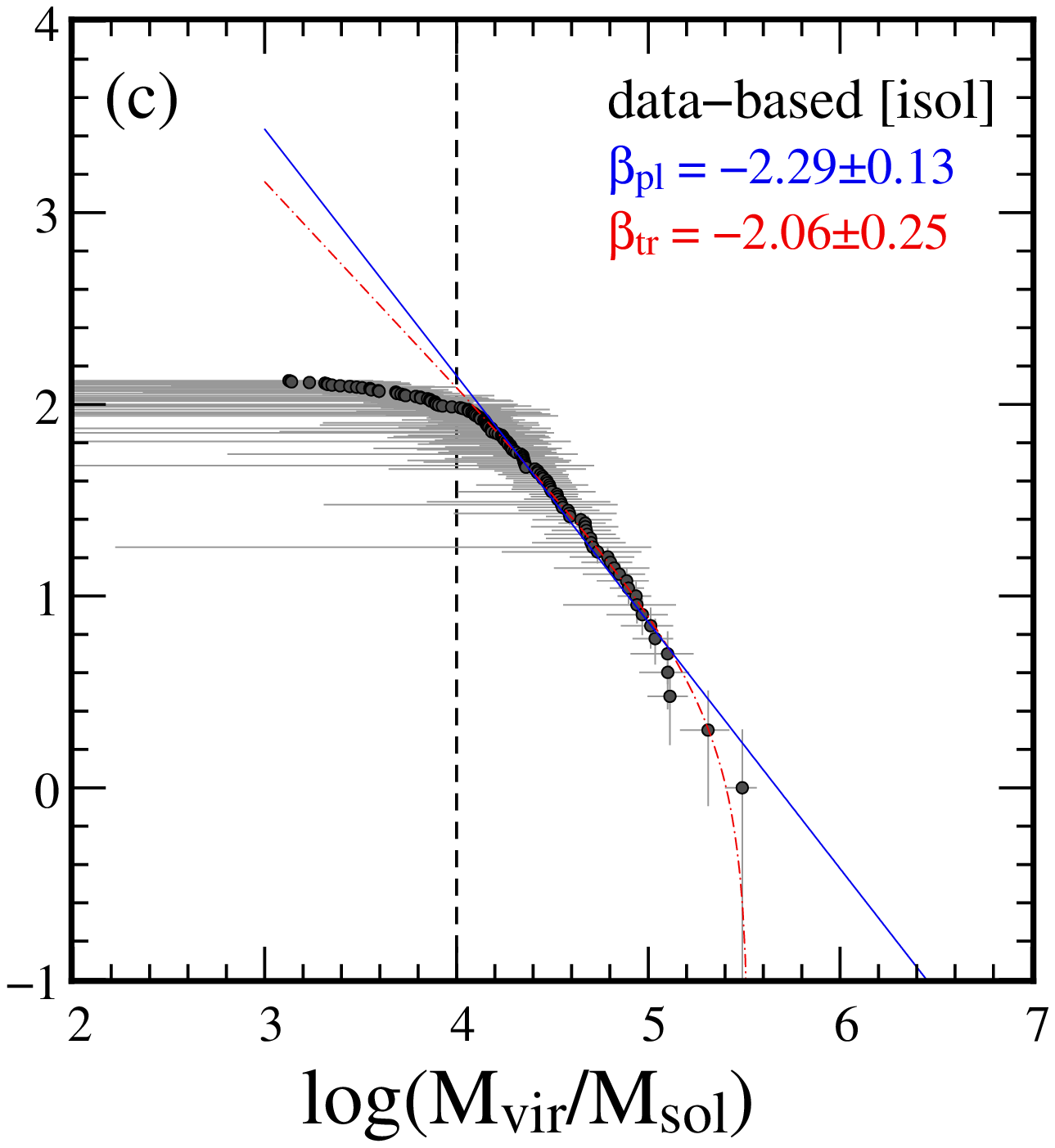}
\end{center}
\caption{
Cumulative virial mass spectra for {\it isolated} clouds defined with the islands ({\it left}), physical ({\it middle}), and data-based ({\it right}) parameter sets.  Blue solid lines are simple power-law fits to points above a completeness limit of $3 \times 10^4$ \Msol\ (vertical dashed line), or $10^4$ \Msol\ for the data-based parameter set.  Red dot-dashed lines lines are truncated power-law fits.
\label{fig:mvcumiso}}
\end{figure*}

\subsection{Distribution of Masses}

The differential cloud mass distribution, expressed as the number of clouds $N$ in a logarithmic mass interval $d\ln M$, is commonly fitted with a power law,
\begin{equation}
\frac{dN}{d\ln M} \propto M^{\beta+1}\;,
\end{equation}
where $\beta$ is the usual power law index for the mass distribution ($dN/dM \propto M^{\beta}$).  For distributions with $\beta > -2$, the total molecular mass of a galaxy is dominated by massive GMCs, while for distributions with $\beta < -2$, most of the galaxy's molecular mass is located in small clouds, and there must be a lower mass limit for the molecular clouds or a turnover in the mass distribution at low masses in order for the total mass to remain finite.  In addition, there may be an upper mass limit for clouds, beyond which the power law is truncated.

The cumulative mass distribution, being the integral of the mass distribution, has the same slope as the differential mass function expressed in logarithmic space,
\begin{equation}
N(M^\prime > M) = \left(\frac{M}{M_0}\right)^{\beta+1}\;,
\label{eqn:plaw}
\end{equation}
where $M_0$ represents the maximum mass in the sample, which generally scales with the total mass in the sample.  However, for a distribution which is {\it truncated} at $M_0$ there is a sharp rolloff in the cumulative mass distribution,
\begin{equation}
N(M^\prime > M) = N_0\left[\left(\frac{M}{M_0}\right)^{\beta+1} - 1\right]\;,
\label{eqn:trunc}
\end{equation}
because $N(M^\prime > M_0)$=0.  In this expression, discussed in more detail by \citet{McKee:97} and \citet{Rosolowsky:05b}, $N_0$ is the number of clouds more massive than $2^{1/(\beta+1)}M_0$, where the distribution begins to turn over (for a meaningful truncation to exist one expects $N_0 \gg 1$).

For this study we focus on the cumulative mass spectra since, as argued by \citet{Rosolowsky:05b}, these do not require the choice of a bin size to generate a histogram.  We apply the ``error in variables'' method employed by \citet{Rosolowsky:05b} to incorporate the uncertainty in the independent variable (cloud mass) into the functional fits.  The method, implemented in an IDL program provided by E. Rosolowsky, uses an iterative maximum-likelihood approach to jointly estimate the ``true'' sets of $M$ and $N$ (i.e.\ without measurement errors) and the best-fit parameters $M_0$, $N_0$, and $\beta$.  Figure~\ref{fig:mlcum} shows the cumulative {\it luminous} mass spectra for the clouds derived from the three different decomposition methods, while Figure~\ref{fig:mlcumiso} shows the same distributions considering only ``isolated'' clouds, as defined in Section~\ref{sec:reliability}.  Truncated and non-truncated fits, using the functions given in Equations~(\ref{eqn:plaw}) and (\ref{eqn:trunc}), are shown as red and blue curves respectively.  Fitting is only performed above the completeness limit of $M_{\rm min} = 3 \times 10^4$ \Msol\ shown as a vertical dashed line, except for the isolated data-based clouds (Figure~\ref{fig:mlcumiso}c) for which the fit was performed above $M_{\rm min} = 10^4$ \Msol\ in order to allow for a sufficient number of points in the fitting.  In most cases the power-law and truncated fits yield similar slopes, indicating that there is little evidence for truncation.  Indeed, none of the fits satisfies the $N_0 \gg 1$ criterion for a meaningful truncation to exist.

Figures~\ref{fig:mvcum} and \ref{fig:mvcumiso} show the corresponding cumulative {\it virial} mass spectra.  The virial mass spectra tend to be less steep than the luminous mass spectra and are generally consistent with a slope of $-2$.  Again, we find little clear evidence for a truncation at the high-mass end.
The power-law fit results, with and without a high-mass truncation, are summarized in Table~\ref{tbl:mspec}.

\begin{figure}
\begin{center}
\includegraphics[width=0.45\textwidth]{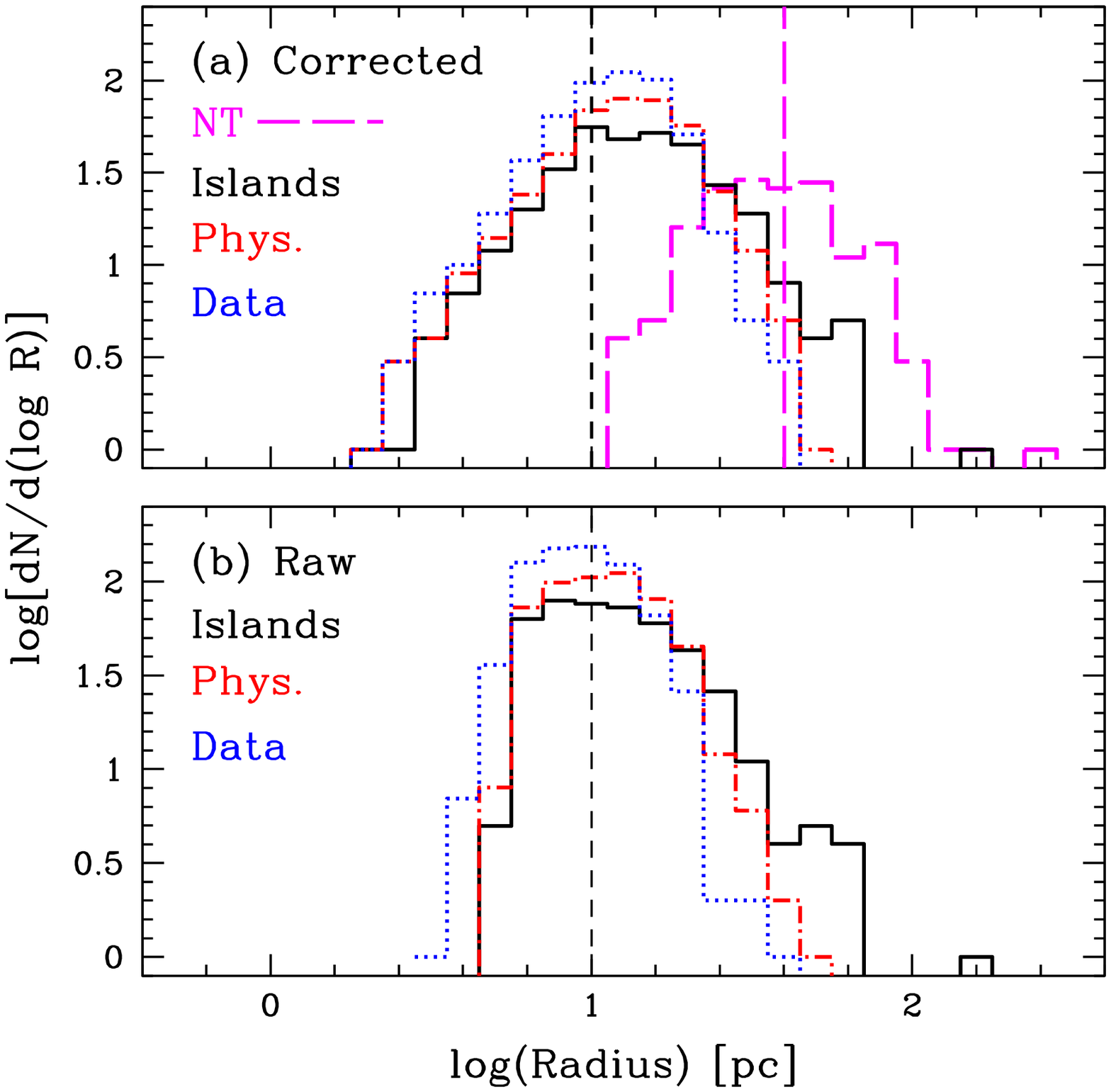}
\end{center}
\caption{
(a) Distributions of the logarithm of the cloud radii, corrected for sensitivity and resolution effects, for the three different decomposition sets: islands (black), physical (red), and data-based (blue).  The NANTEN GMCs of Fu08 are also shown as a magenta histogram.  Vertical lines indicate the resolutions of the MAGMA (black) and NANTEN (magenta) surveys respectively.  (b) Distributions of radii, uncorrected for sensitivity or resolution effects, for the three different decomposition sets.
\label{fig:rhist}}
\end{figure}

\subsection{Distribution of Radii}

Figure~\ref{fig:rhist}(a) shows the distribution of cloud radii for the three different decomposition sets.  The NANTEN GMCs of Fu08 are also shown as a magenta histogram.  There appears to be a preferred scale in each distribution corresponding to the observational resolution (vertical dashed lines at $\sim$10 pc for MAGMA and $\sim$40 pc for NANTEN).  A straightforward interpretation is that clouds near or below the resolution limit are incompletely sampled, while clouds larger than the resolution limit tend to be resolved into smaller clouds.  Figure~\ref{fig:rhist}(b) shows the distributions before sensitivity and resolution corrections are performed.  As expected, these distributions show sharp cutoffs below the resolution limit.

The tendency for CO cloud sizes to be close to the observing resolution has been noted in a previous study of LMC clouds by \citet{Israel:03}, and makes it difficult to recover the intrinsic properties of molecular clouds from even automated decomposition methods.  Undoubtedly this is simply a reflection of the hierarchical structure of molecular clouds \citep[e.g.,][]{Falgarone:91}.  We therefore place greater emphasis on the results derived for the non-decomposed islands, since these represent the largest connected CO structures.  Although identification of these structures is still subject to resolution and sensitivity biases (due to blending in crowded regions and possible failure to detect weak bridging structures), the islands span the largest ranges in size, linewidth, and flux, making them useful to study trends in cloud properties.

\begin{figure*}
\begin{center}
\includegraphics[width=0.31\textwidth]{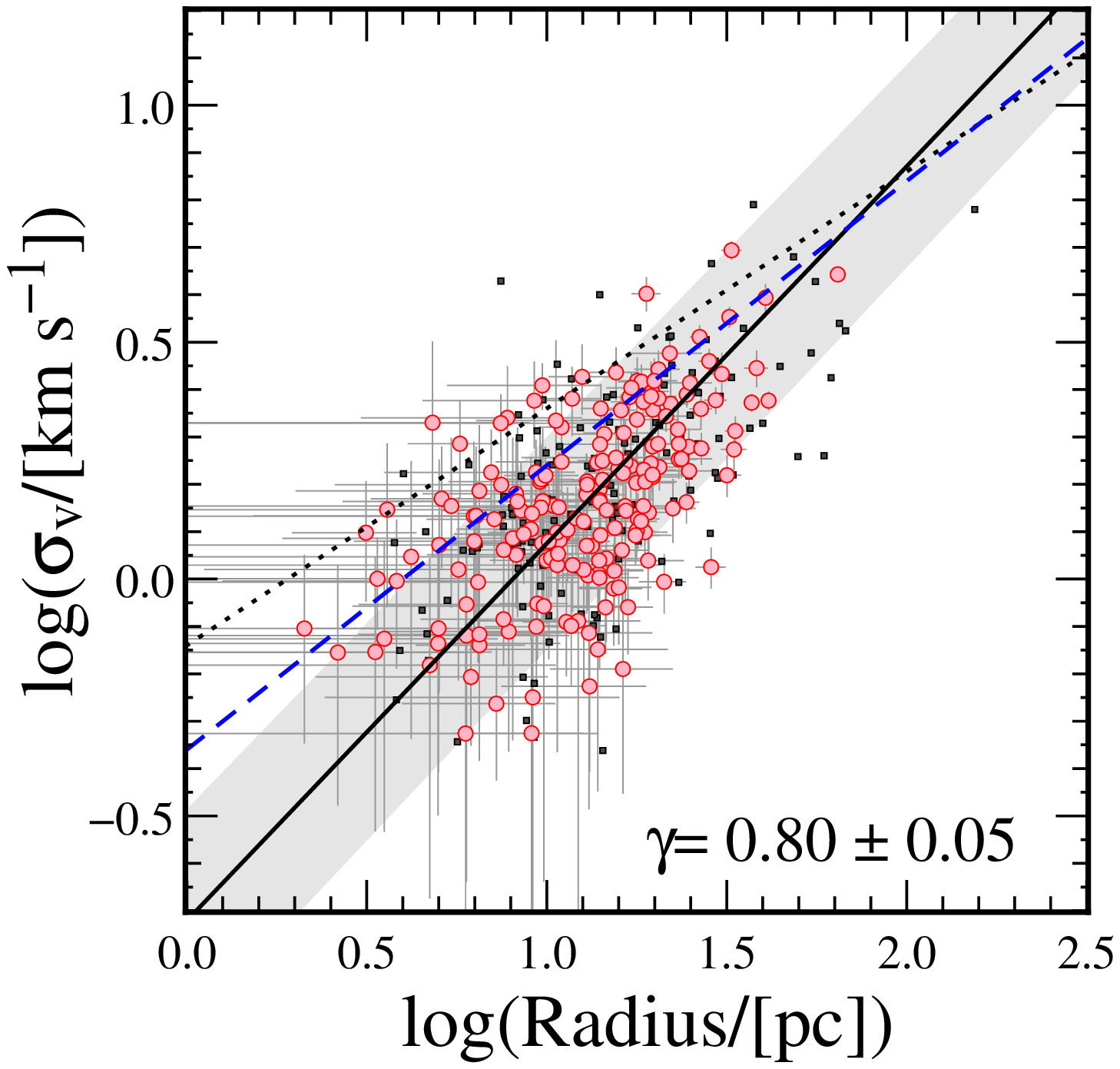}\quad
\includegraphics[width=0.31\textwidth]{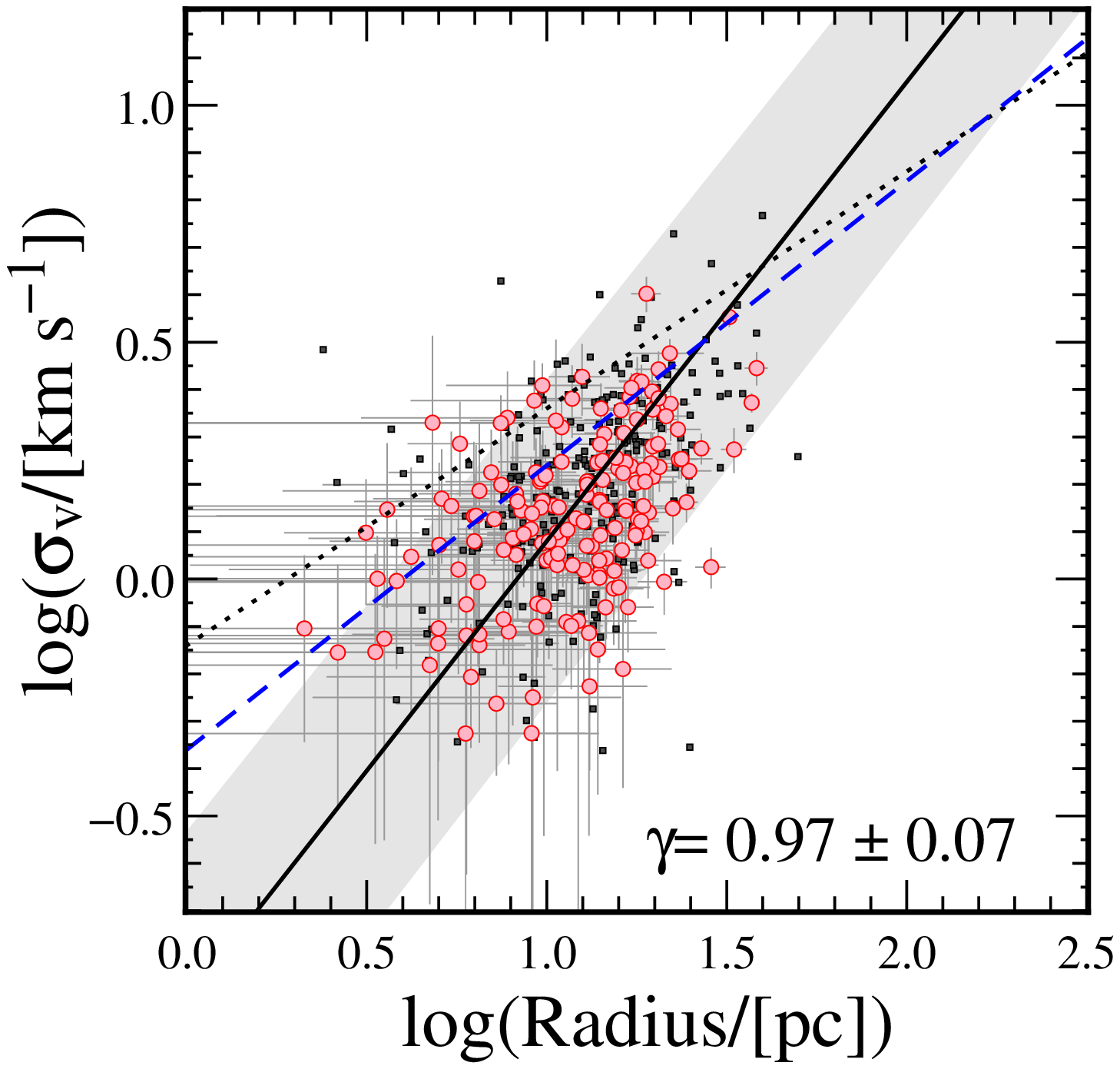}\quad
\includegraphics[width=0.31\textwidth]{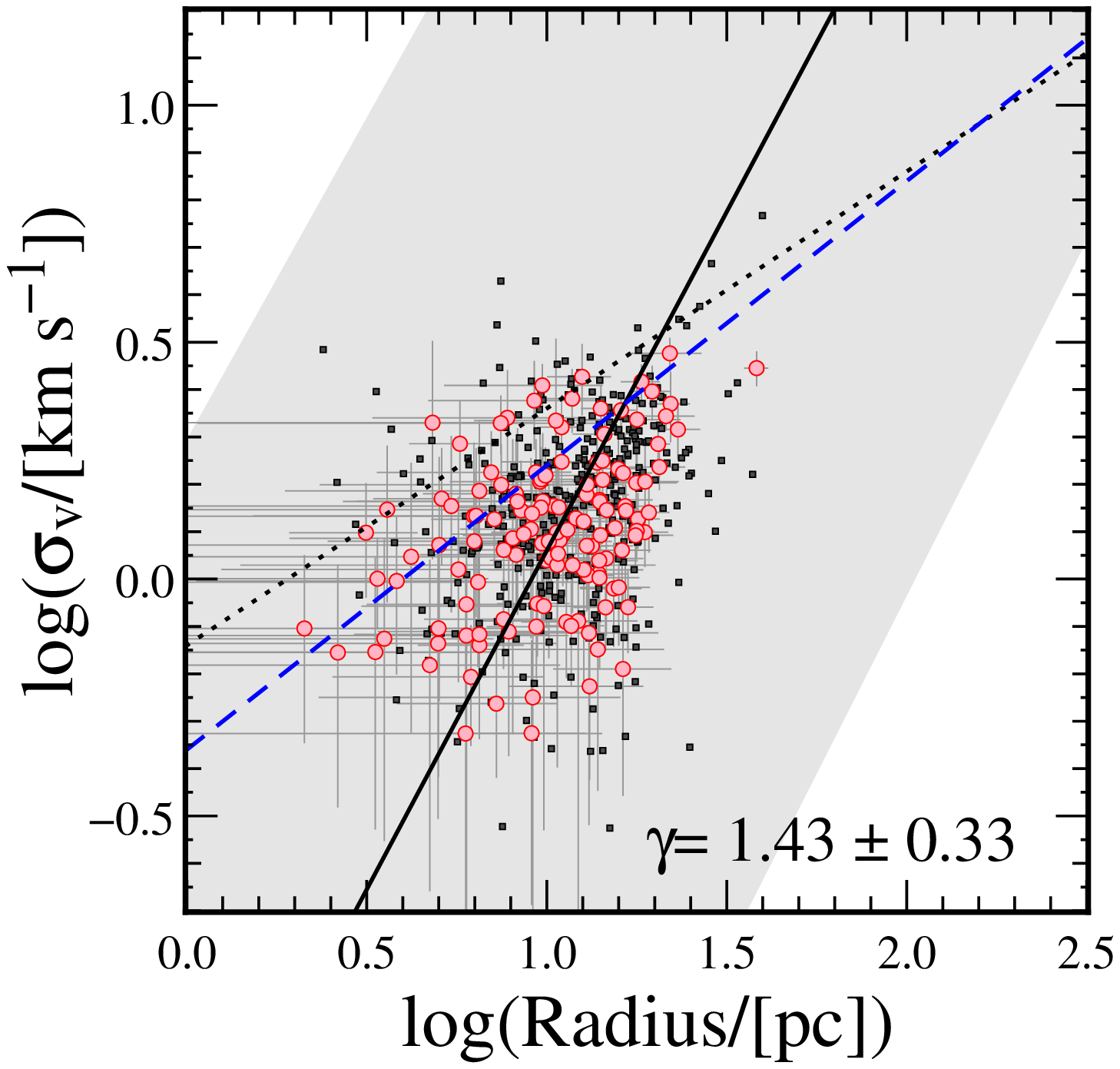}
\end{center}
\caption{
Size-linewidth relation for the islands ({\it left}), physical ({\it middle}), and data-based ({\it right}) parameter sets.  The solid black line is the fitted relation using the BCES bisector, with the gray regions indicating the uncertainty in slope.  The mean relations derived for the inner Galaxy by \citet{Solomon:87} and for nearby galaxies by \citet{Bolatto:08} are shown as a black dotted line and dashed blue line respectively.  Larger (red) symbols represent isolated clouds, with the remaining clouds shown as small black symbols.
\label{fig:rdv}}
\end{figure*}

\begin{figure*}
\begin{center}
\includegraphics[width=0.31\textwidth]{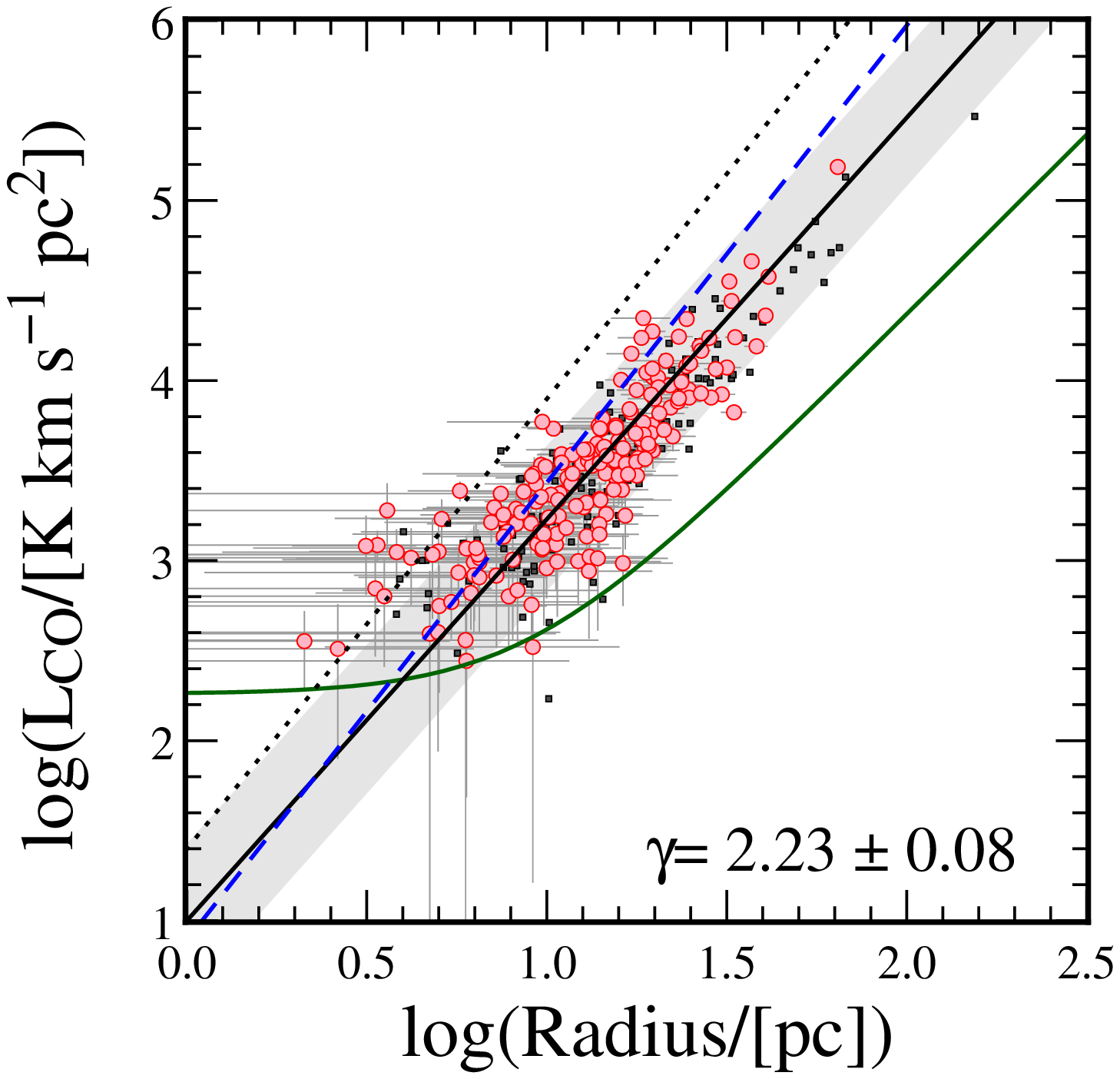}\quad
\includegraphics[width=0.31\textwidth]{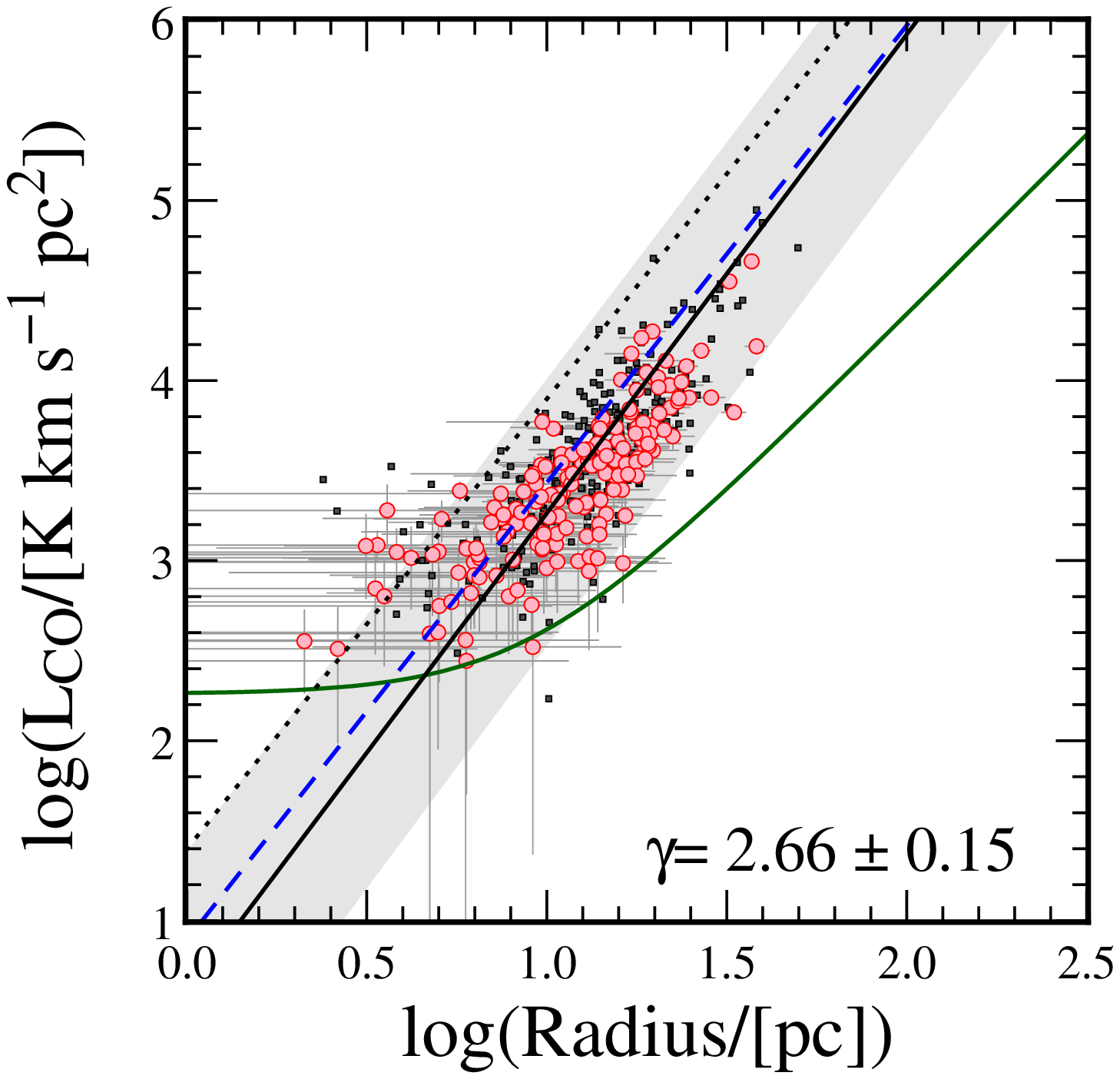}\quad
\includegraphics[width=0.31\textwidth]{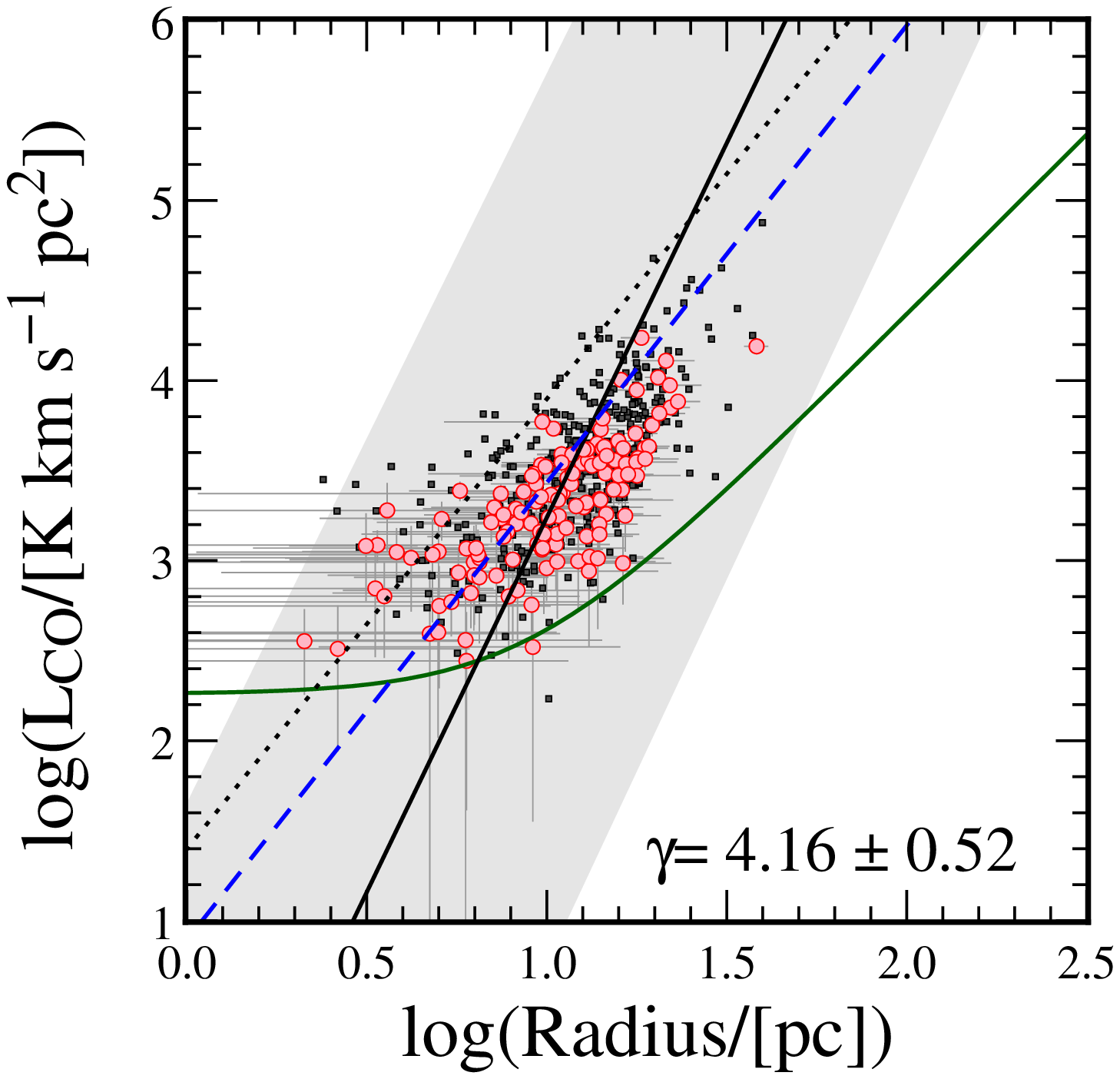}
\end{center}
\caption{
Size-luminosity relation for the islands ({\it left}), physical ({\it middle}), and data-based ({\it right}) parameter sets.  The solid black line is the fitted relation using the BCES bisector, with the gray regions indicating the uncertainty in slope.  The mean relations derived for the inner Galaxy by \citet{Solomon:87} and for nearby galaxies by \citet{Bolatto:08} are shown as a black dotted line and dashed blue line respectively.  Larger (red) symbols represent isolated clouds.  A solid green curve represents the estimated sensitivity limit.
\label{fig:rml}}
\end{figure*}

\subsection{Larson-type Relations}

\begin{figure*}
\begin{center}
\includegraphics[width=0.31\textwidth]{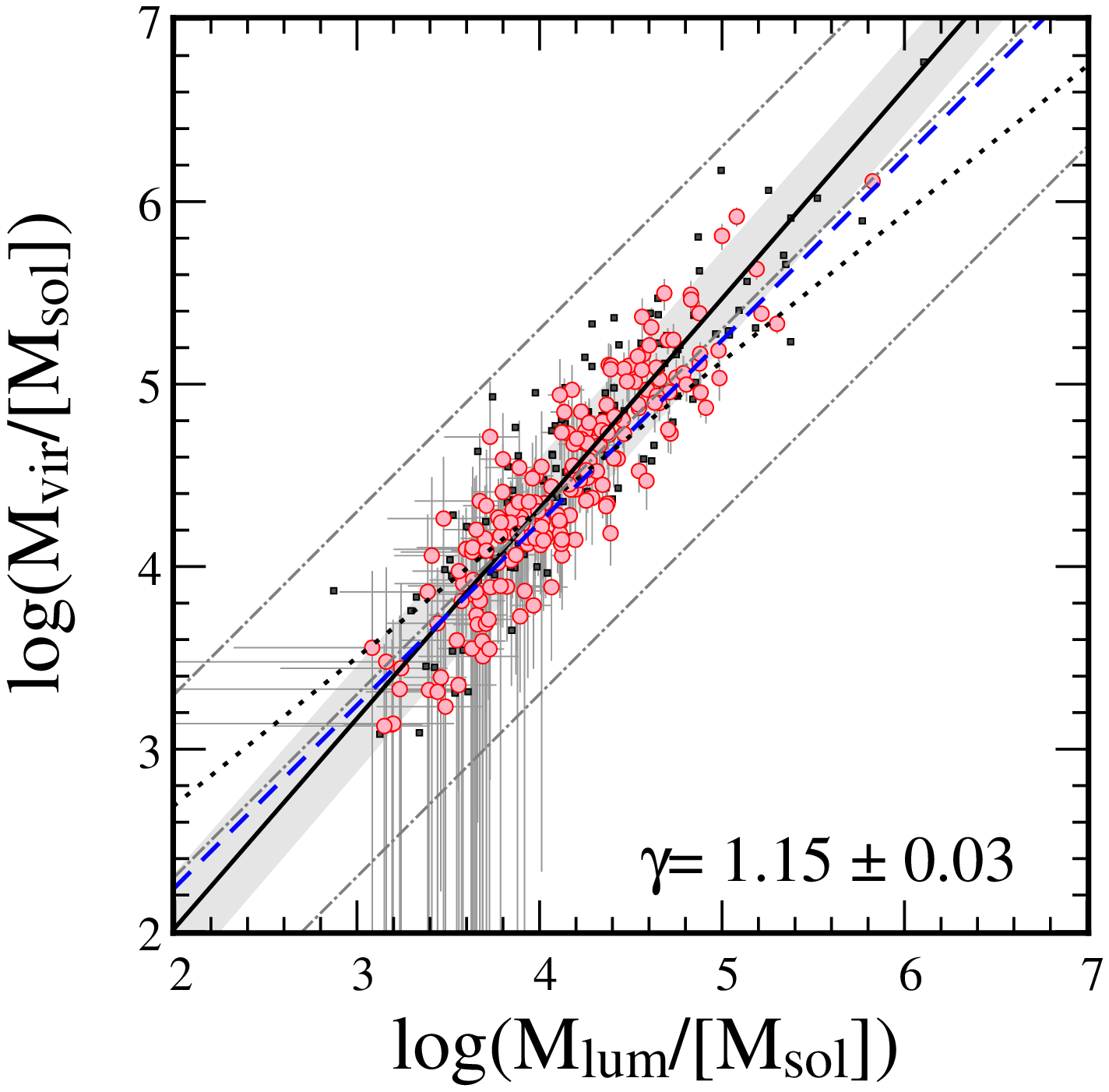}\quad
\includegraphics[width=0.31\textwidth]{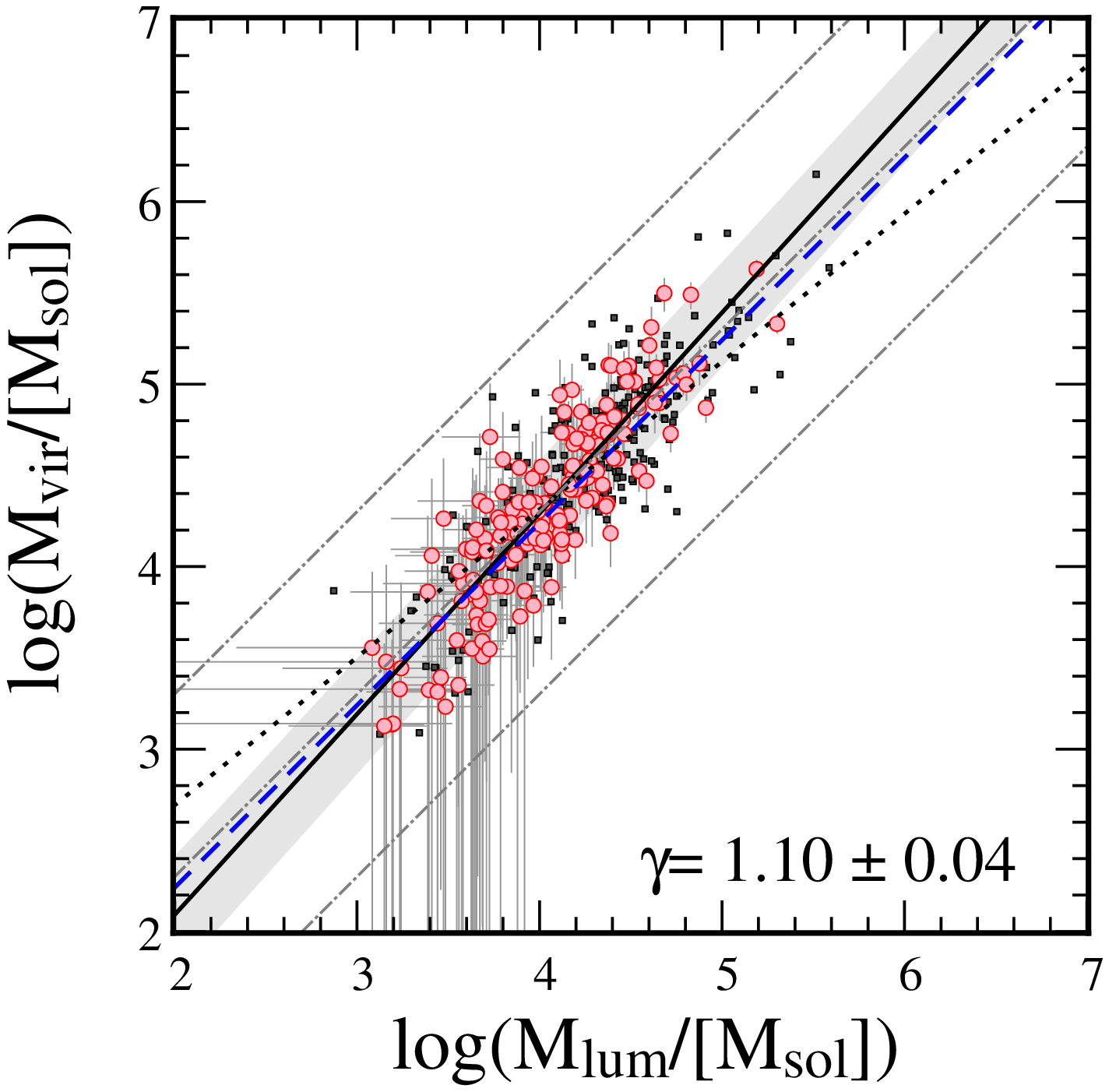}\quad
\includegraphics[width=0.31\textwidth]{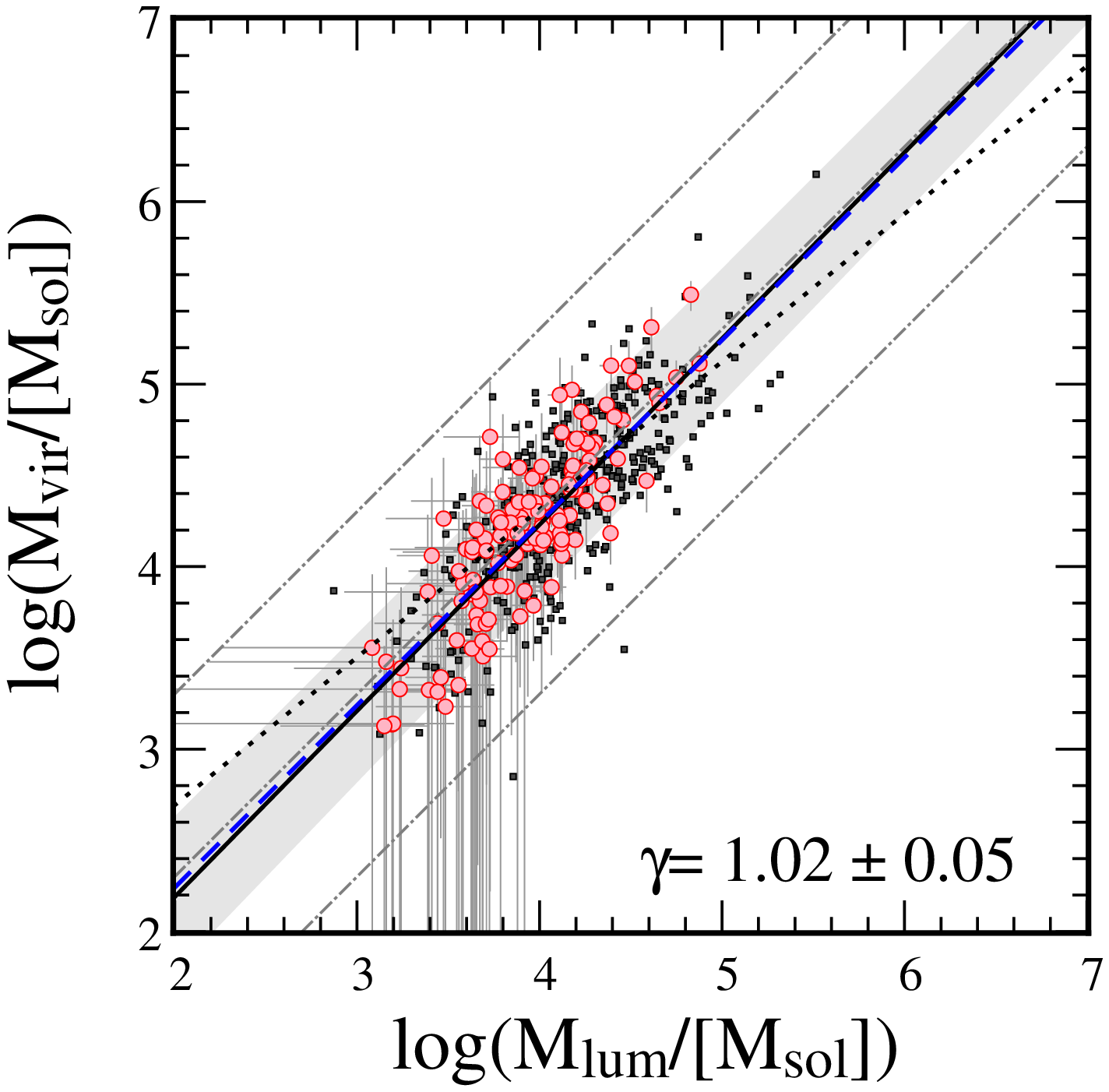}
\end{center}
\caption{
Correlation between luminous and virial mass for the islands ({\it left}), physical ({\it middle}), and data-based ({\it right}) parameter sets.  The solid black line is the fitted relation using the BCES bisector, with the gray regions indicating the uncertainty in slope.  The mean relations derived for the inner Galaxy by \citet{Solomon:87} and for nearby galaxies by \citet{Bolatto:08} are shown as a black dotted line and dashed blue line respectively.  Larger (red) symbols represent isolated clouds.  Gray diagonal lines indicate \xco=0.4, 4, and $40 \times 10^{20}$ \xcou.
\label{fig:mlmv}}
\end{figure*}

Correlations among cloud properties are frequently discussed as indicators of their surface densities and dynamical state \citep[][Hu10]{Bolatto:08}.  We have plotted the correlations between size and linewidth, size and luminosity, and luminosity and virial mass in Figures~\ref{fig:rdv}, \ref{fig:rml}, and \ref{fig:mlmv} respectively.  Each of the three cloud sets is plotted in a separate panel.  Red symbols highlight ``isolated'' clouds, which can be uniquely matched to an island defined out to a 1.5$\sigma$ edge.  Note, however, that many ``isolated'' clouds will break into multiple substructures upon further decomposition, so the number of isolated clouds decreases in going from panel (a) to (c) even as the total number of identified clouds increases (Table~\ref{tbl:cprops_summ}).

Following Hu10, we fit a linear regression to each panel using the `BCES' (bivariate, correlated errors with intrinsic scatter) method of \citet{Akritas:96}, as implemented in a FORTRAN program distributed by M. Bershady.  
This method takes into account the uncertainty in each individual measurement, as estimated by CPROPS.  
We have assumed that measurement errors in the two plotted variables are uncorrelated, although some pairs of parameters (e.g., $M_{\rm lum}$ and $M_{\rm vir}$) are expected to have substantial covariance.  
We plot in each panel of Fig.~\ref{fig:rdv}--\ref{fig:mlmv} a line representing the BCES bisector fit, which bisects the results obtained by choosing either the $X$ or $Y$ axis as the independent variable.  
The gray shaded region around the fit provides a rough estimate of the uncertainty in the slope $\gamma$ of the bisector fit, as delimited by two lines of slope $\gamma-\sigma_\gamma$ and $\gamma+\sigma_\gamma$ constrained to pass through the center of the scatterplot (as defined by the mean values of the logarithms of the two parameters).

For the ``islands'' decomposition set, the best-fit relation for the size-linewidth relation is given by
\begin{equation}
\log \sigma_v = (-0.72 \pm 0.05) + (0.80 \pm 0.05)\log R
\end{equation}
where $\sigma_v$ is measured in \kms\ and $R$ is measured in pc.  The size-luminosity relation can be fit with
\begin{equation}
\log L_{\rm CO} = (1.00 \pm 0.10) + (2.23 \pm 0.08)\log R
\end{equation}
where $L_{\rm CO}$ is measured in \kkms\ pc$^{-2}$.  Finally, the correlation between CO and virial mass can be expressed as
\begin{equation}
\log M_{\rm vir} = (-0.28 \pm 0.14) + (1.15 \pm 0.03)\log M_{\rm lum}
\end{equation}
where $M_{\rm lum}$ is the CO-derived mass assuming a Galactic value for the \xco-factor.  Figures~\ref{fig:rdv}--\ref{fig:mlmv} indicate that the slopes of the size-linewidth and size-luminosity relations tend to become steeper and more scattered as one decomposes the CO emission further (``physical'' and ``data-based'' sets).  Further decomposition primarily affects the size distribution, concentrating it near the observational resolution [see also Fig.~\ref{fig:rhist}(a)] and resulting in steeper relationships with linewidth and luminosity.  Restricting the fits to ``isolated'' clouds also leads to a slight steepening of the size-linewidth and size-luminosity relations, since the largest clouds again tend to be excluded, although the differences in slope are at the 1--1.5$\sigma$ level.

Hu10 obtained quite similar fits to the MAGMA clouds using the BCES bisector, although their size-luminosity relation is somewhat less steep ($\gamma=1.88\pm0.08$) than derived here.  The Hu10 analysis differs from the present one in that an earlier version of the MAGMA cube with somewhat poorer sensitivity is used, and slightly different criteria are employed to identify true emission.  Hu10 also filter out many of the weakest clouds based on signal-to-noise considerations, resulting in few or no clouds with deconvolved linewidths $<$1 \kms\ or radii $<$10 pc, whereas the catalogs presented here contain substantial numbers of such clouds, due in part to the improved sensitivity.

\begin{figure*}
\begin{center}
\includegraphics[width=0.31\textwidth]{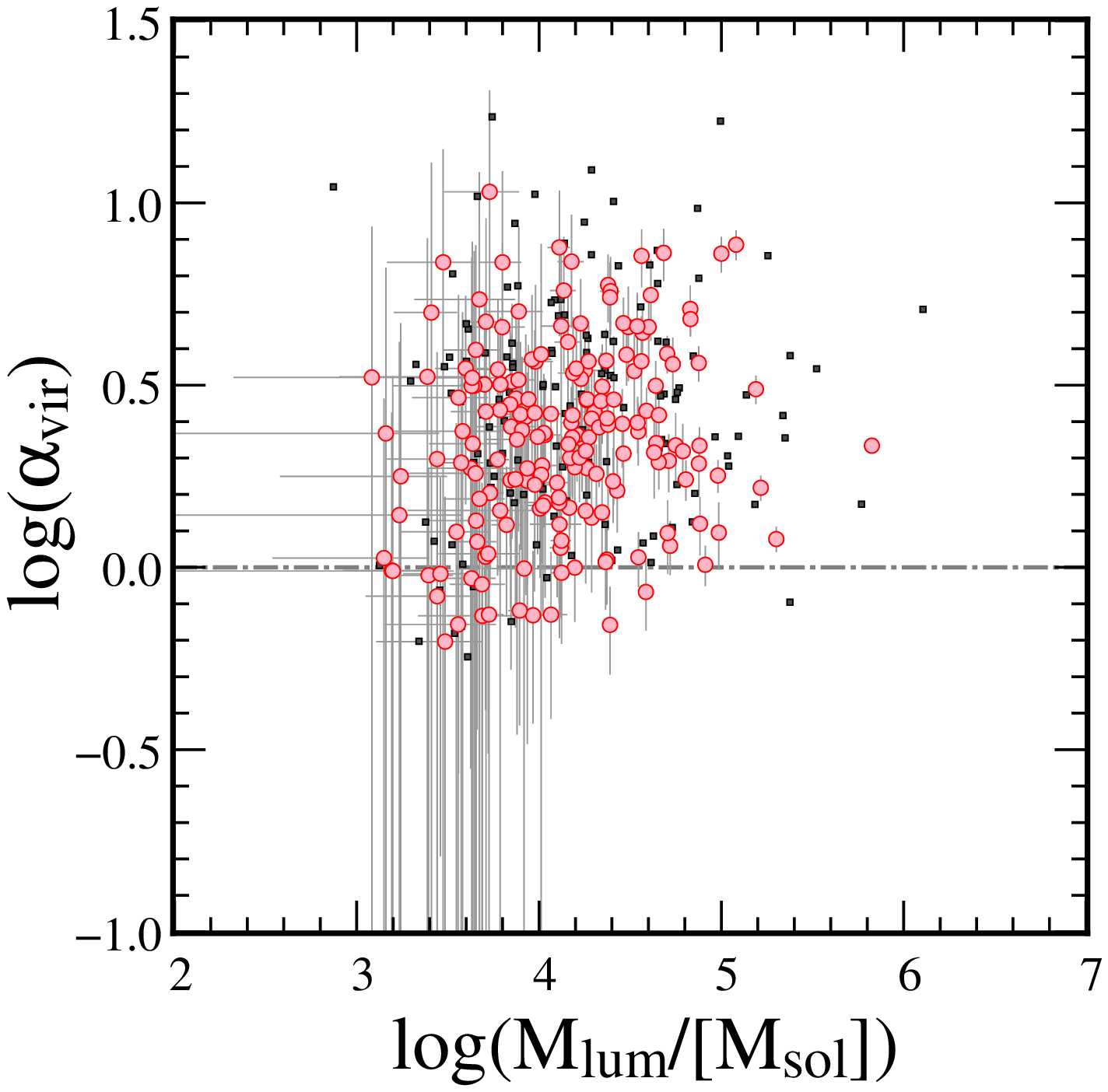}\quad
\includegraphics[width=0.31\textwidth]{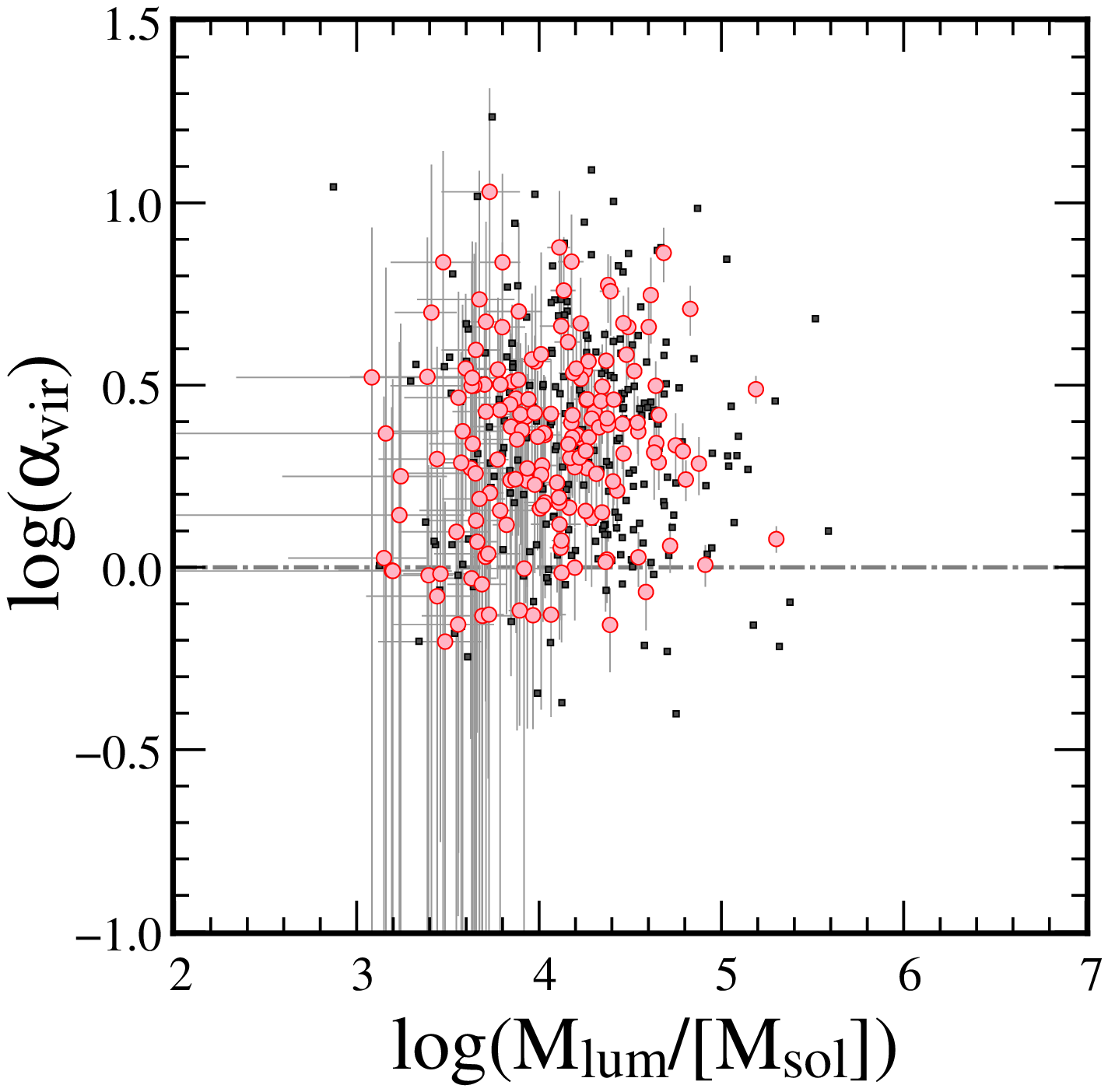}\quad
\includegraphics[width=0.31\textwidth]{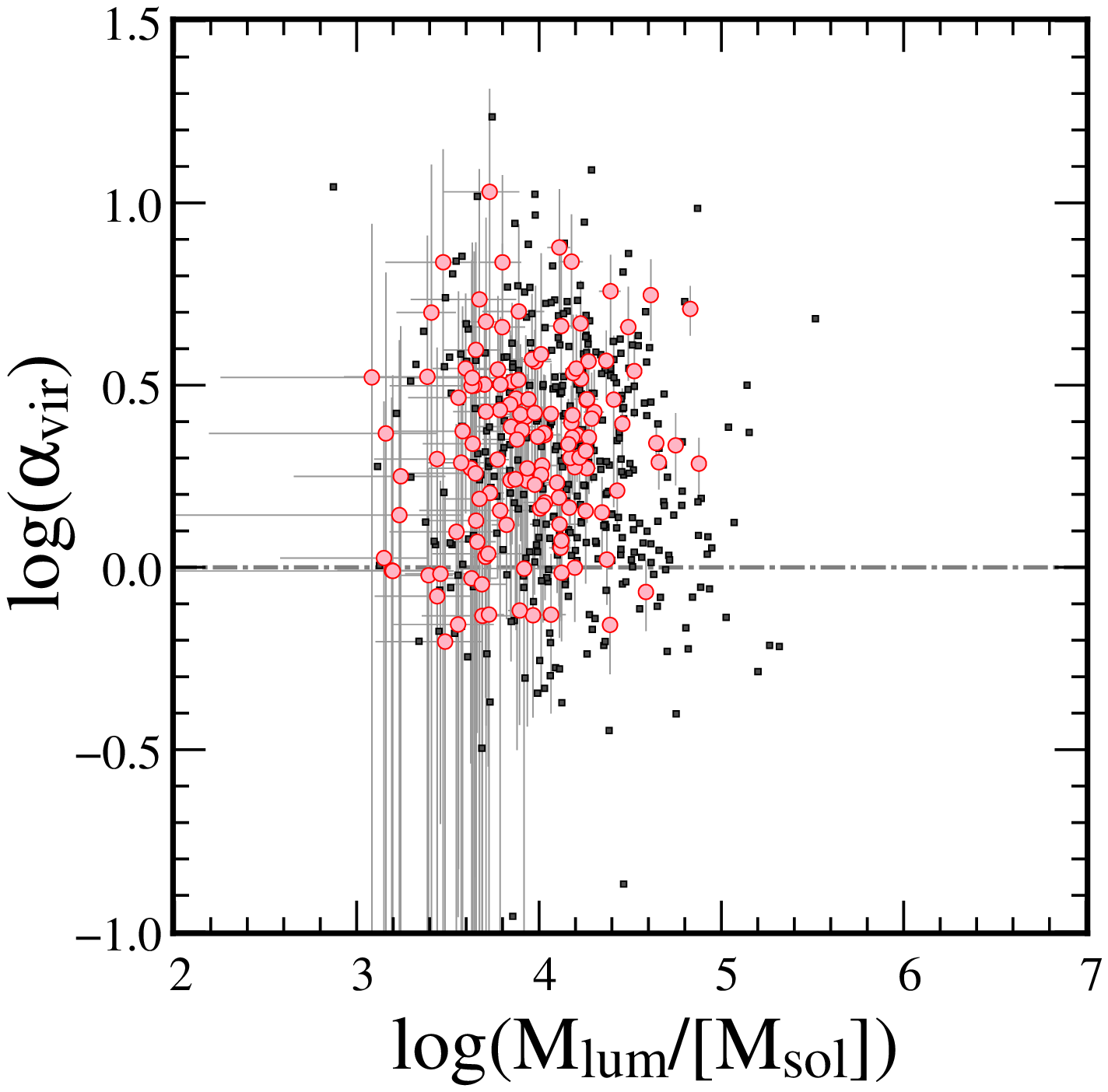}
\end{center}
\caption{
Virial parameter as a function of luminous mass for the islands ({\it left}), physical ({\it middle}), and data-based ({\it right}) parameter sets.  Larger (red) symbols represent isolated clouds.  The gray horizontal line represents $\alpha$=1.
\label{fig:alpha}}
\end{figure*}

\subsection{Virial Parameter}

Figure~\ref{fig:alpha} shows the virial parameter, defined as
\begin{equation}
\alpha_{\rm vir} = \frac{5\sigma_v^2R}{GM_{\rm lum}} = 1.12\frac{M_{\rm vir}}{M_{\rm lum}}\;,
\label{eqn:alpha}
\end{equation}
plotted as a function of luminous mass for the three decomposition sets, with the gray horizontal line representing $\alpha_{\rm vir}$=1.  Under the assumption that CO luminosity faithfully traces cloud mass, the condition $\alpha_{\rm vir} \sim 1$ is usually interpreted as being gravitationally bound \citep[cf.][]{Bertoldi:92}, and $\alpha_{\rm vir} \gg 1$ as being confined by external pressure (which is not included in the calculation of $\alpha_{\rm vir}$).  
The normalization of $\alpha_{\rm vir}$ depends on the appropriate value for \xco, so a value of \xco\ is often chosen to bring the median value of $\alpha_{\rm vir}$ close to 1 (e.g., Hu10).
In this paper we do not calibrate \xco\ in this way, but simply examine Figure~\ref{fig:alpha} for evidence of decreasing $\alpha_{\rm vir}$ with increasing cloud mass, as has been previously identified in the outer Galaxy by \citet{Heyer:01}.
However, we find no clear tendency for more luminous clouds to be more gravitationally bound (with $\log\alpha_{\rm vir} \sim 0$), although many of the ``new'' clouds that arise from the decomposition of larger clouds have values of $\log\alpha_{\rm vir} \lesssim 0$.

\begin{table}
\begin{center}
\caption{CO Properties of Candidate YSOs\label{tbl:yso}}
\begin{tabular}{lcccc}
\hline
 & \multicolumn{2}{c}{GC09} & \multicolumn{2}{c}{W08}\\
Property & All & [8.0]$<$8 & All & [8.0]$<$8\\\hline
Number of sources & 1172 & 247 & 1197 & 170\\
NANTEN observed & 1147 & 246 & 1141 & 164\\
NANTEN detected &  669 & 191 &  367 &  60\\
 MAGMA observed &  722 & 196 &  421 &  67\\
 MAGMA detected &  552 & 171 &  259 &  46\\
 MAGMA non-det, NANTEN det. & 99 & 15 & 75 & 9\\
\hline
\end{tabular}
\end{center}
\end{table}

\begin{figure*}
\begin{center}
\includegraphics[angle=-90,width=\textwidth]{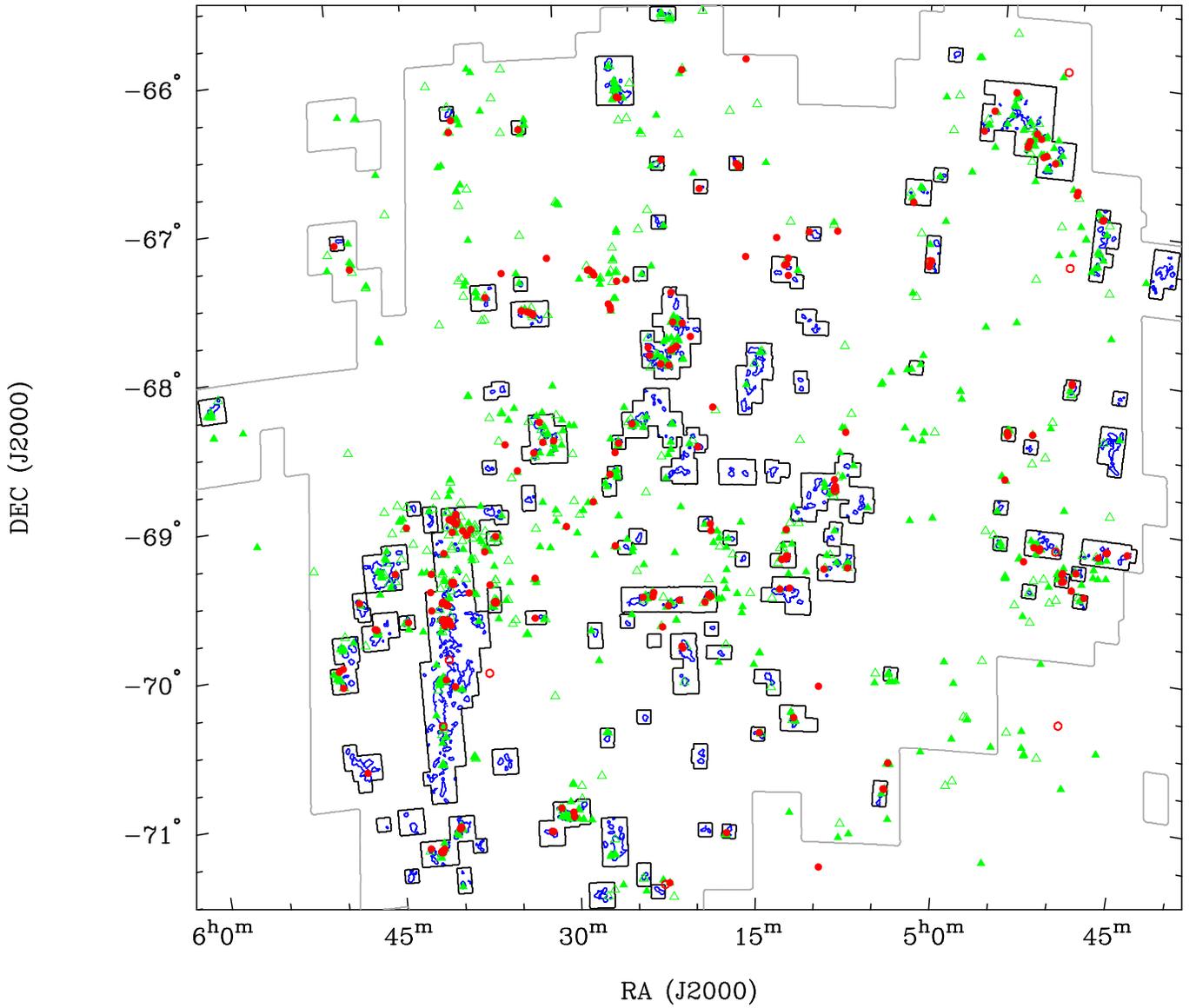}
\end{center}
\caption{
Comparison of CO emission with young stellar populations in the LMC.  The jagged gray contour at the periphery of the map shows the region surveyed in CO by NANTEN.  Multiple rectangular black contours show the regions mapped in CO with Mopra, based on the NANTEN map.  Blue contours show integrated CO intensity at a level of 1 K km s$^{-1}$.  Filled and open symbols represent ``definite'' and ``probable'' YSOs respectively from Gruendl \& Chu (2009), red circles denoting the high-mass sample and green triangles denoting the intermediate-mass sample.
\label{fig:magmayso}}
\end{figure*}

\begin{figure*}
\begin{center}
\includegraphics[width=0.4\textwidth]{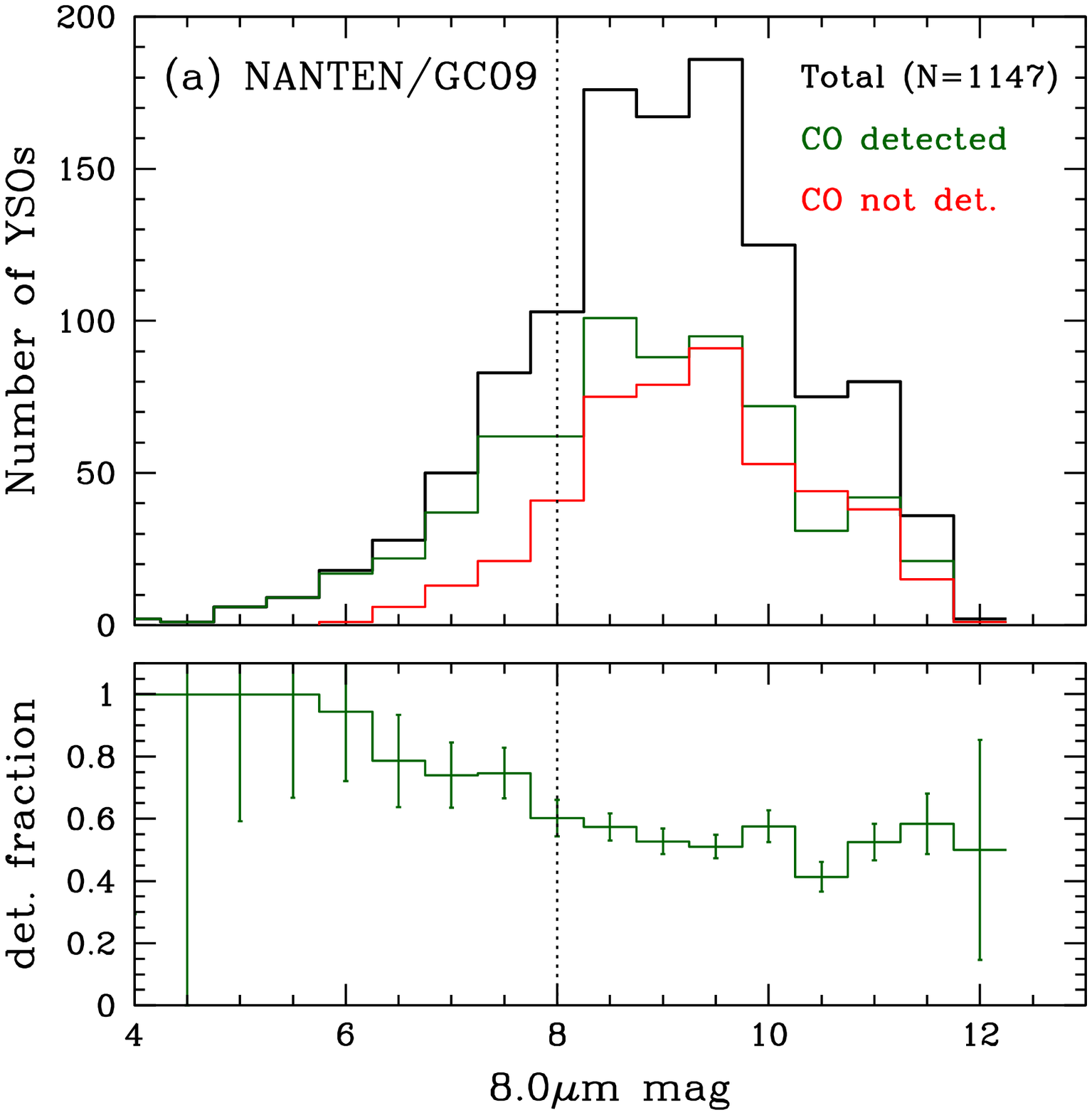}\quad\quad
\includegraphics[width=0.4\textwidth]{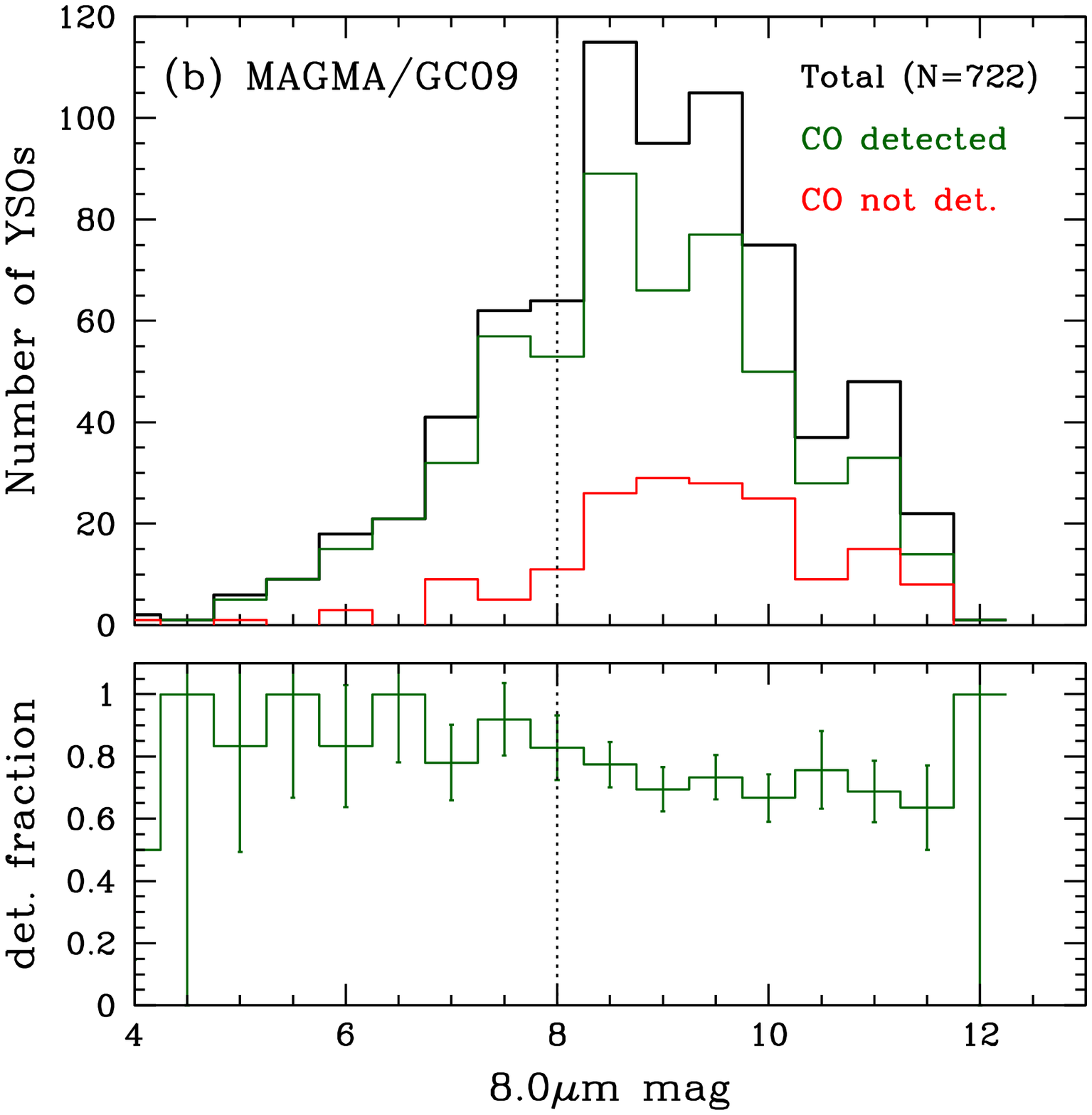}\\[4ex]
\includegraphics[width=0.4\textwidth]{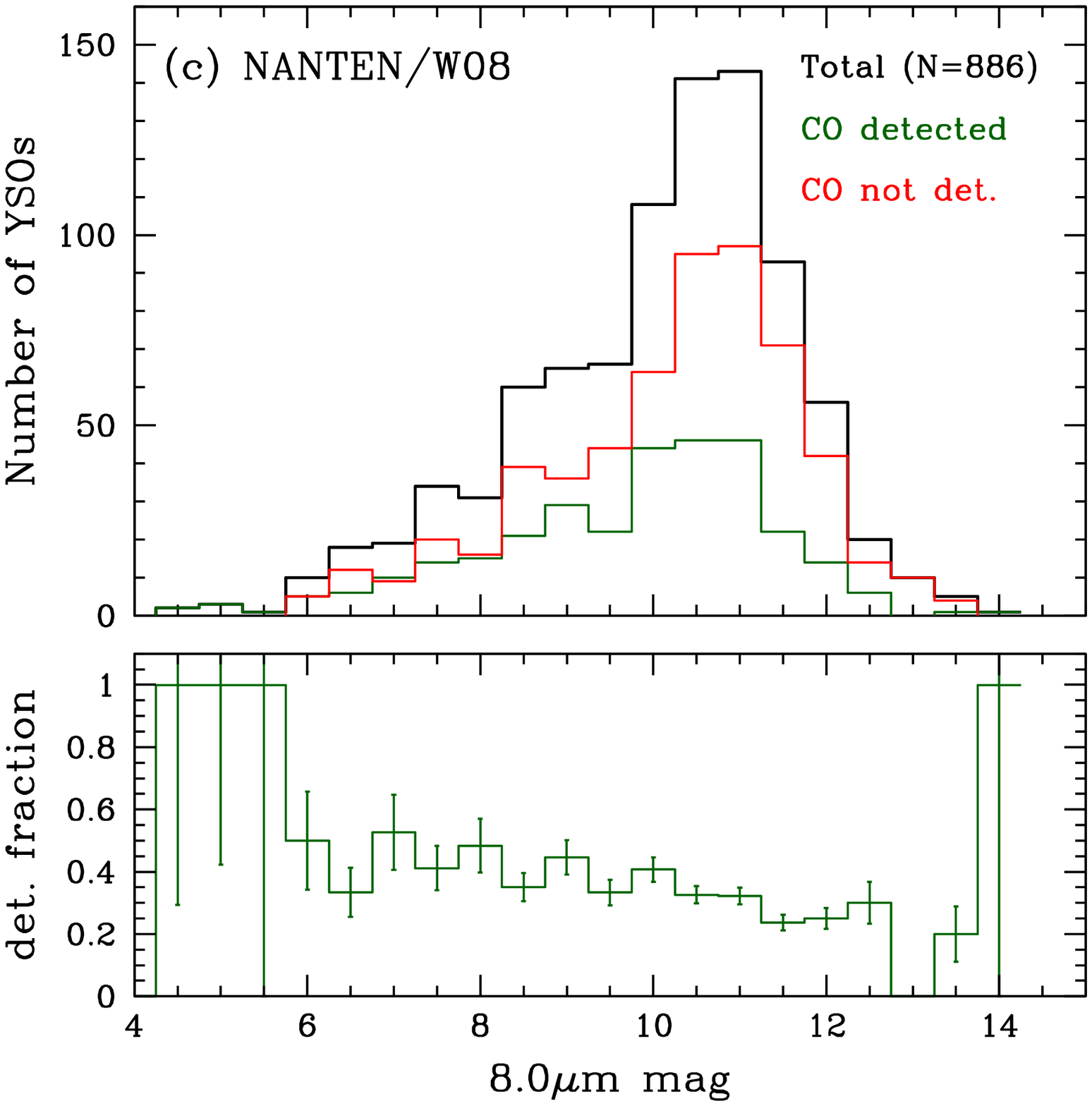}\quad\quad
\includegraphics[width=0.4\textwidth]{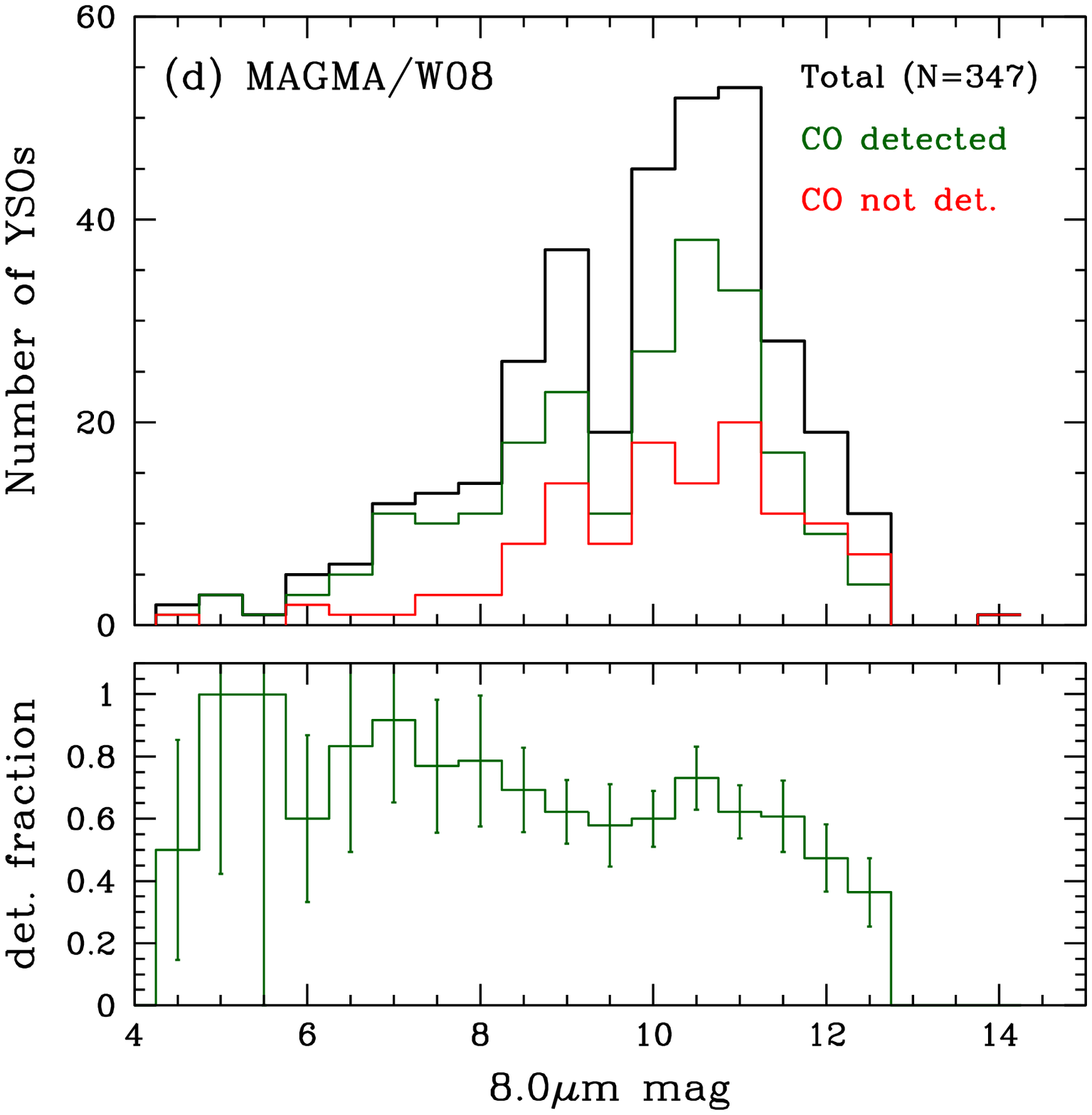}\\
\end{center}
\caption{
Correspondence between YSO catalogs and CO emission.  Histograms show the distribution of 8$\mu$m fluxes for (a) CO from NANTEN, YSO list from GC09; (b) CO from MAGMA, YSO list from GC09; (c) CO from NANTEN, YSO list from W08; (d) CO from MAGMA, YSO list from GC09.  Only YSOs covered by the specific CO survey are shown, with green and red histograms corresponding to detections and non-detections respectively.  Vertical dotted lines in panels (a) and (b) indicate the separation between ``high-mass'' and ``intermediate-mass'' sources for the GC09 sample.  The detection fraction is shown in the lower panel, with $\sqrt{N}$ uncertainties in the bin totals indicated.
\label{fig:ysohist}}
\end{figure*}

\subsection{Comparison with YSO Catalogs}

Infrared fluxes and colors of sources in the LMC derived from Two Micron All-Sky Survey (2MASS) and {\it Spitzer Space Telescope} observations have been used by \citet{Whitney:08} (hereafter W08) and \citet{Gruendl:09} (hereafter GC09) to obtain lists of candidate young stellar objects (YSOs).  
Being less sensitive to dust extinction, these observations likely trace star formation at earlier stages than optical surveys of \HII\ regions and O \& B stars.  
However, the masses and ages of the YSOs are highly uncertain, as they rely on comparisons with models that are sensitive to the dust geometry and are primarily calibrated on Galactic data \citep{Robitaille:07}.  
In addition, YSOs can have similar broadband colors to asymptotic giant branch (AGB) stars, planetary nebulae, and background star-forming galaxies, which will therefore tend to contaminate the candidate lists.  
W08 and GC09 used different sets of criteria to identify YSOs and reject likely contaminants, and thus their candidate lists, although of similar length, differ substantially.

We examined both the W08 list of candidate YSOs (1197 sources) and the GC09 list of ``definite'' and ``probable'' YSOs (1172 sources) for spatial coincidence with CO emission.  
In practice, this was done by assuming the YSO position was precisely known and examining the overlapping pixel in the CO map for significant emission.
The results are summarized in Table~\ref{tbl:yso}. 
We used both the NANTEN and Mopra maps for identifying coincident CO emission since the NANTEN map is more spatially complete whereas the Mopra map offers better angular resolution.  
Figure~\ref{fig:magmayso} shows the locations of the GC09 YSOs relative to the MAGMA CO clouds, with red symbols denoting luminous sources (with magnitudes [8.0]$<$8.0).
It is clear that the GC09 sources show a high likelihood of being associated with CO emission---especially the luminous sources, a large fraction of which (97\%) were subsequently confirmed to have YSO-like spectra by \citet{Seale:09}.  
In particular, detection rates in the NANTEN survey are 58\% for the entire GC09 sample and 78\% for the luminous subsample, compared to 32\% for the entire W08 sample and 37\% for the luminous subsample.
On the other hand, the GC09 list is likely to be biased against YSOs that lie outside of GMCs, since the criteria for identifying YSOs may favor deeply embedded objects and takes into account proximity to NANTEN-detected GMCs (although the latter is not generally the determining factor).  
Thus, while the GC09 list may contain a larger fraction of {\it bona fide} YSOs, the W08 list may contain a more representative sample of YSOs across a range of evolutionary stages (but see discussion in GC09 regarding possible contaminants in the W08 sample).

CO detection rates when using the MAGMA image are significantly higher than for the NANTEN CO image, which is not surprising since the MAGMA observation region is biased towards bright CO emission: the fraction of map pixels with detected CO emission is 18\% for MAGMA compared to 9\% for NANTEN.  It is noteworthy, nonetheless, that very few luminous YSOs are found to coincide with CO at the NANTEN resolution while being unassociated with CO at the improved MAGMA resolution.  Only 8\% (15/191) of the luminous YSOs meet this criterion, whereas the corresponding fraction is 15\% (99/669) for the GC09 YSOs as a whole.  Given the coarser resolution of the NANTEN data and its better sensitivity to low-brightness emission, this suggests that more luminous YSOs tend to be confined within the brightest CO emission whereas less-luminous YSOs are distributed both within and around the brightest CO emission.

Figures~\ref{fig:ysohist}(a)--(d) show the flux distributions of YSOs in the NANTEN and MAGMA observation areas and how the probability of CO detection varies with 8.0$\mu$m magnitude.  
We began by selecting, for the W08 sample, only those sources which are labeled as ``highly probable'' YSOs and for which [8.0] magnitudes were measured (932 sources); for the GC09 sample all ``definite'' and ``probable'' YSOs were considered.  
We then eliminated sources that fell outside the NANTEN or MAGMA observation regions and classified the remainder as CO detections or non-detections using the appropriate signal detection masks from CPROPS (these correspond to the areas covered by CO islands before extrapolation).
The most luminous sources, towards the left of the histograms, show the highest rate of CO detection, consistent with extreme youth and formation within GMCs.
The peaks in the distributions near magnitude 9 and falloff at fainter magnitudes suggests substantial incompleteness in the YSO catalogs for magnitudes [8.0]$>$8.0, which may explain the high fraction of CO islands that do not harbor any YSO candidates (62\% in the case of the MAGMA survey, when compared to the GC09 catalog).

\begin{figure*}
\begin{center}
\includegraphics[width=0.4\textwidth]{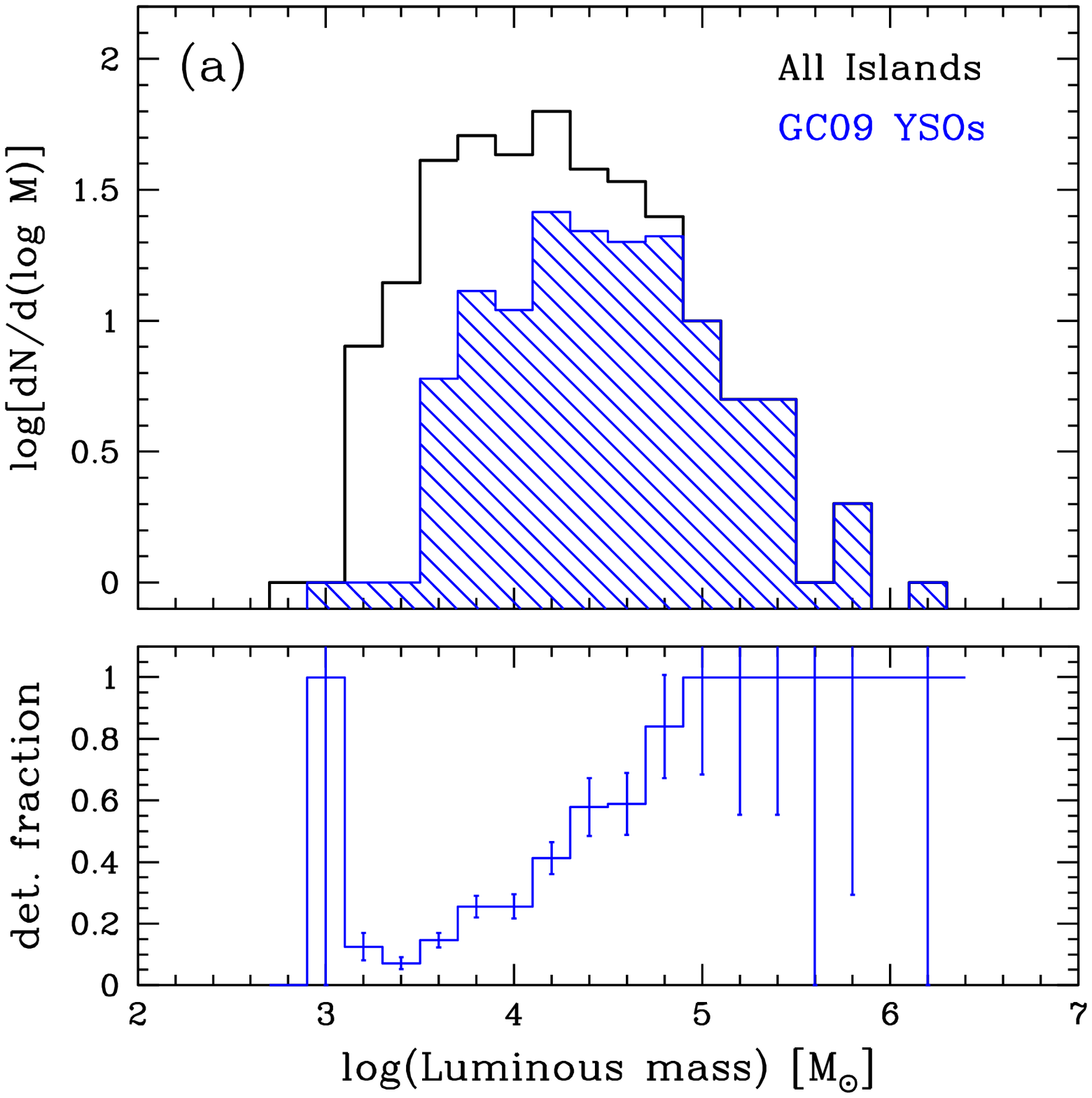}\quad\quad
\includegraphics[width=0.4\textwidth]{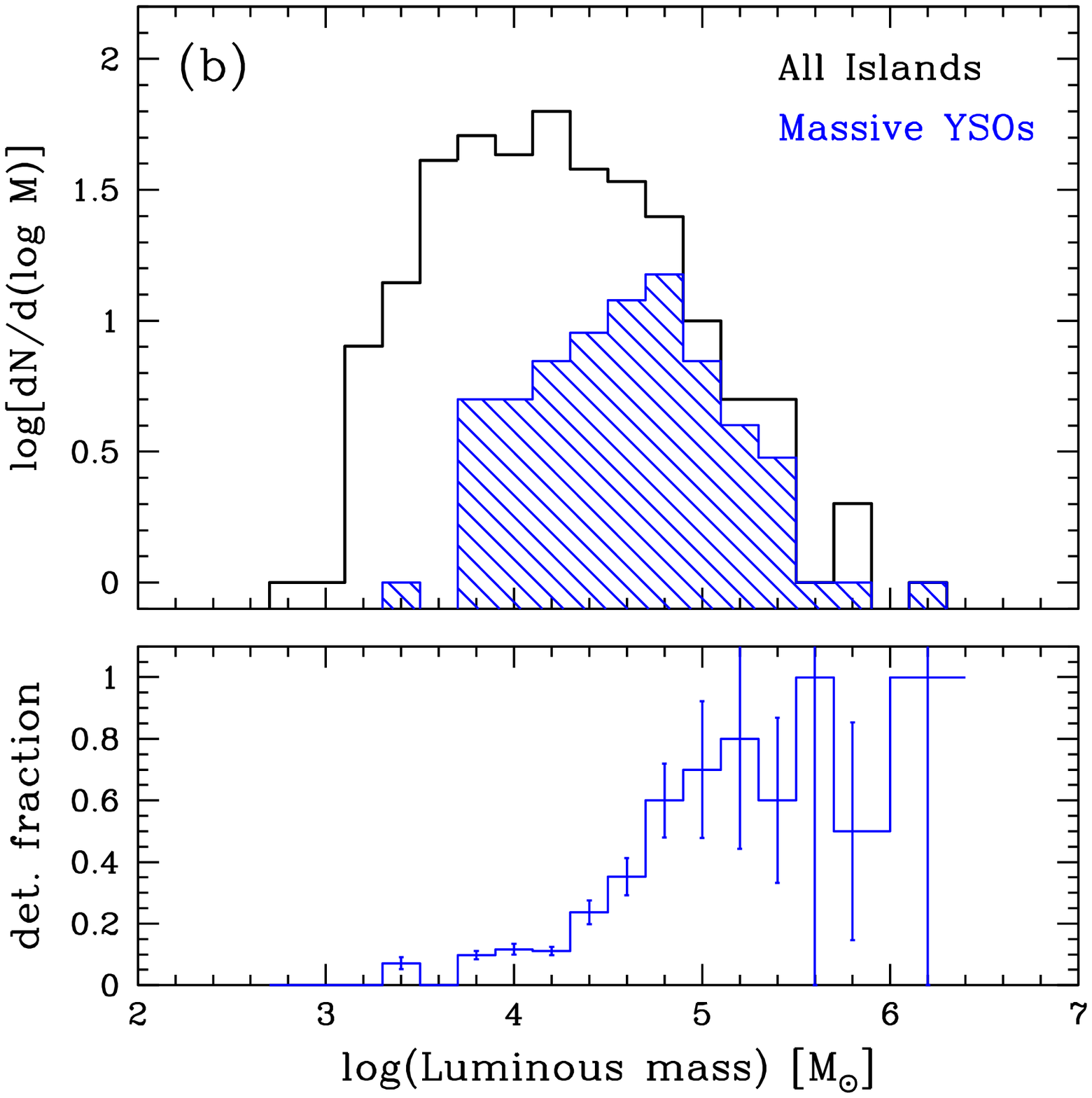}\\[4ex]
\includegraphics[width=0.4\textwidth]{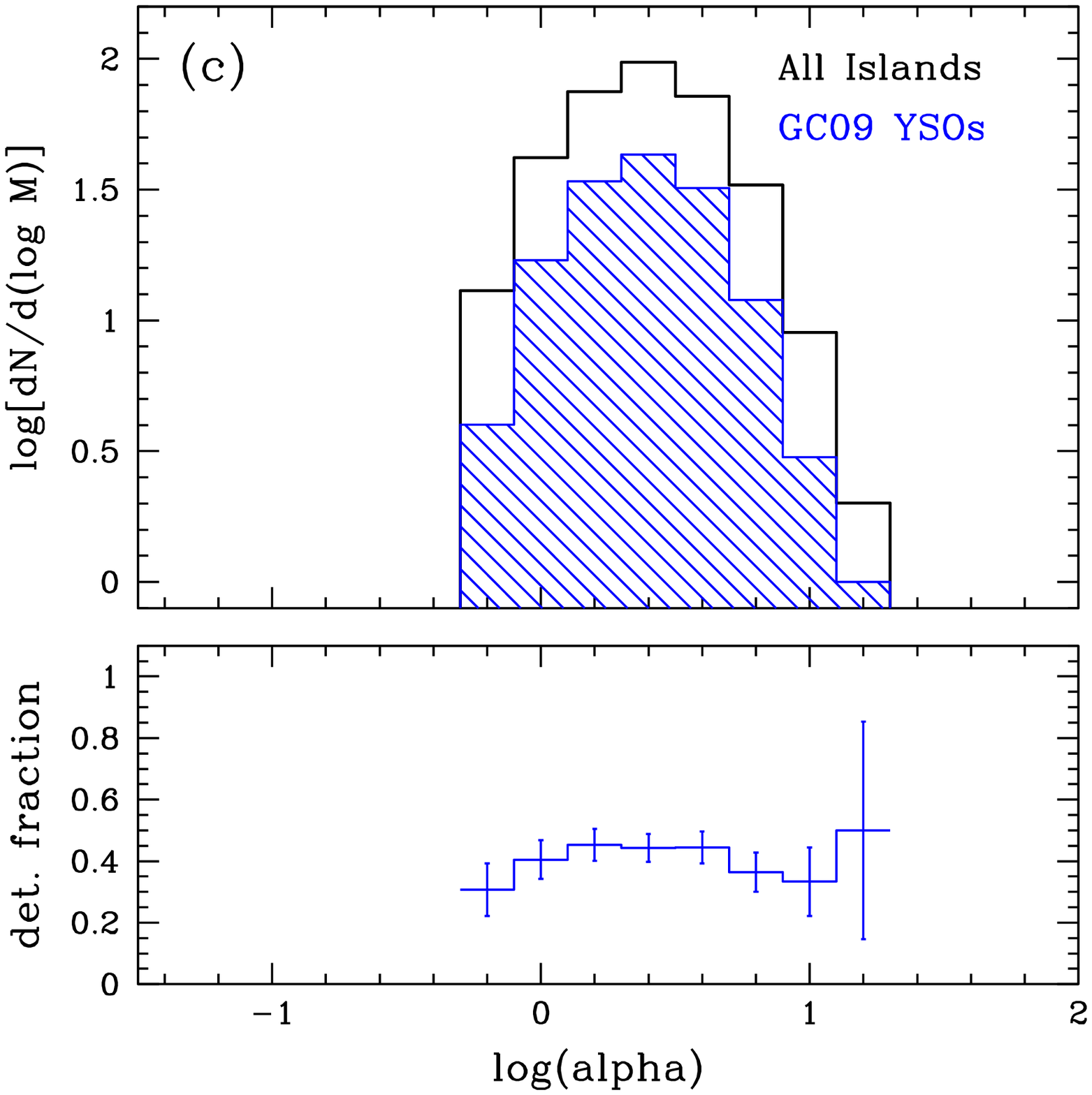}\quad\quad
\includegraphics[width=0.4\textwidth]{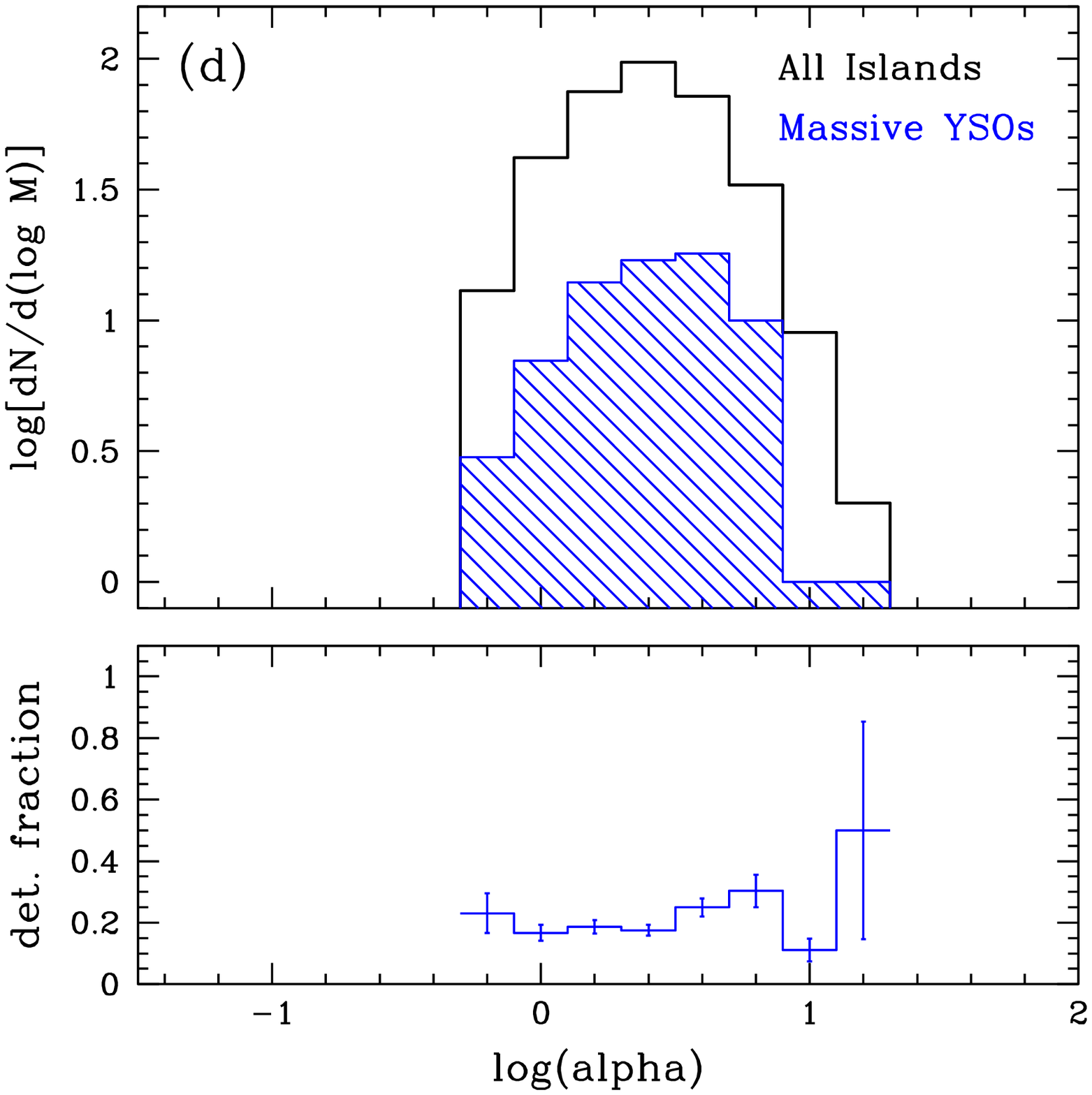}\\
\end{center}
\caption{
Distribution of islands containing (a) YSOs as a function of CO-derived mass; (b) luminous YSOs as a function of CO-derived cloud mass; (c) YSOs as a function of virial parameter; (d) luminous YSOs as a function of virial parameter.  The upper panels show the distribution of values while the lower panels show the fraction of islands containing YSOs, with $\sqrt{N}$ uncertainties in the bin totals indicated.
\label{fig:ysomlum}}
\end{figure*}

Figures~\ref{fig:ysomlum}(a)--(d) show how the probability that a GMC contains a YSO depends on GMC luminosity and virial parameter.  There is a monotonic increase in the probability of containing a YSO at higher CO luminosities.  The probability of containing a luminous YSO ([8.0]$<$8.0 mag) also increases with $M_{\rm lum}$.  On the other hand, the probability of hosting a YSO is uncorrelated with the virial parameter $\alpha_{\rm vir}$.  This raises doubts about the ability of $\alpha_{\rm vir}$ to diagnose whether a cloud can collapse to form stars.  On the other hand, clouds hosting detectable YSOs may be already undergoing collapse motions and/or feedback effects that could modify their line widths (and hence virial parameters).

\section{Discussion}\label{sec:disc}

\subsection{The GMC Mass Spectrum}

The mass spectrum of GMCs in the inner Galaxy has been estimated to have a slope of $\beta = -1.5$ \citep{Solomon:87,Rosolowsky:05b}.  A similar slope has been found for M31 \citep{Rosolowsky:07,Blitz:07}, whereas a much steeper slope ($\beta = -2.6$) has been determined for M33 \citep{Engargiola:03}.  In the LMC we find slopes ranging from $\beta = -2.3$ to $-2.9$ for the luminous mass spectrum and $\beta = -2.0$ to $-2.6$ for the virial mass spectrum, becoming steeper (more negative) as one approaches ``data-based'' decomposition parameters.  If a steep power-law slope continues below our completeness limit, then a significant or even dominant fraction of the molecular mass may exist in structures that we cannot detect individually.  Even the fraction of the total luminous mass in clouds found in {\it identified} structures below the adopted completeness limit is substantial ($\gtrsim$20\%; see Table~\ref{tbl:cprops_summ}), especially for data-based parameters.
On the other hand, placing too much CO flux in unidentified clouds that fall below the completeness limit would violate the integrated flux measurements presented in Section \ref{sec:nanten}, which are a factor of $\sim$2 larger than the extrapolated flux in the identified MAGMA clouds (Table~\ref{tbl:coflux}).

For comparison, inferring mass for the NANTEN GMCs from CO luminosity, Fu08 measured a slope of $\beta = -1.75$ using a completeness limit of $1.4 \times 10^4$ \Msol\ (corrected to our assumed \xco-factor).  As noted by Fu08, the fitted mass spectrum is quite sensitive to the adopted completeness limit, becoming steeper for higher values of $M_{\rm min}$.  The fact that we find a steeper slope using a higher completeness limit of $M_{\rm min} = 3 \times 10^4$ \Msol\ is consistent with this trend; if we reduce $M_{\rm min}$ to $10^4$ \Msol\ then the slope of the luminous mass spectrum for islands flattens to $-2.0$, more similar to the Fu08 result.

In addition, as noted by \citet{Rosolowsky:05b}, blending effects will tend to flatten the observed mass spectrum, as several less massive clouds can appear to be a single massive cloud.  The degree of blending depends on both angular resolution (which determines whether two objects to be separated) and sensitivity (which determines whether two objects can be merged at a lower contour level), as well as being sensitive to the decomposition technique employed.  The MAGMA maps have better resolution but somewhat poorer brightness sensitivity than the NANTEN maps.  Both effects will tend to separate objects which appear blended in the NANTEN maps, steepening the mass spectrum.  Of course, even the MAGMA data likely suffer from some residual blending: Figure~\ref{fig:mlcumiso}(a) suggests a steeper power-law slope ($\beta \approx -2.6$ rather than $-2.3$) when islands that connect with other islands at the 1.5$\sigma$ contour level are excluded from the fit.

Even stronger variations in the mass spectrum result when islands are decomposed into substructures.  Not performing decomposition always yields the flattest spectra, while more aggressive decomposition tends to depopulate the high-mass end and steepen the spectra.  There is little doubt that the size spectrum following decomposition bears the strong imprint of the angular resolution of the observations (Figure~\ref{fig:rhist}), and thus the mass spectrum would be expected to as well.  Unless decomposition parameters can be chosen that are tuned to size scales that are both physically meaningful (e.g., related to a Jeans mass) and well-resolved, it seems doubtful that any decomposition method can produce mass spectra that are free from observational biases, and the spectrum of ``islands'' may thus prove to be the more useful diagnostic.

We therefore concur with previous authors \citep[e.g.,][]{Sheth:08,Reid:10} that comparisons of mass spectra obtained with different techniques and from different observational data sets must be made with caution.  The number and properties of the structures identified depend on the dynamic range (both spatial and intensity) of the observations, and deciding which substructures should be considered distinct introduces additional ambiguity.  Choosing an appropriate completeness limit is also critically important---too low a completeness limit causes low-mass clouds to be missed, while too high a limit leads to sampling and truncation effects at the high-mass end.  Yet, in the presence of blending, any claim to have constructed a complete sample of clouds above a certain mass is open to question.  Fortunately, the low incidence of cloud overlap suggests that line-of-sight blending is not a severe issue for molecular gas in the LMC, so future maps with improved resolution and sensitivity can be expected to minimize blending.  In addition, the use of hierarchical decomposition methods \citep[e.g.][]{Rosolowsky:08,Kauffmann:10a} will help to characterize the observed structures more robustly.

We note that the distribution of column densities, as shown in Figure~\ref{fig:nhist}, can be derived without resorting to a cloud decomposition, although it is still very dependent on angular resolution: at a scale of 11 pc, we are likely smoothing out small-scale fluctuations.  A lognormal distribution of volume densities is expected for isothermal turbulence with a given Mach number $M$, as a result of successive density jumps $\rho_1/\rho_0 \propto M^2$ produced by shocks \citep{Passot:98}.  This may translate into a lognormal distribution of {\it column} densities if the the correlation length for the density field is comparable to the cloud size, so the largest fluctuations occur on the largest scales \citep{Vazquez:01}.  Detailed mapping of Galactic clouds, however, has revealed evidence for departures from the lognormal distribution in star-forming clouds, with a power-law tail appearing at large column densities \citep{Kainulainen:09,Froebrich:10}.  In contrast, non-star-forming clouds appear to lack these power-law tails.  We see no power-law excess in our CO-based column density distribution, suggesting that the scales probed by our observations are still dominated by turbulence.  In fact, the column density appears to be truncated at the high end---a possible indication of opacity effects in the CO line.  We plan to investigate this issue in further detail using the $^{13}$CO MAGMA data (Ott et al., in preparation).

\subsection{The Larson Scaling Relations}

Correlations between size, density, and line width for molecular clouds were discussed in a seminal paper by \citet{Larson:81}, and have been come to be known as ``Larson's laws.''  One of these laws, a roughly linear relationship between density and size, implies a constant column density, and may be an artifact of the limited range of column densities that CO observations are sensitive to, coupled with the tendency of most lines of sight to sample gas at column densities near the observational sensitivity limit \citep[e.g.,][]{Scalo:90,Ballesteros:02}.  Consistent with these findings, we find the slope of the $L_{\rm CO}$--$R$ relation to lie fairly close to our estimated sensitivity limit [Figure~\ref{fig:rml}(a)], with a slope for the islands decomposition set that is only slightly steeper than the constant column density slope of 2.  Further decomposition tends to remove the large $R$ clouds and increase the scatter in CO surface brightness, leading to a steeper fitted relation [Figure~\ref{fig:rml}(b)--(c)].

On the other hand, the size-linewidth relation identified by \citet{Larson:81} is not subject to obvious observational biases and has frequently been discussed as a real property of interstellar turbulence.  A relation of the form $\sigma_v \propto R^{0.5}$ was derived by \citet{Solomon:87} for the inner Galaxy, and a similar relation with slope of $0.60 \pm 0.10$ was derived by \citet{Bolatto:08} for extragalactic clouds.  Both studies have found the normalization of the size-linewidth relation, in combination with the mass surface density estimated from CO observations, to be consistent with gravitational equilibrium ($M \approx 5\sigma_v^2R/G$).  This has been taken as strong evidence that molecular clouds are gravitationally bound, although this situation does not apply to lower mass clouds in the outer Galaxy \citep{Heyer:01}.

For the MAGMA islands we fit a size-linewidth relation of $\sigma_v \propto R^{0.80 \pm 0.05}$ (Figure~\ref{fig:rdv}), significantly steeper than previous studies.  Nonetheless there is considerable scatter around the mean relationship: in fact many of the MAGMA islands are consistent with the \citet{Bolatto:08} relation, although the majority have smaller linewidths than would be predicted for their size.  Decomposition of the islands leads to a further steepening of the relationship, as large clouds are preferentially eliminated from the right side of the diagram.  In addition, we find a greatly increased scatter in the size-linewidth relation: the decomposed substructures often scatter well away from the mean relationship.  At face value this seems inconsistent with the relation having its origin in a turbulent cascade spanning a wide range of scales \citep{Maclow:04}.  On the other hand, a CO-based decomposition may fail to properly separate structures in three-dimensional space that appear superposed in velocity, creating additional scatter in the relation \citep[see discussion in][]{Ballesteros:02}.  While distinct velocity components along the line of sight are rarely seen in our data, blended velocity profiles would be much more difficult to distinguish.


Figure~\ref{fig:mlmv} shows a nearly linear relationship between CO luminosity and virial mass, as has been noted in previous studies \citep{Solomon:87,Bolatto:08}.  Although our virial masses tend to be higher than our luminous CO masses, this may be due to our choice of a Galactic value for \xco.  Doubling the value of \xco\ to $4 \times 10^{20}$ \xcou\ would place the two mass scales in reasonable agreeement.  We are hesitant to use the virial mass to derive a value for \xco, given that $M_{\rm vir} \propto \sigma_v^2R$ and $M_{\rm lum} \propto \sigma_v R^2 T_{\rm CO}$, so for a modest range in CO temperature there is an intrinsic correlation between $M_{\rm vir}$ and $M_{\rm lum}$.  The scatter in the $M_{\rm vir}$--$M_{\rm lum}$ relationship, also apparent as a variation in the virial parameter (Figure~\ref{fig:alpha}), is quite large, and appears to correlate with the scatter in the size-linewidth relation, consistent with the $M_{\rm vir}$--$M_{\rm lum}$ relation arising algebraically, as originally discussed by \citet{Maloney:90}.  Moreover, Figure~\ref{fig:alpha} does not reveal a trend that more massive clouds are more strongly bound, as had been found by \citet{Heyer:01}.

A recent study by \citet{Heyer:09} noted a correlation between $\sigma_v/R^{0.5}$ and mass surface density $\Sigma$ in the Boston University-FCRAO Galactic Ring Survey, with a slope that appears consistent with that expected for gravitationally bound clouds, $\sigma_v/R^{0.5} = (\pi G/5)^{1/2}\Sigma^{1/2}$.  While such a correlation arises trivially when using virial masses to derive $\Sigma$, the fact that \citet{Heyer:09} use masses from an LTE analysis of $^{13}$CO leads them to argue in favor of clouds being gravitationally bound.  However, not unlike the correlation between $M_{\rm vir}$ and $M_{\rm lum}$ found in this paper, a correlation between $\sigma_v/R^{0.5}$ and $\Sigma$ arises naturally because both quantities involve $\sigma_v/R$.  The use of dust emission or extinction to estimate GMC surface densities may circumvent this issue.  While difficult within the Galactic plane due to line-of-sight projection, dust-based approaches applied to the LMC will provide an important test of whether the slope {\it and} normalization of $M_{\rm vir}$--$M_{\rm lum}$ correlation are consistent with virial equilibrium.

\subsection{Star Formation in LMC GMCs}

It is well known that the correlation between CO emission and star formation tracers such as H$\alpha$ and radio continuum emission is rather poor in the LMC \citep[e.g.,][]{Hughes:06}.  Figure~\ref{fig:magmaha} confirms that many luminous \HII\ regions show little or no CO emission.  On the other hand, Figure~\ref{fig:magmayso}, along with the detection statistics presented in Figures~\ref{fig:ysohist} and \ref{fig:ysomlum}, indicate that obscured star formation, as traced by IR-bright YSOs, is spatially well-correlated with CO emission.  This suggests that the differences between H$\alpha$ and CO distributions can be attributed to feedback effects that lead to a rapid dispersal of GMCs during the lifetime of the ionizing O \& B stars ($\sim$10 Myr).  We see possible indications for such evolution in the comparison with GC09 YSOs in Figures~\ref{fig:ysohist}(a)--(b): less luminous (and thus likely longer-lived) YSOs are less likely to be associated with CO emission than more luminous YSOs.  A similar trend is seen for the W08 sources, although with lower overall CO detection rates.  However, this interpretation is not unique, as the less luminous sources may simply lie in clouds that are too weak to be detected in CO at our current sensitivity.  Deeper CO observations are being pursued to investigate this possibility.

Recently, \citet{Kawamura:09} compared the locations of CO clouds with optical catalogs of young stellar clusters, finding that $\sim$2/3 of the $\lesssim$10 Myr old clusters are associated with GMCs, the remainder having apparently dispersed their natal clouds.  This implies a typical duration of $\sim$7 Myr for a GMC to host young clusters (although the GMC lifetime may be longer if GMCs go through non-cluster-bearing stages).
Assuming the YSOs detected by {\it Spitzer} are much younger than 7 Myr, it is thus not surprising that we find nearly all of the luminous GC09 YSOs lying within CO clouds.  
Importantly, we do not see evidence that any {\it massive} GMCs lack YSOs, arguing against a long gestation period before star formation commences (although the specific rate of star formation may vary among GMCs).  At present, the limited sensitivity of the CO data and the poor characterization of YSO ages and masses limit our ability to strongly constrain the GMC lifetime, but further work along these lines clearly holds tremendous promise for constraining GMC lifetimes, thus shedding light on the question of whether GMCs are able to achieve gravitational equilibrium.  


\section{Conclusions}\label{sec:conc}

We have presented in this paper a new high-resolution imaging survey of CO in the LMC, representing a principal outcome of the Magellanic Mopra Assessment (MAGMA).  The survey targets include the most luminous GMCs detected in the second NANTEN survey \citep{Fukui:08}.  We present three separate catalogs of CO emission structures identified using the automated CPROPS algorithm \citep{Rosolowsky:06}.  The ``islands'' catalog selects out all contiguous emission structures, the ``physical'' catalog decomposes the islands into structures designed to resemble Galactic GMCs, and the ``data-based'' catalog decomposes the islands into structures to nearly the resolution limit of the survey.  Based on an analysis of these catalogs, we draw the following conclusions.

\begin{enumerate}

\item CO ``islands'' show a steeper mass spectrum than inferred from previous studies of GMCs in the Galaxy \citep{Solomon:87} and of GMCs in the LMC using the NANTEN survey.  The spectral slope of $\lesssim -2$ suggests a significant amount of mass in low-luminosity clouds.  The discrepancy with the earlier NANTEN result may reflect uncertainty in the appropriate completeness limit above which to fit the spectrum, as well as the impact of map resolution and sensitivity in breaking apart blended structures into smaller units.  We caution that the mass spectrum for decomposed emission is very sensitive to the chosen decomposition parameters.

\item CO ``islands'' show a roughly constant CO surface brightness, as seen in Galactic clouds, but a somewhat steeper dependence of linewidth on radius.  Decomposition of the islands increases the scatter in both relations as large clouds are broken into smaller clouds spanning a wide range in linewidth and surface brightness.

\item The overall distribution of CO column densities in the MAGMA field is roughly consistent with a lognormal distribution, as expected for isothermal turbulence, although the distribution may be cut off at the high end due to saturation of the CO line.  However, these column densities are measured over much larger scales (11 pc) than analogous measurements for Galactic molecular clouds.

\item The nearly linear correlation between virial mass and CO luminosity may arise trivially because both quantities scale with the product of linewidth and size: virial mass scales as $\sigma_v^2R$ and luminosity as $\sigma_v R^2 T_{\rm CO}$.  The correlation is tightest on scales where the size-linewidth relation is tightest, as expected.  There is no tendency for luminous clouds to show smaller values of the virial parameter which would indicate an increasing dominance of self-gravity.

\item We find an increased likelihood for more massive GMCs to contain YSO candidates, and an increased likelihood for more luminous YSO candidates to be associated with CO emission.  These trends confirm the close link between giant molecular clouds (as traced by CO emission) and massive star formation: one is rarely seen without the other.  However, the likelihood of a GMC containing a YSO appears unrelated to the virial parameter.

\item We speculate that low-luminosity YSOs may be outliving their natal GMCs, based on their lower association with CO and tendency to be offset from the brightest CO emission.  Deeper CO observations and more reliable estimates for YSO masses and ages will be needed to confirm this interpretation.

\end{enumerate}

Future papers in this series will investigate the velocity gradients across GMCs, the relationship between atomic and molecular gas, and the optical depth of the CO emission (using the $^{13}$CO data obtained in parallel).  We will also perform a comparative analysis of GMCs in the SMC.\@  We expect the MAGMA CO maps to have significant legacy value until more sensitive and spatially complete mapping surveys are possible with multi-pixel receivers, and are releasing the data products on web sites hosted by the University of Illinois and CSIRO.

\acknowledgments

We are deeply indebted to the ATNF for the generous allocation of time for this project and for assistance with planning and executing the project over its duration of several years.  
Particular thanks go to Michael Kesteven and Mark Calabretta for making the OTF mode at Mopra possible, and to Balthasar Indermuehle for assistance with Mopra.
TW thanks Ned Ladd for assistance with developing the OTF mode.  
Lister Staveley-Smith and Sungeun Kim furnished the \HI\ map of the LMC that proved valuable for planning the observations.  
Erik Rosolowsky, Adam Leroy, and Alberto Bolatto provided useful advice and suggestions on the identification of clouds.  
We also benefitted from stimulating discussions with Remy Indebetouw and Rosie Chen.
We thank the anonymous referee for a number of helpful suggestions.
Research by TW was supported by NSF grant 08-07323, the University of Illinois, and NASA grant 10-ADAP10-0137.
This research has been carried out in part at the Jet Propulsion Laboratory, California Institute of Technology.
AK acknowledges support from the Japan Society for the Promotion of Science (KAKENHI No.\ 22540250).


\bibliographystyle{apj}
\bibliography{merged}

\end{document}